\pdfoutput=1
\documentclass[review]{elsarticle}
\usepackage{lineno}
\modulolinenumbers[1]
\journal{Physics Reports}

\usepackage{graphicx}  % Include figure files
\usepackage{amssymb}
\usepackage{amsmath}
\usepackage{dcolumn}  % Align table columns on decimal point
\usepackage{bm}          % bold math
\usepackage{color}

\newcommand{\hla}{$\mathrm{^{3}_{\Lambda}H}$}
\newcommand{\hala}{$\mathrm{^{3}_{\bar{\Lambda}}\overline{{H}}}$}
\newcommand{\sNN}{$\sqrt{s_{\rm NN}}$ }
\newcommand{\he}{$\mathrm{^{3}He}$}
\newcommand{\ahe}{$\mathrm{^{3}\overline{{He}}}$}

\newcommand{\dbar}{$\overline{d}$}

\begin{document}
\begin{frontmatter}

\title{
Antinuclei in Heavy-Ion Collisions
}

\author[SINAP]{Jinhui Chen}
\author[KSU]{Declan Keane\corref{cor1}}
\author[SINAP,UCAS]{Yu-Gang Ma\corref{cor2}}
\author[BNL]{Aihong Tang}
\author[BNL,SDU]{Zhangbu~Xu}

\address[SINAP]{Shanghai Institute of Applied Physics, Chinese Academy of Sciences, Shanghai 201800, China}
%\address[HNU]{College of Physics and Material Sciences, Henan Normal University, Xinxiang 453007, China}
\address[KSU]{Kent State University, Kent, Ohio 44242, USA}
\address[UCAS]{University of Chinese Academy of Sciences, Beijing 100049, China}
\address[BNL]{Brookhaven National Laboratory, Upton, New York 11973, USA}
\address[SDU]{Shandong University, Jinan, Shandong 250100, China}

\cortext[cor1]{Corresponding author, keane@kent.edu}
%\ead{keane@kent.edu}
\cortext[cor2]{Corresponding author, ygma@sinap.ac.cn}
%\ead{ygma@sinap.ac.cn}

\begin{abstract}
We review progress in the study of antinuclei, starting from Dirac's equation and the discovery of the positron in cosmic-ray events. The development of proton accelerators led to the discovery of antiprotons, followed by the first antideuterons, demonstrating that antinucleons bind into antinuclei. With the development of heavy-ion programs at the Brookhaven AGS and CERN SPS, it was demonstrated that central collisions of heavy nuclei offer a fertile ground for research and discoveries in the area of antinuclei. In this review, we emphasize recent observations at Brookhaven's Relativistic Heavy Ion Collider and at CERN's Large Hadron Collider, namely, the antihypertriton and the antihelium-4, as well as measurements of the mass difference between light nuclei and antinuclei, and the interaction between antiprotons. Physics implications of the new observations and different production mechanisms are discussed. We also consider implications for related fields, such as hypernuclear physics and space-based cosmic-ray experiments. \\
\end{abstract}


\end{frontmatter}

%Contents
\tableofcontents

%\linenumbers
%\begin{linenumbers}
\realpagewiselinenumbers
\setlength\linenumbersep{0.10cm}  %% 0.10cm is often but not always the best number.

\section{Historical Introduction}

\subsection{The Dirac Equation
}\label{Dirac}

The earliest reference to antimatter in the scientific literature dates back to the end of the nineteenth century, when Arthur Schuster coined this term in a letter to Nature \cite{Schuster:1898}. However, Schuster's antimatter idea was largely speculative, and differed in significant ways from the modern concept.  Antimatter as it is understood today has its origin in Paul Dirac's seminal 1928 work in which he formulated his relativistic version of the Schr\"odinger equation \cite{Dirac:1928hu, Dirac:1928ej}.  Dirac's original relatively modest goal was to investigate whether relativistic corrections would improve the explanation of spectroscopic measurements.  The Dirac equation for the electron advanced our understanding of fundamental physics far beyond that initial motivation, and as a result of the combination of quantum mechanics and special relativity (notably the relationship between energy ($E$) and momentum ($p$) as reflected in $E = \pm \sqrt{p^2 + m^2}$, where $m$ is rest energy) the concept of a particle with negative energy and opposite electric charge was introduced for the first time. 

Two years later, in 1930, Dirac proposed the interpretation that the holes in the negative-energy electron sea are protons \cite{Dirac:1930ek}. At the time, very few fundamental particles were known, and postulating a new and undiscovered type of elementary particle would have been a very radical hypothesis; moreover, the proton interpretation offered the appealing prospect of a unified theory of the electron and proton. Also in 1930, Dirac advanced the idea that positive-energy electrons and holes in the negative-energy sea could simultaneously disappear and be converted into electromagnetic radiation \cite{Dirac:1930bga}, although this paper still envisaged the holes as protons. 

Meanwhile, Robert Oppenheimer \cite{Oppenheimer:1930} pointed to various difficulties raised by the proton interpretation, and Hermann Weyl \cite{Weyl:1927vd, Weyl:1931} emphasized that the holes cannot be protons and must have the same mass as electrons. By the year 1931, Dirac had laid out an updated and remarkably prescient picture, in which the ``anti-electron" was explicitly named as a particle that was yet to be discovered experimentally and would have the same mass as an electron \cite{Dirac:1931kp}. This paper also anticipated and named the antiproton. 

It is noteworthy the Dirac himself attached considerable importance to the observation of antinuclei. He did not consider the existence of antimatter to be properly verified until it had been experimentally demonstrated that antinucleons could bind together and form antinuclei \cite{Dirac1983, Zichichi:2009zza}.

%%%%%%%%%%%%%% Fig. 1 %%%%%%%%%%%%%%%%%%%
\begin{figure}[!bt]
\centering
\centerline{\includegraphics[scale=0.22, bb=1300 -50 -230 950]{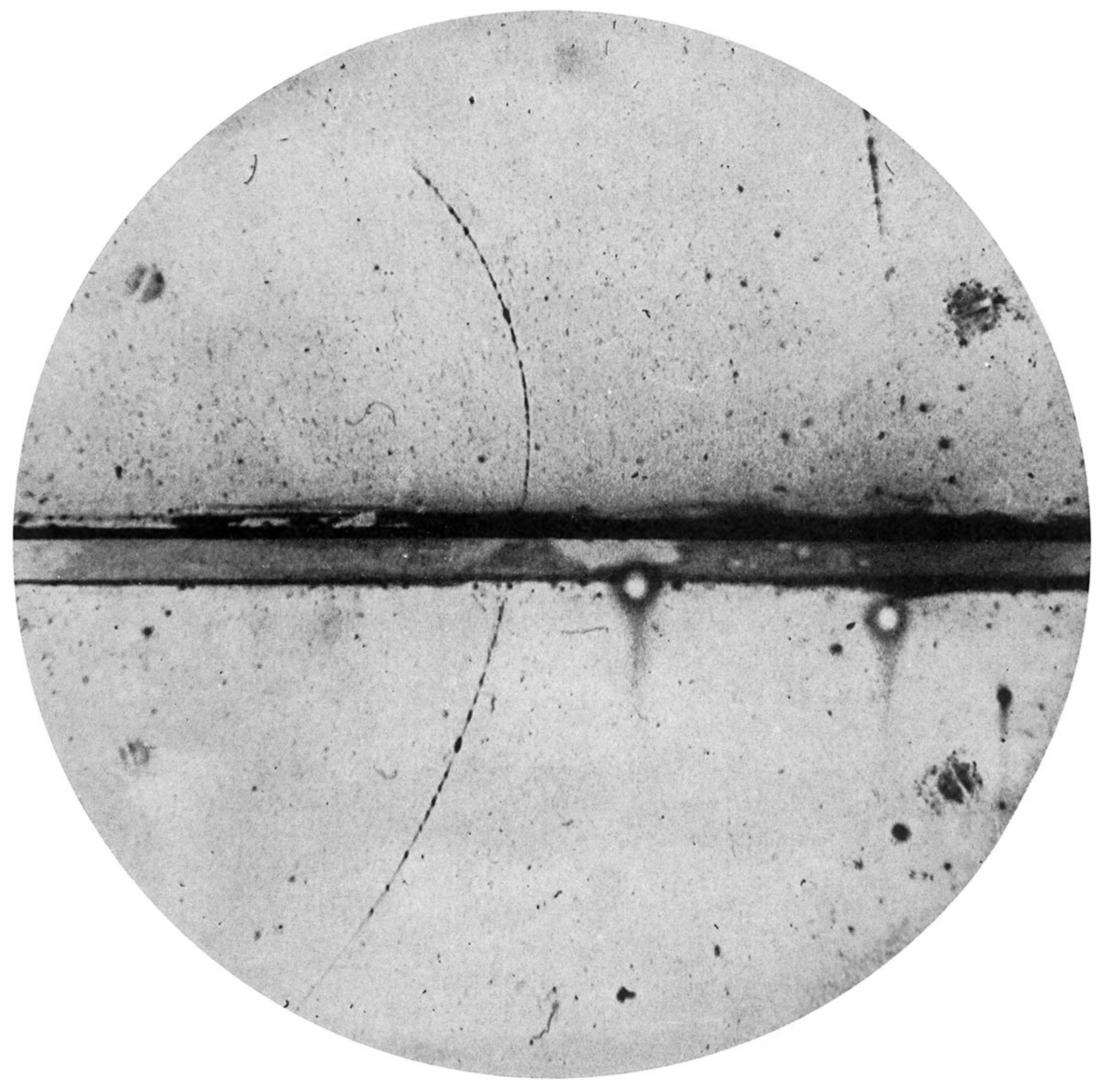}}
\caption{A positively-charged particle, identified as a positron of energy 63 MeV, enters from the bottom, loses energy in a 0.6 cm lead plate, and emerges in the upper region with an energy of 23 MeV \cite{Anderson:1932zz, Anderson:1933mb}. A proton with the observed upper curvature would have a total range in the cloud chamber of $\sim 0.5$ cm, whereas the observed track is consistent with a constant curvature over a distance of more than 5 cm.}
\label{fig:Positron}
\end{figure}
%%%%%%%%%%%%%%%%%%%%%%%%%%%%%%%%%%%%

\subsection{The Experimental Discovery of Positrons and Antiprotons }

Just one year after Dirac's ``fully formed" prediction of the anti-electron in 1931, Carl Anderson's study of cosmic radiation using a cloud chamber led to the experimental discovery of a new positively-charged particle. Its curvature in a 1.5T magnetic field and energy loss in a lead plate, as illustrated in Fig.~\ref{fig:Positron}, could only be understood if its mass was at most 20 times the electron mass and was also consistent with having the same mass \cite{Anderson:1932zz, Anderson:1933mb}. Although Anderson's chamber was triggered at random times, he was able to observe 15 candidates among 1300 chamber photographs. %%He proposed the name positron, and also advocated (without success) that the normal $e^-$ be thereafter called the ``negatron". 
He proposed the name positron, but his two initial publications \cite{Anderson:1932zz, Anderson:1933mb} did not make any reference to Dirac's theoretical predictions. 

Within a few months of Anderson's first paper, Blackett and Occhialini \cite{Blackett:1933} had confirmed his discovery using a cloud chamber in a 0.3T magnetic field. Their paper included extensive discussion of ``Dirac's theory of holes'', including calculations related to the annihilation properties of Dirac's anti-electron.  The next step forward in the investigation of positrons was made possible when it was discovered that they are produced by exposing target materials to various radioactive sources, resulting in production under more controlled conditions than in cosmic ray experiments.  This work led to verification with better precision that the positron and electron masses are equal, as well as a determination that the combined rest and kinetic energies of the $e^+$ and $e^-$ agree within uncertainties with the energy of the incoming photon \cite{Chadwick:1933, Meitner:1933, Curie:1933a, Curie:1933b, Anderson:1933, Anderson:1933zz, Chadwick235, Blackett:1933b}.

%%%%%%%%%%%%%% Fig. 2 %%%%%%%%%%%%%%%%%%%
\begin{figure}[!htb]
\centering
\centerline{\includegraphics[scale=0.40, bb=300 -50 230 450]{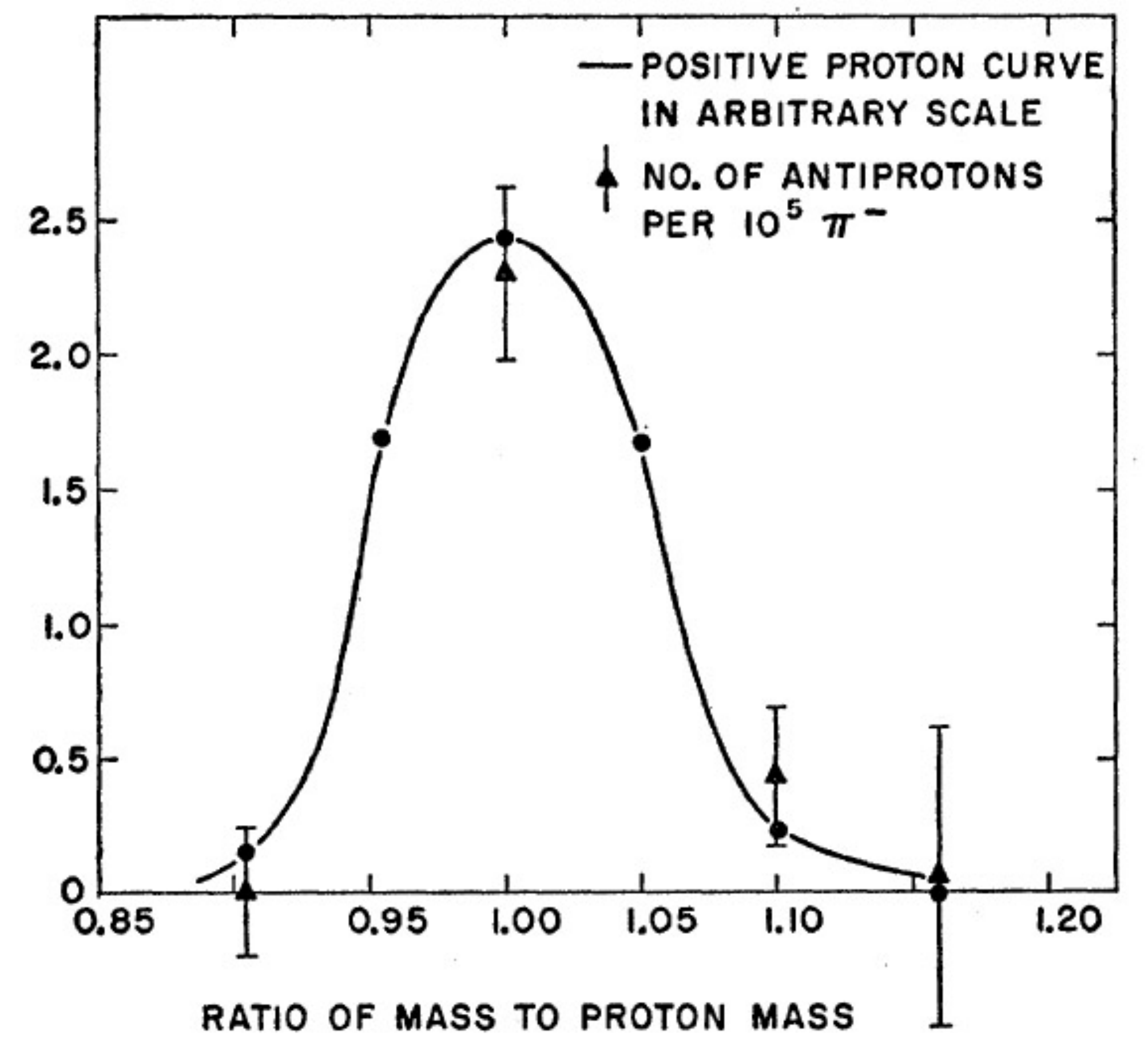}}
\caption{The mass resolution of the magnetic spectrometer of Chamberlain {\it et al.}, demonstrating that the mass of the new negatively-charged particle was within 5\% of the proton mass \cite{Chamberlain:1955ns}. }
\label{fig:antiproton1}
\end{figure}
%%%%%%%%%%%%%%%%%%%%%%%%%%%%%%%%%%%%

By the 1950s, the frontier of experimental research on antimatter had moved away from the domain of cosmic rays and radioactive sources. The first unambiguous observation of antiprotons in cosmic ray experiments did not occur until 1979 \cite{Golden:1979bw}, long after antiprotons were routinely produced at accelerator laboratories. The Bevatron at the University of California Radiation Laboratory (now Berkeley Lab) was designed with the search for antiprotons in mind. It was capable of generating protons of kinetic energy up to 6.5 GeV, a value that exceeded the threshold for antinucleon production in $pp$ collisions. 

The discovery of the antiproton in a magnetic spectrometer experiment at the Bevatron was reported in 1955 by Chamberlain, Segr\`e, Wiegand and Ypsilantis \cite{Chamberlain:1955ns}. The accelerated protons interacted with a copper target located inside the Bevatron ring, with negatively-charged particles being deflected outward by the magnetic field of the accelerator. The main task of the detector system, which consisted of bending and focusing magnets, time-of-flight counters and \v{C}erenkov detectors, was to distinguish antiprotons from negative pions, which were more abundant by a factor on the order of $10^5$. Figure~\ref{fig:antiproton1} illustrates the good mass resolution of the spectrometer of Chamberlain {\it et al.} \cite{Chamberlain:1955ns}, which allowed the high rate of $\pi^-$ background to be rejected. 

\begin{figure}[!htb]
\centering
\centerline{\includegraphics[scale=0.52,bb=420 -50 230 450]{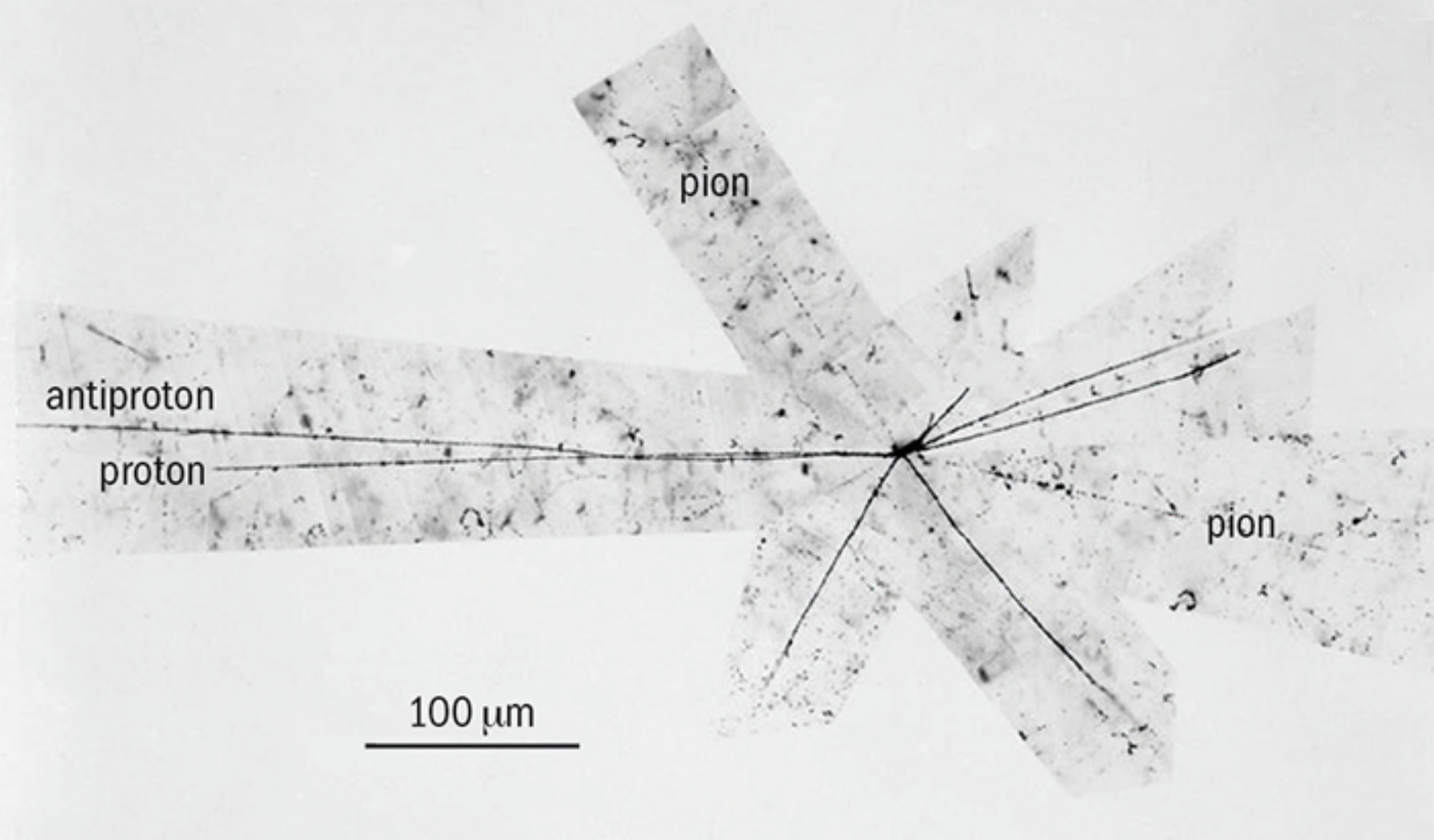}}
\caption{An antiproton annihilation vertex in nuclear emulsion.  In the microscopic image reproduced here, the emitted charged particles include two pions, a proton, and several unidentified charged particles \cite{Chamberlain:1956}.}
\label{fig:antiproton2}
\end{figure}

The original antiproton discovery paper was followed soon afterwards by an independent $\bar{p}$ identification in nuclear emulsion.  In order to maximize the rate of detection and identification of antiproton annihilation vertices in a nuclear emulsion stack, it was necessary to slow-down the antiprotons in a passive absorber.  This increased the probability that antiprotons which did not interact in flight would come to rest inside the emulsion and then annihilate.  In the measurement of Chamberlain {\it et al.} \cite{Chamberlain:1956}, antiprotons with a momentum of 1.09 GeV/$c$, selected by a magnetic spectrometer, traversed a copper absorber of thickness 132 g/cm$^2$, after which they entered and stopped inside the emulsion; see the composite microscopic image of a sample annihilation vertex in Fig.~\ref{fig:antiproton2} \cite{Chamberlain:1956}.  

%%%%%%%%%%%%%% Fig. 3 %%%%%%%%%%%%%%%%%%%
%%%%%%%%%%%%%%%%%%%%%%%%%%%%%%%%%%%%

%%%%%%%%%%%%%% Fig. 4 %%%%%%%%%%%%%%%%%%%
\begin{figure}[!htb]
\centering
\centerline{\includegraphics[scale=0.80,bb=100 -50 230 250]{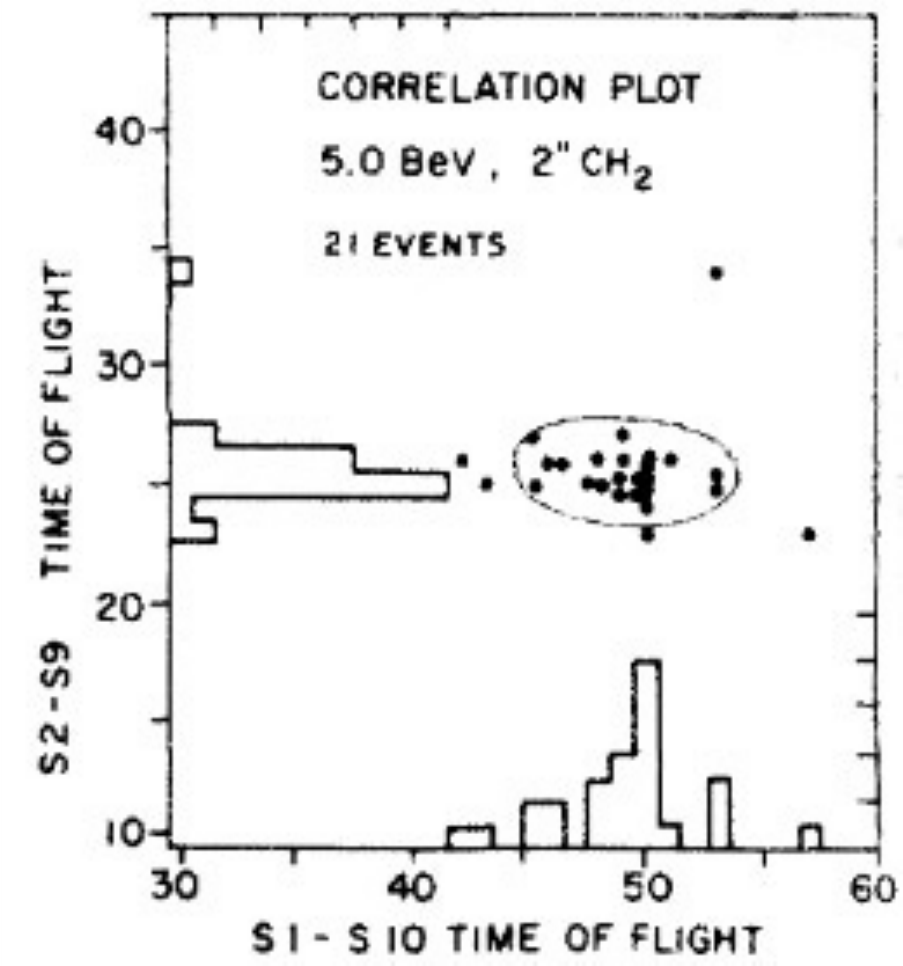}}
\caption{Dorfan {\it et al.} \cite{Dorfan:1965zz, Dorfan:1965uf} observed 21 $p$+Be collisions at 30 GeV in which negatively-charged secondary particles had time-of-flight correlations consistent with these fragments having a mass within 3\% of the deuteron mass.}
\label{fig:antideuteron}
\end{figure}
%%%%%%%%%%%%%%%%%%%%%%%%%%%%%%%%%%%%

\subsection{Early Experimental Discoveries of Antinuclei at Proton Accelerators
}\label{pA}

\begin{figure*}[!htb]
\centering
\centerline{\includegraphics[scale=0.4,bb=800 -50 230 650]{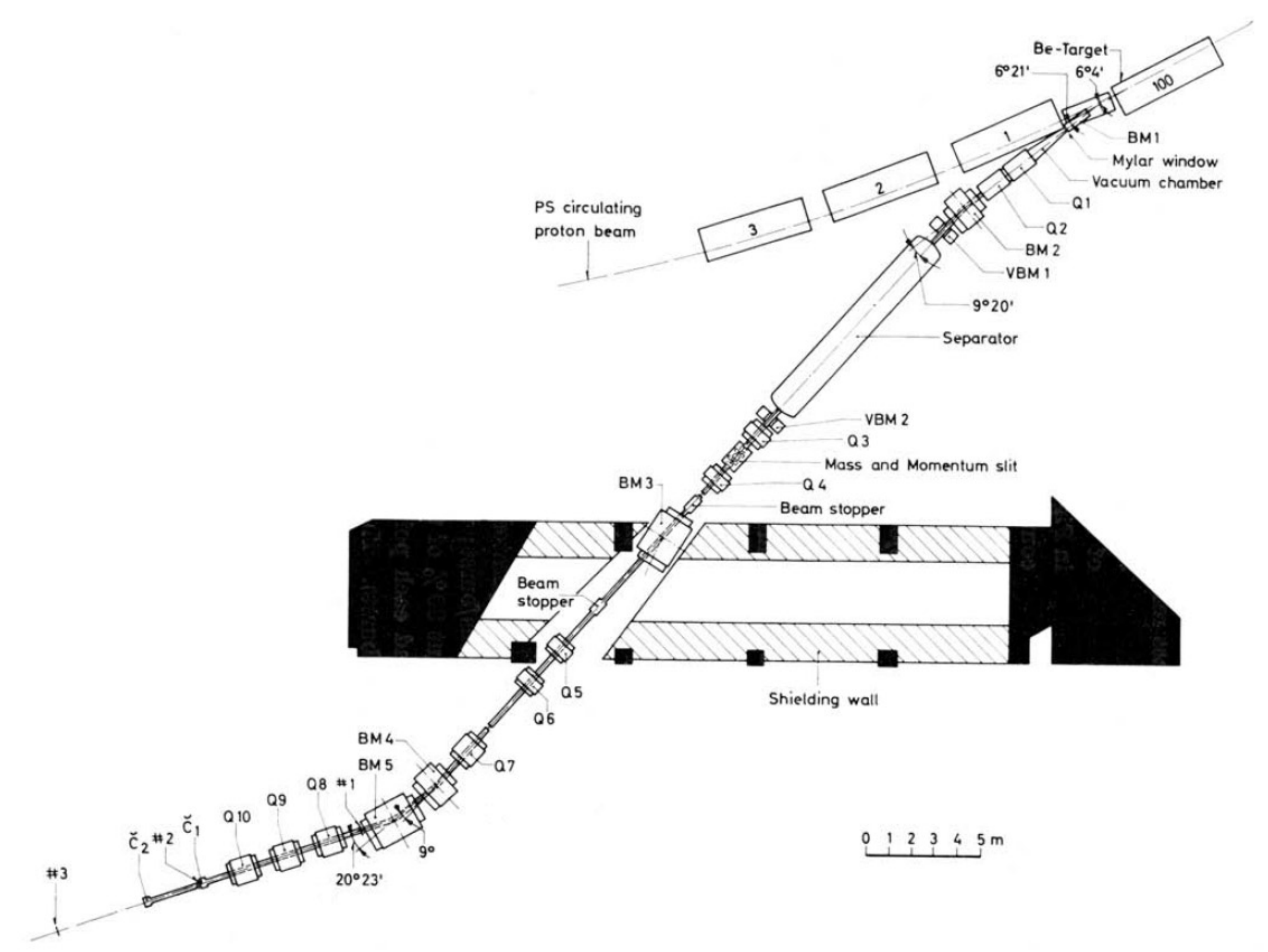}}
\caption{The layout of the 61 meter-long beamline \cite{Brautti1965} at the CERN Proton Synchrotron, used to identify antideuteron candidates in 1965. }
\label{fig:CERN1965}
\end{figure*}

The next fundamental question in the early history of antimatter was whether antinucleons could form antinuclei with properties identical to those of their matter partners.  At the Alternating Gradient Synchrotron (AGS) at Brookhaven National Laboratory, which began operation in 1960 and could accelerate protons to 33 GeV, Dorfan, Eades, Lederman, Lee and Ting used a high-transmission mass-analyzing beamline to search for various new states of positive and negative electric charge \cite{Dorfan:1965zz}.  In 1965, while studying 30 GeV protons on a beryllium target, they observed particles with negative electric charge, and mass consistent with that of the deuteron. In Fig.~\ref{fig:antideuteron}, the plotted time-of-flight correlations demonstrate the detection of 21 antideuteron candidates within the acceptance of the detector.  In a paper received earlier but published later than that of Dorfan {\it et al.}, Massam, Muller, Righini, Schneegans and Zichichi \cite{Massam1965} presented lower-statistics results from a similar $p$+Be experiment \cite{Brautti1965} at the CERN Proton Synchrotron, also reporting an observation of antideuterons. Figure \ref{fig:CERN1965} shows the layout of the beamline spectrometer used by the authors of Refs.~\cite{Massam1965} and \cite{Brautti1965}. 

%%%%%%%%%%%%%% Fig. 5 %%%%%%%%%%%%%%%%%%%
%%%%%%%%%%%%%%%%%%%%%%%%%%%%%%%%%%%%

Models that were used in the early 1960s to calculate the production rate of antideuterons (and heavier antinuclei) had much in common with those used to describe measurements of deuterons (and heavier nuclei) in proton-nucleus collisions \cite{Cocconi:1960zz, Fitch:1962fq, Schwarzschild:1963zz, Amaldi1963, Diddens1964}. These experiments dealt with forward kinematic regions, where composite nuclei could not have originated from multifragmentation of an excited target spectator.  One such model, proposed by Dmitrii Blokhintsev \cite{Blokhintsev:1958}, attributed an important role to density fluctuations in the target nucleus, while Rolf Hagedorn proposed a statistical approach in which a deuteron could be directly produced in an elementary nucleon-nucleon interaction \cite{Hagedorn:1960zz}. Hagedorn subsequently employed this model to calculate the rate of antideuteron production in $p$-nucleus interactions \cite{Hagedorn1962}, but the resulting prediction differed greatly from the observations of Dorfan {\it et al.} \cite{Dorfan:1965zz}, namely, the prediction exceeded the measured rate of antideuteron production by a factor of about $2 \times 10^3$.  Butler and Pearson were the first to propose the idea of statistical coalescence of nucleons produced independently, an approach that was immediately found to offer better agreement with the experimental measurements available at the time \cite{Butler:1961pr, BUTLER196277, Butler:1963pp}.  Statistical coalescence predicts that the invariant momentum-space density $\rho_A$ for deuterons or heavier composite nuclei with mass number $A$ and momentum $Ap$ can be obtained from the nucleon density $\rho_{A=1}$ at momentum $p$ to the power of $A$.  Some contemporary papers referred to this as the ``sticking model", but this terminology has since fallen out of use, and behavior of the type 
\begin{equation}
\rho_A(Ap) \propto \rho_{A=1}^A(p)
\end{equation}
is generally referred to as the coalescence power law in the current literature.  

The threshold kinetic energy in a fixed-target configuration for production of an antinucleon is 5.6 GeV; for two antinucleons, the threshold is 15 GeV, and for three antinucleons, it is 28 GeV.  However, to detect an antinucleus with mass number $A = -3$ in practice involves a search over a large sample of events with a beam energy well above the threshold energy.  For instance, the maximum proton beam energy at the AGS at Brookhaven National Laboratory exceeded the threshold for antitritium production, but Dorfan {\it et al.} reported, with 90\% confidence, that no antitritium candidates were detected down to a level of about $3 \times 10^{-10}$ times the rate of pion production with a 30 GeV proton beam \cite{Dorfan:1965zz}. 

Beginning in the late 1960s, various collaborations reported results of searches for new negatively-charged heavy states using the higher energies of new proton accelerators.  The first of these new facilities was the 70 GeV synchrotron at the Institute of High Energy Physics, Protvino (Serpukhov), USSR, which began operation in 1967.  Next came the CERN Intersecting Storage Rings; the first proton-proton collisions at the ISR took place in 1971, and it initially operated at $\sqrt{s} =$ 53 GeV.  Meanwhile, the US National Accelerator Laboratory (NAL) in Illinois (later renamed Fermilab) accelerated protons to 200 GeV in 1972, and soon afterwards reached 300 GeV. 

The first results from Serpukhov, published in 1969, investigated proton-aluminum collisions at beam energies of 43 and 52 GeV as well as the top energy of 70 GeV \cite{Binon:1969qz, Antipov:1971zs}, and provided greatly improved antideuteron statistics compared with Refs.~\cite{Dorfan:1965zz} and \cite{Massam1965}, including measurement of differential cross sections, but reported only an upper limit for antitritons.  The first ISR results followed in 1973, when Alper {\it et al.} \cite{Alper:1973my} reported a substantial antideuteron signal and an unprecedentedly-low deuteron to antideuteron ratio of $3.7 \pm 1.2$.  However, Alper {\it et al.} found no signal for antitritons or anti-$^3$He.  The initial search for heavy negatives at Fermilab, published in 1974 by Appel {\it et al.} \cite{Appel:1974fs}, used 300 GeV protons on a tungsten target, with a mass-identifying beam line of length 1.1 km.  As in previous experiments, antideuterons were produced at rates that agreed well with the coalescence power law, but again, no antitriton or anti-$^3$He signal was found.

%%%%%%%%%%%%%% Fig. 6 %%%%%%%%%%%%%%%%%%%
\begin{figure}[!hbt]
\centering
\centerline{\includegraphics[scale=0.50,bb=300 -50 230 450]{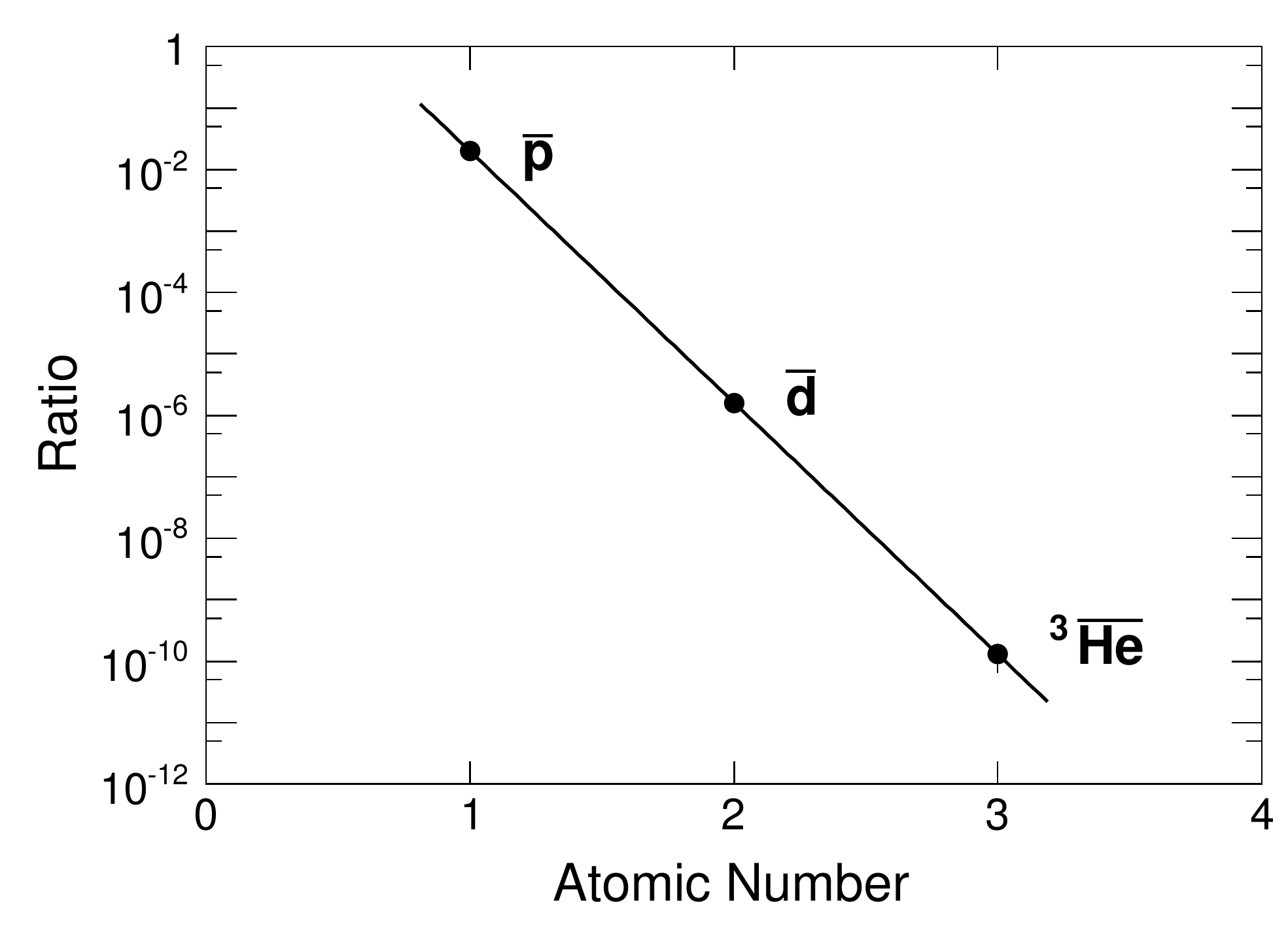}}
\caption{ The ratio of antinucleus to pion production as a function of antinucleus mass number, under the conditions of the experiment of Antipov {\it et al.} \cite{Antipov:1970uc, Antipov:1971iq}. }
\label{fig:antiHe3}
\end{figure}
%%%%%%%%%%%%%%%%%%%%%%%%%%%%%%%%%%%%

The 70 GeV Serpukhov facility reached the milestone of the first detection of antihelium-3 \cite{Antipov:1970uc, Antipov:1971iq} in the year 1970.  The apparatus of Antipov {\it et al.}, which sampled $2.4 \times 10^{11}$ forward-going tracks emitted from $p$ + Al collisions, had similarities to that used for antideuteron discovery at Brookhaven and CERN.  It consisted of a long magnetic spectrometer which used \v{C}erenkov counters, $dE/dx$ and time-of-flight to identify five $^3 \overline{\rm He}$ tracks. The Protvino authors were also the first to draw attention to the exponentially decreasing probability of antinucleus formation as the mass number increases; under the conditions of their detector, as demonstrated in Fig.~\ref{fig:antiHe3}, the probability dropped about 4 orders of magnitude going from antiprotons to antideuterons, and another 4 orders going from antideuterons to antihelium-3 \cite{Antipov:1970uc, Antipov:1971iq}.  This observation supports the statistical coalescence picture first proposed by Butler and Pearson \cite{Butler:1961pr, BUTLER196277, Butler:1963pp}, and at least in the environment of the measurements under discussion, disfavors the above-mentioned alternative approaches of Blokhintsev \cite{Blokhintsev:1958}, and of Hagedorn \cite{Hagedorn:1960zz}.  

In 1974, a different collaboration working at the same Protvino facility reported the observation of four antitriton candidates \cite{Vishnevsky:1974ks}.  Four years later, the CERN WA33 collaboration carried out an experiment using the 200 GeV proton beam from the Super Proton Synchrotron (SPS) on beryllium and aluminum targets.  Improved statistics for deuterons, tritons, $^3$He and their antimatter partners were reported \cite{Bozzoli:1978ud}.  Production of light nuclei and antinuclei was presented in the form of ratios of nuclei and antinuclei to positive and negative pions, respectively, at various laboratory momentum settings of the WA33 detector.  The raw data were corrected for pion decays and for secondary interactions in the target and in the detector material~\cite{Bozzoli:1978ud}. 

%%%%%%%%% Fig. 7 %%%%%%%%%%%%%%%%%
\begin{figure}[!htb]
\centering
\centerline{\includegraphics[scale=0.85,bb=20 -30 220 260]{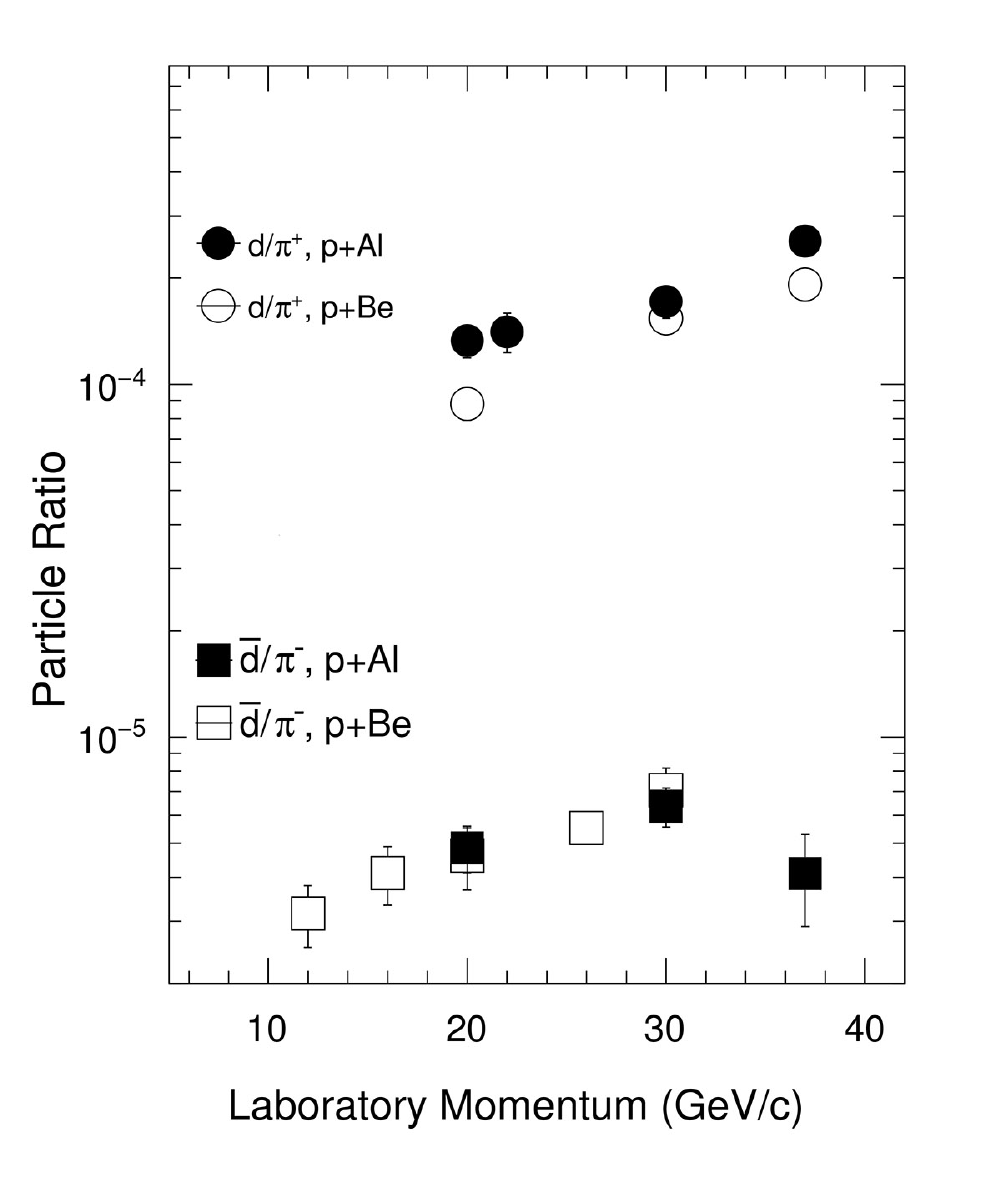}}
\caption{Deuteron to $\pi^+$ and antideuteron to $\pi^-$ ratios versus laboratory momentum of the detected secondaries at 0$^\circ$. These produced particles were from collisions of the 200 GeV CERN SPS proton beam on an aluminum target (solid points) and on a beryllium target (open points)~\cite{Bozzoli:1978ud}.}
\label{fig:SPS_Fig1}
\end{figure}
%%%%%%%%%%%%%%%%%%%%%%%%%%%%

Figure \ref{fig:SPS_Fig1} shows the ratios $d/\pi^+$ and $\bar{d}/\pi^-$ versus the laboratory momentum of the secondaries detected at zero degrees. These produced particles were from collisions of the 200 GeV proton beam of the CERN SPS on an aluminum target (solid points) and on a beryllium target (open points).  The $d/\pi^+$ ratio increases with laboratory momentum, and the ratio with an aluminum target is about 1.4 times larger than with a beryllium target.  The data indicate that deuteron production in the WA\,33 experiment depends on the target mass number and may be related to the leading particle effects of the scattered protons.  In contrast, the antideuteron to $\pi^-$  ratio is the same for the beryllium and aluminum targets.  This ratio has a maximum at about 30 GeV/$c$.  Bozzoli {\it et al.} argue that the measurements favor a scenario where antideuteron production takes place in elementary proton-nucleon collisions~\cite{Bozzoli:1978ud}.  They further point out that sufficient statistics were not available to address whether or not the same scenario is valid for antitriton and antihelium-3 production.  The WA\,33 authors also corroborated the observation by Antipov {\it et al.} \cite{Antipov:1970uc, Antipov:1971iq} of an exponential decrease in the probability of production of both nuclei and antinuclei as mass number increases from 1 to 3. 

%%%%%%%%% Fig. 8 %%%%%%%%%%%%%%
\begin{figure}[!htb]
\centering
\centerline{\includegraphics[scale=0.5,bb=320 -30 220 460]{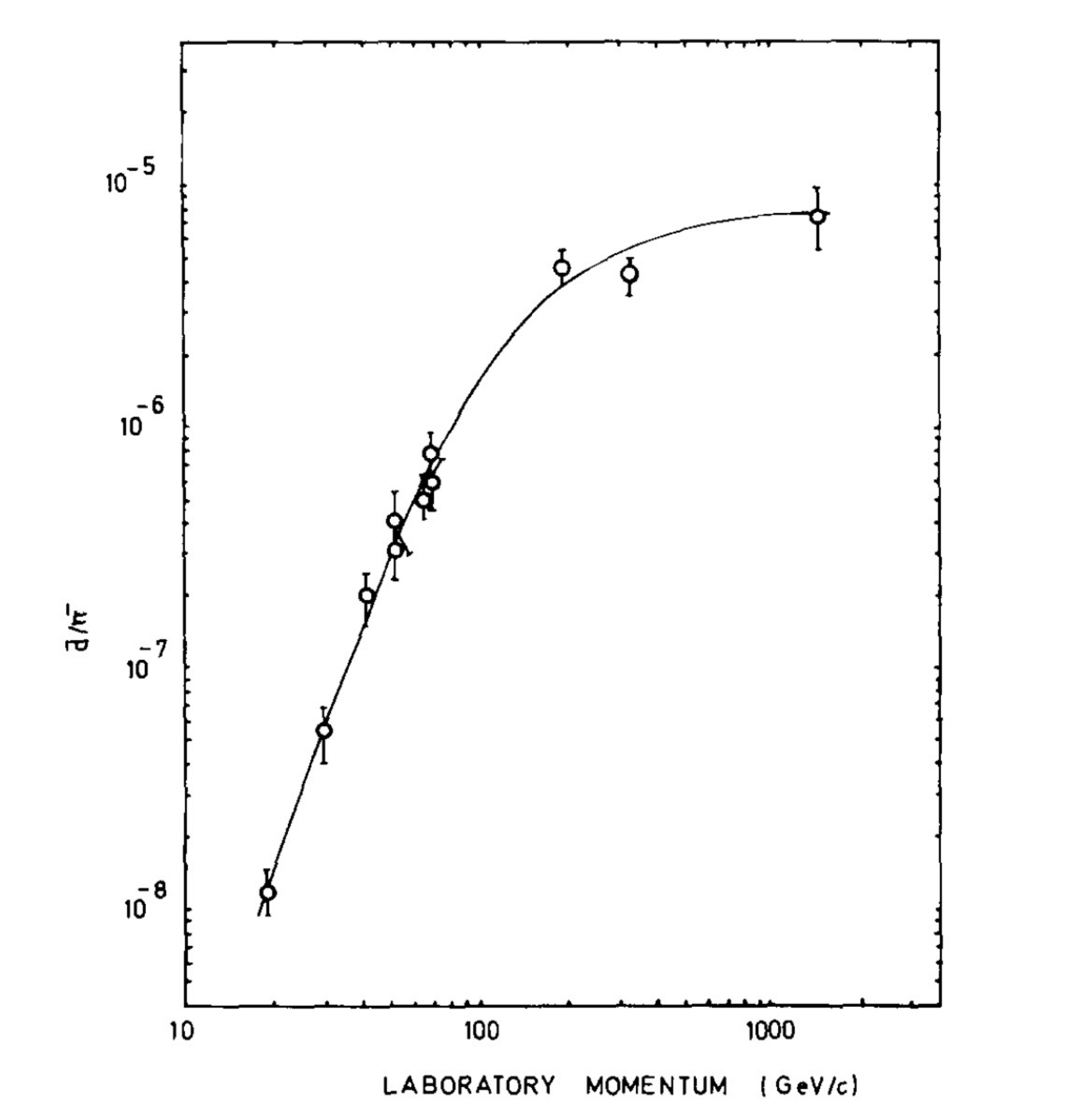}}
\caption{ A compilation of antideuteron to $\pi^-$ ratios versus laboratory momentum, as published by WA33 \cite{Bozzoli:1978ud}, combining results from Refs.~\cite{Dorfan:1965zz, Binon:1969qz, Antipov:1971zs, Alper:1973my, Appel:1974fs, Albrow:1975gr} (but not including WA\,33's own measurements).  The various experiments in the compilation have somewhat different kinematic conditions, but the plotted results generally correspond to Bjorken $|x| \lesssim$ 0.2 and $p_T \lesssim$ 0.2 GeV/$c$.}
\label{fig:SPS_Fig3}
\end{figure}
%%%%%%%%%%%%%%%%%%%%%%%%%%%

Another valuable insight offered in the paper of Bozzoli {\it et al.} \cite{Bozzoli:1978ud} is a compilation of antideuteron to $\pi^-$ ratios versus laboratory momentum, combining results from many prior experiments, as reported in Refs.~\cite{Dorfan:1965zz, Binon:1969qz, Antipov:1971zs, Alper:1973my, Appel:1974fs, Albrow:1975gr}; this compilation is reproduced in Fig.~\ref{fig:SPS_Fig3}.  The various experiments whose results are summarized in Fig.~\ref{fig:SPS_Fig3} (which does not include measurements from WA\,33) have somewhat different kinematic conditions, but the plotted measurements generally correspond to Bjorken $|x| \lesssim$ 0.2 and $p_T \lesssim$ 0.2 GeV/$c$.  These conditions also apply in the case of the WA\,33 measurements.  The compiled $\bar{d}/\pi$ ratios have a strong dependence on laboratory momentum over most of the plotted range, with evidence of flattening above 100 GeV/$c$.  However, the WA\,33 measurements reproduced in Fig.~\ref{fig:SPS_Fig1} indicate considerably higher rates of antideuteron production than the rates compiled in Fig.~\ref{fig:SPS_Fig3}, well beyond the reported errors.

\subsection{Antimatter and Symmetry
}\label{sym}

Symmetries play a major role in our understanding of many aspects of the structure of matter at the most elementary level, and play an essential role in our understanding of antimatter \cite{Robinson2011, Costa2012}.  The charge conjugation operator $C$ changes a particle into its antiparticle; thus, it reverses the sign of electric charge, baryon number, lepton number, as well as strangeness and heavy quark flavor quantum numbers. The parity operator $P$ applies an inversion, i.e., it reverses the sign of all three spatial coordinates in a Cartesian system; a plane mirror reflection reverses one coordinate only, and is a useful proxy for an inversion, since it is equivalent to an inversion and a $180^\circ$ rotation. 

After the 1956 proposal by Lee and Yang \cite{Lee:1956qn} to search for parity violation in weak decays, and the subsequent experimental discovery of such a violation in the beta decay of $^{60}$Co \cite{Wu:1957my}, it was initially speculated that the combined operator $CP$ might be conserved without exception.  However, even before the discovery of parity violation, Gell-Mann and Pais pointed out that the $K^0$ and its antiparticle $\overline{K}^0$ must have remarkable properties by virtue of their common decay modes and the consequent second-order weak coupling between them \cite{GellMann:1955jx}. These ideas led to the 1964 discovery by Christenson, Cronin, Fitch and Turlay of a small CP violation through the decay (with a probability of about $1/500$) of $K^0_L$ to two pions instead of the parity-conserving three pion decay \cite{Christenson:1964fg}.  

Another related operator is time reversal, $T$. This operator is relevant for antimatter in part due to the Stueckelberg-Feynman \cite{Stueckelberg:1941rg, Feynman:1948ur} interpretation of a hole in the negative-energy sea (an antiparticle) being equivalent to a particle propagating backward in time.  Time reversal symmetry is especially difficult to test directly for weak interactions, but there are compelling reasons to believe that it is violated when $CP$ is not conserved --- see below.  

%%%%%%%%%%%%%% Fig 9%%%%%%%%%%%%%%%%%%%
\begin{figure}[!hbt]
\centering
\centerline{\includegraphics[scale=0.35,bb=820 -30 220 460]{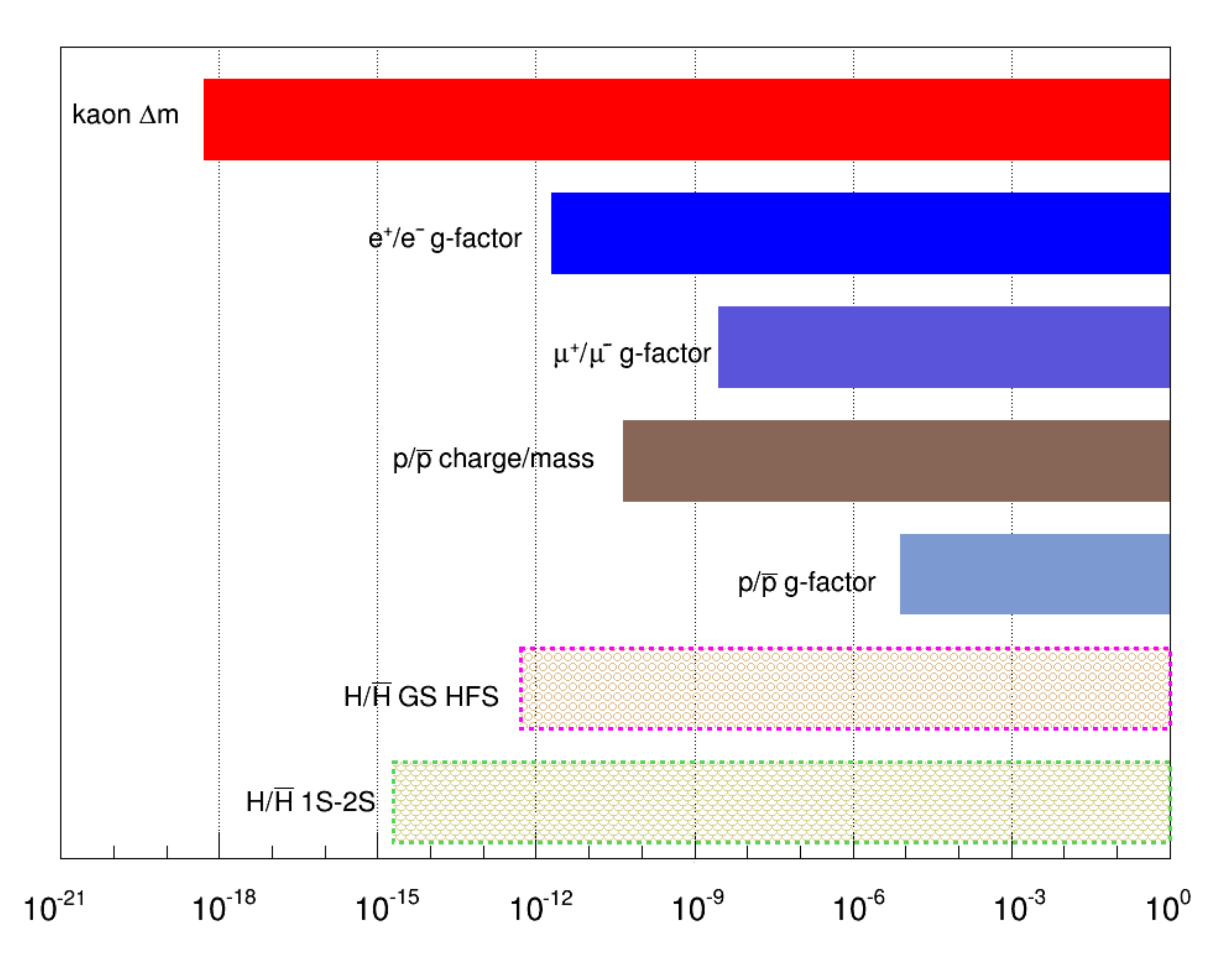}}
\caption{ A subset of {\it CPT} tests and their precision in dimensionless units. The tests represented here involve the mass difference between $K^0$ and $\overline{K}^0$ \cite{Abouzaid:2010ny, Olive:2016xmw}, the $g$-2 for the positron \cite{Mittleman:1999it, Dehmelt:1999jh, Hanneke:2008tm} and muon \cite{Bennett:2007aa, Grange:2015fou}, the charge-to-mass ratio and $g$-factor of the antiproton \cite{Ulmer:2015jra, Smorra:2015cuj}, and two forthcoming comparisons between neutral atoms of antihydrogen and hydrogen \cite{Amole:2014vna}: the ground-state hyperfine-splitting (GS-HFS) and the 1S-2S transition frequency \cite{Amole:2012zza, Kuroda:2014yya}.}
\label{fig:CPT}
\end{figure}
%%%%%%%%%%%%%%%%%%%%%%%%%%%%%%%%%%%%

In parallel with the above theoretical and experimental developments in the area of parity and $CP$, independent work by Schwinger \cite{Schwinger:1951xk}, L\"uders \cite{Luders:1954zz, Lueders:1992dq, Luders:1957zz}, Pauli \cite{Pauli:1955} and Bell \cite{Bell:1996nh}, based on general considerations of quantum field theory, locality and Lorentz invariance, established the $CPT$ theorem. This theorem indicates that every process in nature exactly conserves the three combined operators $C$, $P$ and $T$. A further consequence is that particle-antiparticle partners have exactly the same mass and lifetime, and exactly opposite magnetic moments \cite{Robinson2011, Costa2012}.  Many tests of these matter-antimatter symmetries have been carried out, of which a subset is represented in Fig.~\ref{fig:CPT}. No $CPT$ violations have ever been observed, and an especially precise test is provided by the magnitude of the mass difference between $K^0$ and $\overline{K}^0$, depicted in terms of a dimensionless fraction by the first (red) bar in Fig.~\ref{fig:CPT}, and known to be $ < 4 \times 10^{-19}$ GeV/$c^2$ at 90\% confidence level~\cite{Carosi:1990ms, AlaviHarati:2002ye, Abouzaid:2010ny, Olive:2016xmw}. Owing to the fundamental importance of $CPT$, many qualitatively different tests of this symmetry are of scientific interest and continue to be investigated and improved.  Overviews of the relevant literature can be found in Refs.~\cite{Kostelecky:2008ts, Smorra:2015cuj, Olive:2016xmw}.  

Baryogenesis refers to the process whereby the universe at the time of the Big Bang evolved into its present baryon-dominated form \cite{Weinberg2008, Sozzi2012, Canetti:2012zc}. For a time after the first observation of $CP$ violation, this asymmetry was considered to be a promising contender to account for the baryogenesis puzzle, namely, the surprising lack of any present-day remnants of the antibaryonic matter produced in the Big Bang. Within the framework on the Standard Model, a number of sources of $CP$ violation have been identified, but all investigations to date have demonstrated that $CP$ violation does not produce a large enough asymmetry to explain the observed universe \cite{Weinberg2008, Sozzi2012, Canetti:2012zc}.  

Apart from $CP$ violation, there are other proposed candidates to explain the baryogenesis puzzle: for example, the possible existence of bulk primordial antimatter in remote regions of the universe.  Cosmic ray experiments have been searching for antihelium in space, since such an observation could support the bulk antimatter hypothesis.  Another focus in the cosmic ray physics community is the investigation of the spectrum of positrons in space, which has connections to bulk primordial antimatter or dark matter hypotheses \cite{Olive:2016xmw, Weinberg2008, Canetti:2012zc, Schlickeiser2002, Stanev2010}.  See Section \ref{CR} for further details.

A variety of extensions beyond the Standard Model have also been proposed in connection with the baryogenesis puzzle, including possible consequences of Grand Unified Theories \cite{Fukugita:2002hu, Huang:2016wwj} or a possible time reversal ($T$) asymmetry and an associated non-zero electric dipole moment for elementary particles that would be asymmetric under $T$.  To date, no candidate electric dipole moment signal has been observed, and recent atomic physics measurements have achieved about a ten-fold improvement in the upper limit on the electron's electric dipole moment that significantly constrains $T$-violating physics \cite{Baron:2013eja}.  Overall, within or beyond the framework of the Standard Model, no known mechanism offers an adequate explanation for the imbalance between matter and antimatter in the observed universe, and the absence today of antibaryonic matter left-over from the Big Bang is invariably counted among the ``major unsolved problems of physics" \cite{Weinberg2008, Sozzi2012, Canetti:2012zc, Olive:2016xmw}.

\section{State-of-the-Art Detectors for Relativistic Heavy-Ion Collisions}
\label{Dect}
The Solenoidal Tracker at RHIC (STAR) and A Large Ion Collider Experiment (ALICE) are currently active experiments dedicated to relativistic heavy-ion physics. Their subsystems undergo periodic upgrades, providing optimum performance in the area of detecting and measuring antinuclei as well as in measuring many other observables of current physics interest.  

\subsection{The STAR Detector}
\begin{figure}[!htb]
\centering
\centerline{\includegraphics[scale=0.32,bb=720 -30 220 700]{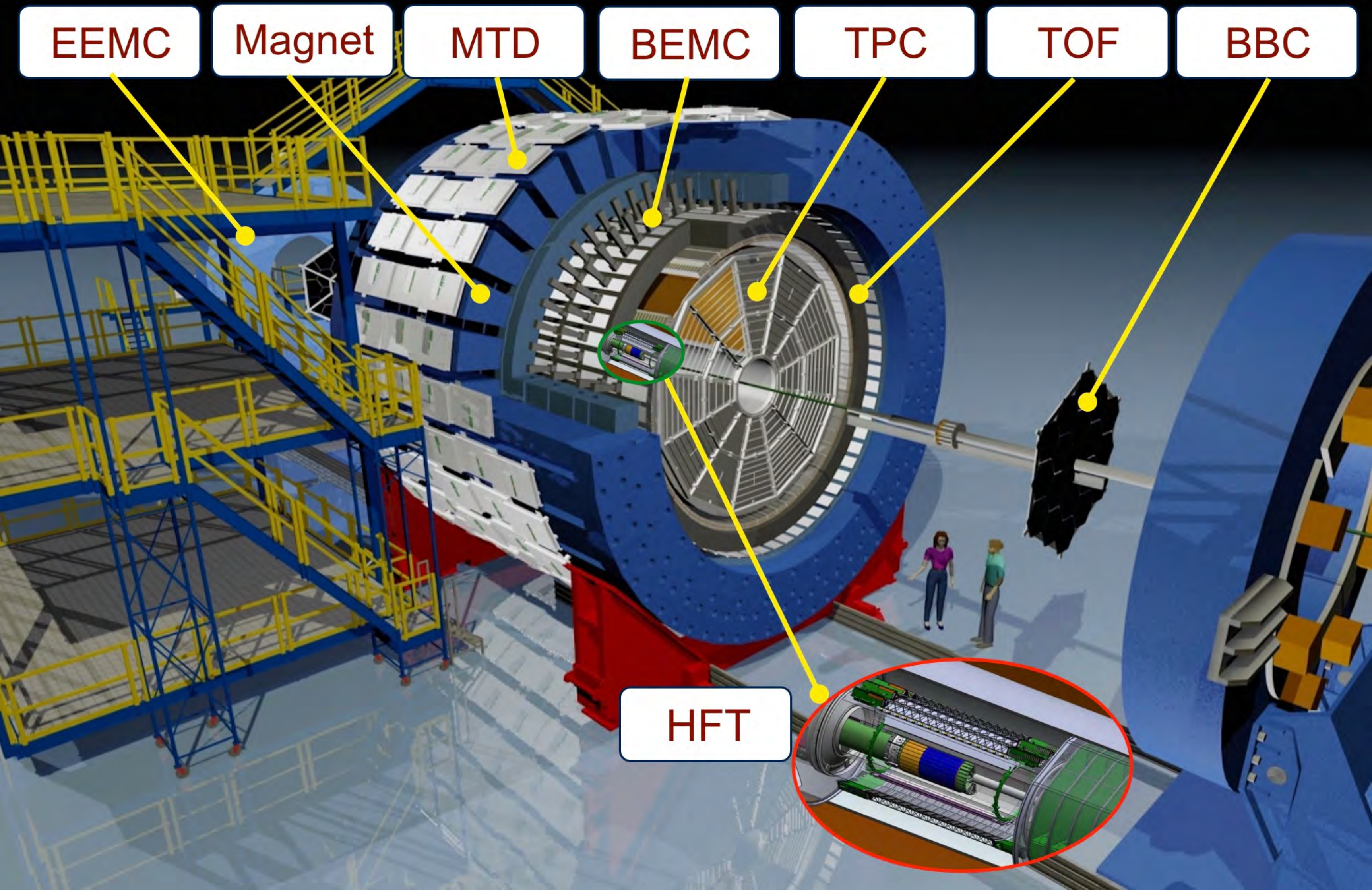}}
\caption{Schematic view of the STAR detector system. Each sub-detector is described in the text.}
\label{fig:STAR_detector}
\end{figure}

STAR \cite{Ackermann:2002ad} is one of four experiments installed at the Relativistic Heavy Ion Collider (RHIC)~\cite{Harrison:2003sb} at BNL, and the only one in operation as of the year 2018. STAR has a complex set of detector subsystems with a broad coverage of difference physics topics. Fig.~\ref{fig:STAR_detector} shows a schematic view of the STAR detector system. 

%%%%%%%%%%%FIGURE%%%%%%%%%%%%%%%%%
%%%%%%%%%%%%%%%%%%%%%%%%%%%%%%%%

The main tracking device in STAR is a gas-filled cylindrical Time Projection Chamber (TPC) \cite{Anderson:2003ur}, which is 4.2 m long and 4 m in diameter. It is located inside a large solenoidal magnet that operates at 0.5 T~\cite{Bergsma:2002ac}. The TPC is filled with P10 gas (10\% methane, 90\% argon) regulated at 2 mbar above atmospheric pressure~\cite{Kochenda:2002zz}. It can record the tracks of particles, measure their momenta, and identify particles via their ionization energy loss ($dE/dx$).  Each track in the TPC is reconstructed using up to 45 hit points. The STAR TPC covers up to $\pm 1.8$ units of pseudorapidity ($\eta$) with a full 360 degrees of azimuthal acceptance~\cite{Ackermann:2002ad}.

The STAR Barrel Electro-Magnetic Calorimeter (BEMC) is used to trigger on high $p_T$ events, including jets, isolated photons and heavy quarkonia. Its acceptance is $|\eta|<1$ and $2\pi$ in azimuthal angle, $\phi$.  The STAR BEMC uses layers of lead and plastic scintillator with 20 times a radiation length at $\eta = 0$ \cite{Beddo:2002zx}. The BEMC includes a Shower Maximum Detector (SMD) which is used to provide fine spatial resolution. The precise spatial information provided by the SMD is essential for $\pi^0$ reconstruction, direct $\gamma$ identification, and electron identification. 

The STAR Time-of-Flight (TOF) detector is located between the TPC and BEMC, with an azimuthal coverage of $2\pi$ and $|\eta| < 0.9$. This TOF detector is based on the technology of Multi-gap Resistive Plate Chambers (MRPC)~\cite{CerronZeballos:1995iy}, which consist of a stack of resistive plates (float glass of 0.54 mm thickness) with five 220 mm gas gaps. There are 120 TOF trays and each contains 32 MRPC modules. The whole TOF system consists of the barrel TOF and the Vertex Position Detector (VPD). The start time of the detected particles is provided by the VPD. The time resolution of the VPD and barrel TOF can achieve $\sim30$ ps and $< 80$ ps, respectively, in heavy-ion collisions.  With this performance, the TOF system allows $\pi/K$ separation up to $\sim$1.6 GeV/$c$ and $(\pi,K)/p$ separation up to $\sim$3 GeV/$c$ \cite{Geurts:2004}. 

The Muon Telescope Detector (MTD)~\cite{Ruan:2009ug, Wang:2011ay} is also based on MRPC technology~\cite{CerronZeballos:1995iy}.  The MTD can trigger on and identify muons based on its precise timing and modest position resolution~\cite{Xu:2016}, allowing measurements of dileptons through the di-muon channel. It provides single-muon and di-muon triggers based on the number of hits within a predefined online timing window. The MTD modules are installed at a radius of about 403 cm, and cover about 45\% in azimuth within $|\eta| < 0.5$. The timing resolution of the MTD is $\sim$100 ps and the spatial resolution is $\sim$1-2 cm in both $r$-$\phi$ and $z$ directions~\cite{Yang:2014xta}.

A new STAR detector, closest to the beam pipe, is the Heavy Flavor Tracker (HFT) which achieves topological reconstruction of secondary decay vertices of open heavy-flavor hadrons~\cite{Qiu:2014dha, Long:2017}. The HFT consists of three sub-detectors: the Silicon Strip Detector (SSD), the Intermediate Silicon Tracker (IST), and the Pixel (PXL) detector. The innermost sub-detector (PXL) has two layers which use state-of-the-art ultra-thin CMOS Monolithic Active Pixel Sensors (MAPS). STAR is the first collaboration to use a CMOS MAPS detector in a collider experiment~\cite{Qiu:2014dha}. The IST has single-sided double-metal silicon pad sensors, and its radius is about 14 cm and thickness is less than 1.5\% radiation lengths. The SSD and IST are fast detectors which can reduce track pile-up.

%%%%%%%%
\subsection{The ALICE Detector}

\begin{figure}[!htb]
\centering
\centerline{\includegraphics[scale=0.15,bb=2020 -30 220 960]{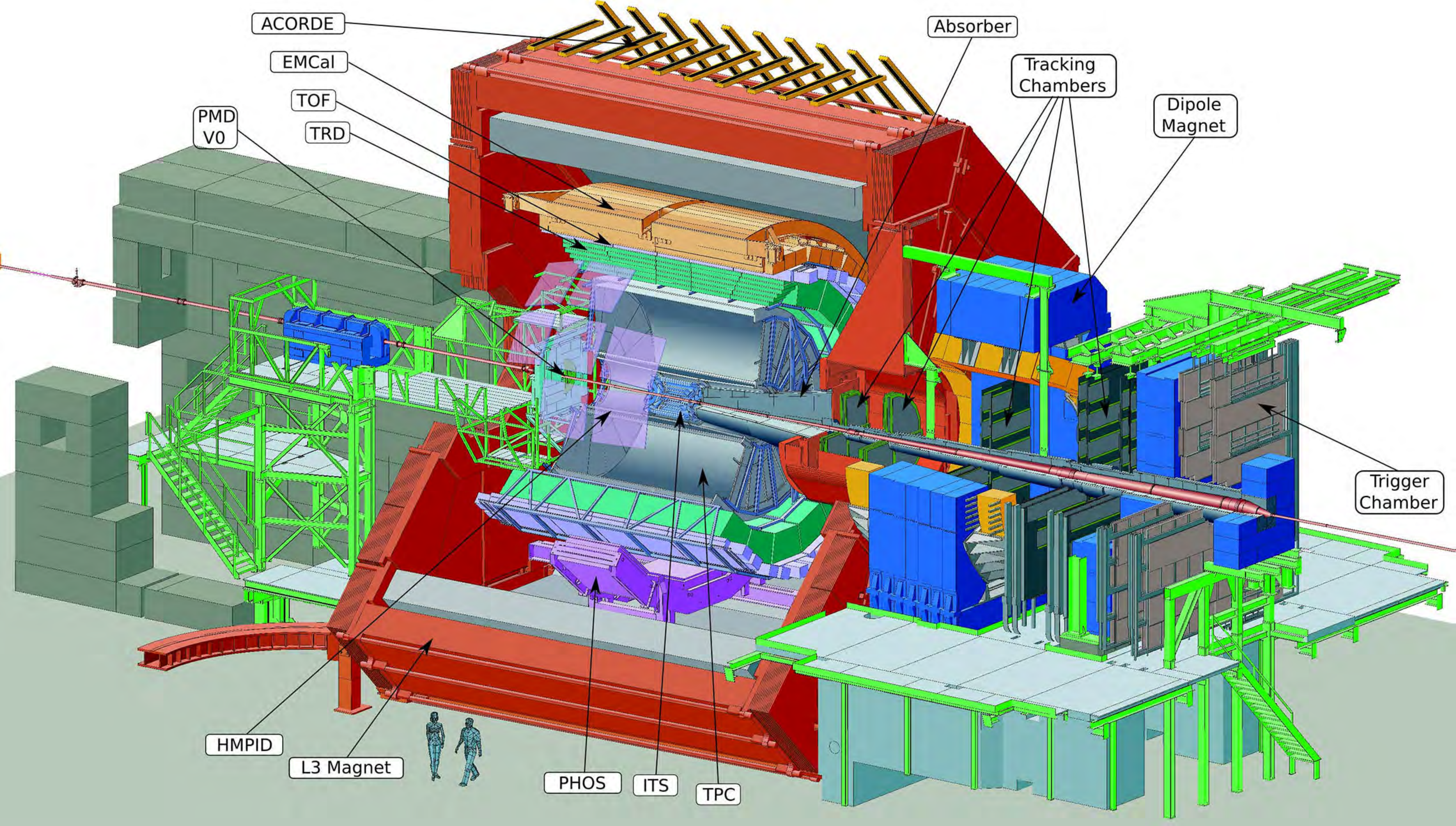}}
\caption{Schematic view of the ALICE detector system.}
\label{fig:ALICE_dectector}
\end{figure}

ALICE is a general-purpose, heavy-ion detector at the Large Hadron Collider (LHC) which focuses on QCD, the strong-interaction sector of the Standard Model. It is designed to address the physics of strongly interacting matter and the quark-gluon plasma at extreme values of energy density ($\mathrm{>}$ 10 $\mathrm{GeV/fm^3}$) and temperature ($\geq$ 0.2 GeV) in nucleus-nucleus collisions~\cite{Aamodt:2008zz, Abelev:2014ffa}. Fig.~\ref{fig:ALICE_dectector} shows a schematic view of the ALICE detector system. The Inner Tracking System (ITS)~\cite{Aamodt:2010aa} is the subsystem that is closest to the beam pipe. The main tasks of the ITS are to reconstruct the primary and secondary vertices, to track and identify particles with momentum below 200 MeV/$c$, and to improve the momentum and angle resolution for particles reconstructed by the TPC \cite{Carminati:2004fp}. The ITS consists of six cylindrical layers of silicon detectors, located at radii between 3.9 cm and 43.0 cm. The two innermost layers are the Silicon Pixel Detectors (SPD). They are followed by a pair of Silicon Drift Detectors (SDD), characterized by a very good multitrack reconstruction capability. Finally, two layers of Silicon Strip Detector complete the ALICE ITS. In addition to participating in the ALICE global tracking, the ITS is capable of performing stand-alone reconstruction with the advantage of restoring tracks lost during global tracking due to spatial acceptance, the intrinsic $p_T$ cutoff of the outer detectors, or due to particle decay~\cite{Aamodt:2010aa}.

%%%%%%%%%%%%%%%%%%%%%%%%%%%%%%%
%%%%%%%%%%%%%%%%%%%%%%%%%%%%%%%

Following the ITS in the radial direction is the TPC, which is the main tracking detector of the central barrel and provides, together with the other central barrel detectors, charged-particle momentum measurements with good two-track separation, particle identification, and vertex determination~\cite{Alme:2010ke}. Its tracking efficiency reaches $\sim 80$\% within $|\eta|< 0.8$ with a momentum resolution $\sigma(p_T)/p_T \sim$5\%, corresponding to $\sim$2.5\% up to $p_T$ = 10 GeV/$c$ (and increasing at higher transverse momenta) when combined with the ITS.  Each track in the TPC is reconstructed using up to 159 space points, with a spatial resolution of 0.8 mm in the $x$-$y$ plane, and 1.2 mm in the $z$ direction.

The TPC is surrounded by the Transition Radiation Detector (TRD). The main purpose of the ALICE TRD is to provide electron identification in the central barrel for momentum above 1 GeV/$c$~\cite{ALICE_TRD1}. The TRD was designed to provide a fast trigger for charged particles with high momentum. It is part of the Level-1 trigger and can significantly enhance the recorded $\gamma$ yields, high $p_T$ $J/\Psi$, the high-mass region of the dilepton continuum, as well as jets. The TRD consists of 540 individual read-out detector modules. They are arranged into 18 super modules, each containing 30 modules arranged in five stacks along $z$ and six layers in radius. In the longitudinal ($z$) direction, the active length is 7 m, and the overall length of an entire super module is 7.8 m with a weight of 1650 kg.

The TRD is surrounded by the TOF, at a radius of 3.7 m from the interaction point. The TOF detector is a large-area array that covers the central pseudorapidity region ($|\eta| < 0.9$) for Particle Identification (PID) in the intermediate-momentum range, below about 2.5 GeV/$c$ for pions and kaons, up to 4 GeV/$c$ for protons, with a $\pi/K$ and $K/p$ separation better than $3\sigma$~\cite{ALICE_TOF1, Alessandro:2006yt}. The TOF, coupled with the ITS and TPC for track and vertex reconstruction and for $dE/dx$ measurements in the low-momentum range (up to about 1 GeV/$c$), provides event-by-event identification of large samples of pions, kaons and protons.

In the central $\eta$ region, ALICE has several sub-detectors, referred to as single-arm detectors, which have a limited acceptance. They consist of a \v{C}erenkov RICH detector (the HMPID), a homogeneous photon spectrometer (PHOS), and a sampling electromagnetic calorimeter (EMCAL). At forward rapidities, there is a Photon Multiplicity Detector (PMD) and a muon spectrometer (MUON).

\subsection{PID Technology in STAR and ALICE}

%%%%%%%%%%%%%%%%%%%%%%%%%%%%%%
\begin{figure}[!htb]
\centering
\centerline{\includegraphics[scale=0.21,bb=1520 -30 220 1060]{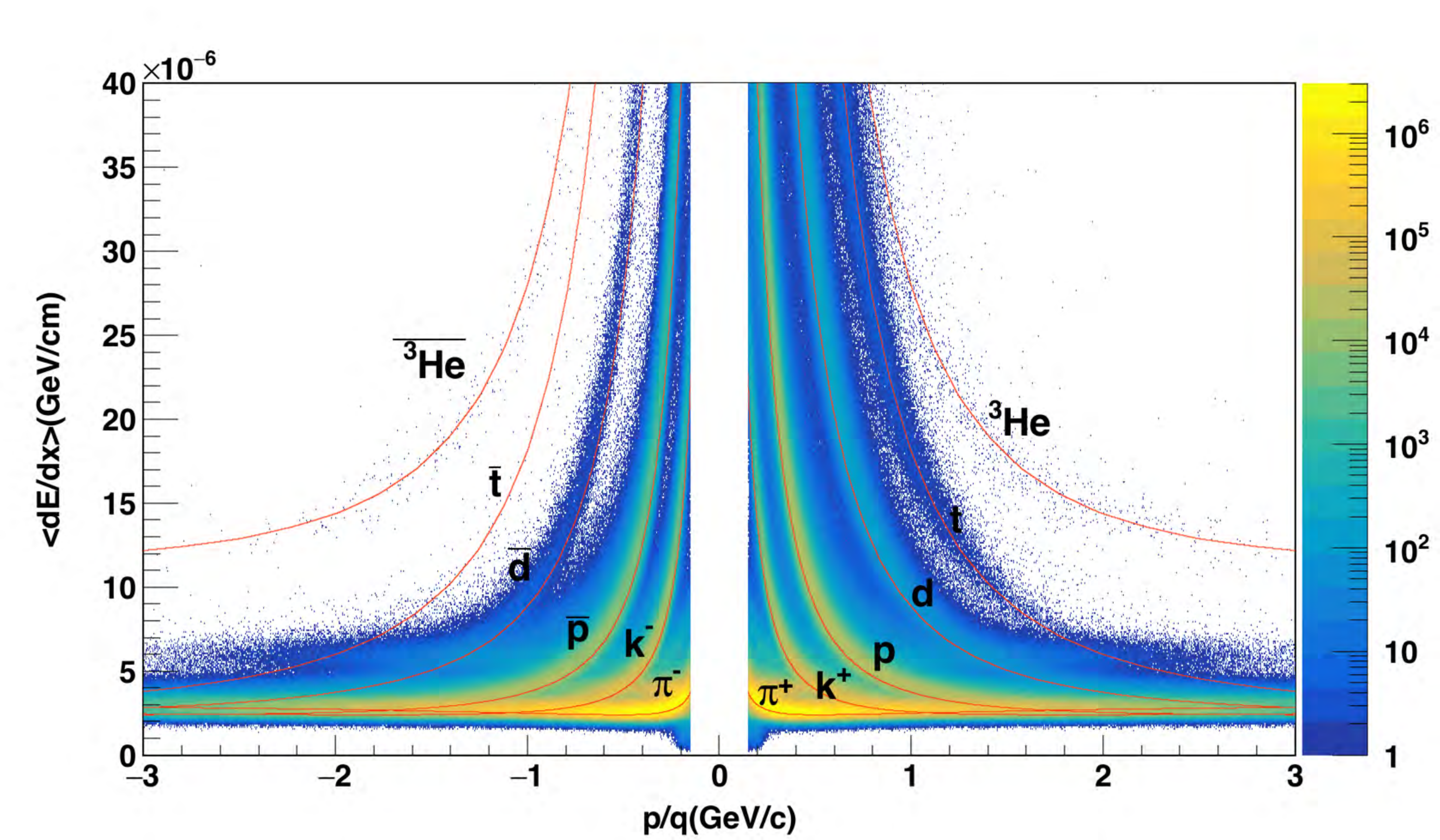}}
\caption{ The $\langle dE/dx \rangle$ for charged tracks at midrapidity ($|y| < 0.5$) as a function of momentum per charge (in units of the electron charge $e$) in Au + Au collisions at $\sqrt{s_{NN}} = 200$ GeV. The curves are based on the Bethe-Bloch formula.}
\label{fig:PID}
\end{figure}

\begin{figure}[!htb]
\centering
\centerline{\includegraphics[scale=0.21,bb=1520 -30 220 720]{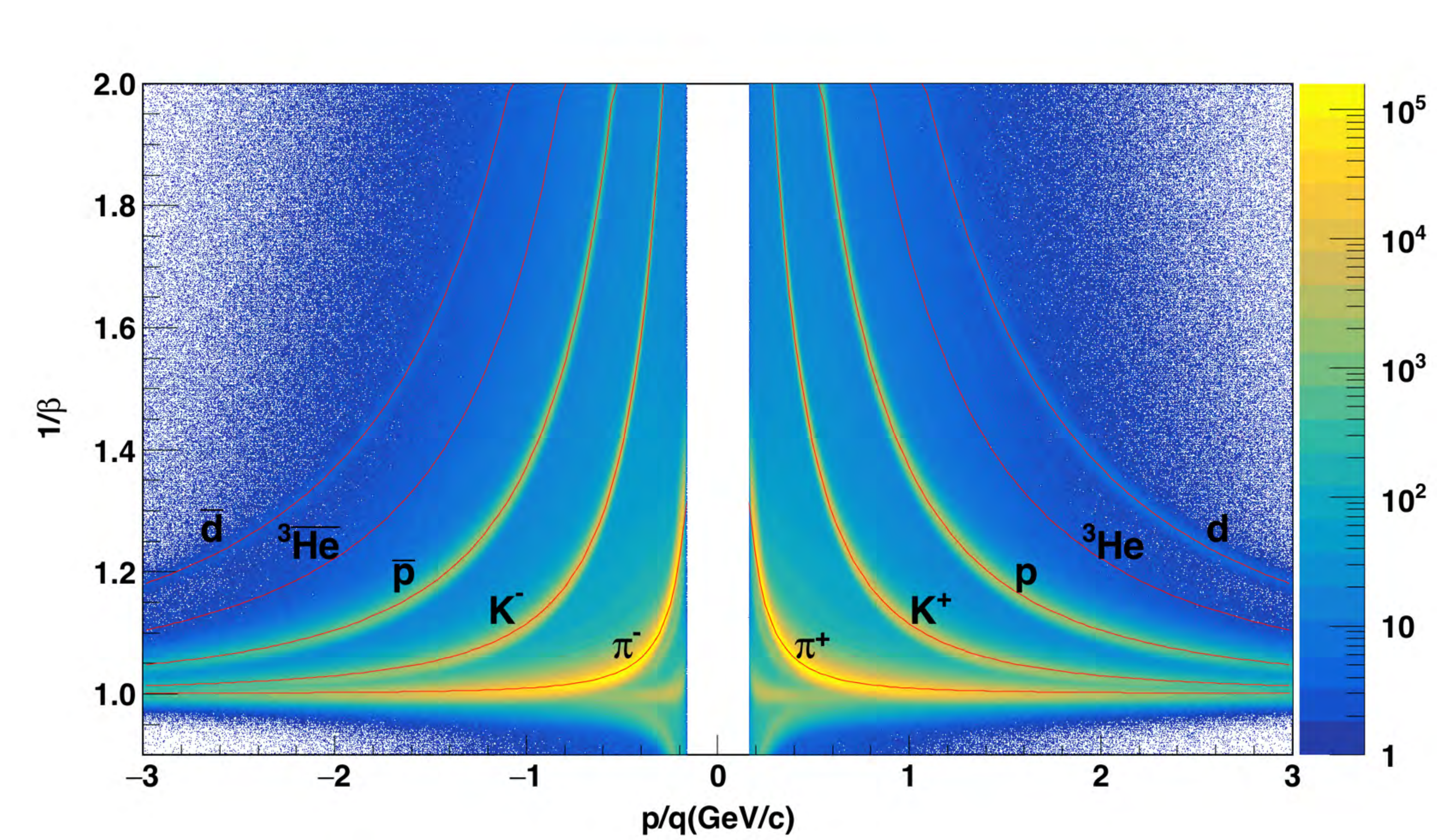}}
\caption{The same data as presented in Fig.~\ref{fig:PID} above, but now with the inverse velocity from TOF on the vertical axis.}
\label{fig:PID2}
\end{figure}

%%%%%%%%%%%%%%%%%%%%%%%%%%%%%%

In high energy nuclear collision experiments such as STAR and ALICE, particle identification is accomplished by measuring the track energy loss in the TPC gas as a function of momentum per charge. As particles traverse the TPC, they lose energy and produce primary ionization in the gas. The ionization electrons drift to the endcap region of the TPC where gas amplification occurs close to the anode wires. The average energy loss per unit track length can then be determined from the endcap pad signals. The Bethe-Bloch formula (Eq. ~\ref{Equ.PID_TPC}) is the theoretical expression for the mean rate of energy loss~\cite{Bichsel:2006cs}:

\begin{equation}
\left\langle\frac{dE}{dx}\right\rangle = 2\pi N_0 q^2 r^2_e m_e c^2\rho\frac{Z}{A}\frac{1}{\beta^2}[\ln\frac{2m_e\gamma^2 v^2 E_M}{I^2}-2\beta^2], 
\label{Equ.PID_TPC}
\end{equation}
where $N_{0}$ is Avogadro's constant, $q$ is the charge of the particle (in units of $e$), $r_{e}$ is the classical radius of the electron, $m_{e}$ is the electron mass, $c$ is the speed of light, $\rho$ is the density of the medium, $Z$ is the charge number of the medium, $A$ is the atomic mass of the medium, $\beta$ ($= v/c$) is the speed of the particle, $\gamma=1/{\sqrt{1-\beta^2}}$, $I$ is the mean excitation energy, and $E_{M}$ is the maximum kinetic energy which can be imparted to a free electron in a single collision. The measured energy loss can be compared with the theoretical expectation from the Bethe-Bloch formula by defining

\begin{equation}
%%%n_{\sigma}=\frac{1}{R}\ \mathrm{ln}\frac{\langle\frac{dE}{dx}\rangle_{\rm data}}{\langle\frac{dE}{dx}\rangle_{\rm theory}}
z = \ln\frac{\langle\frac{dE}{dx}\rangle}{\langle\frac{dE}{dx}\rangle_{\rm theory}}\,,
\label{Equ.PID_TPC2}
\end{equation}
where $\langle dE/dx \rangle_{\rm theory}$ is the expected energy loss for a given particle species. We define $\sigma_z$ as the root mean squared width of the $z$ distribution, and $n_\sigma$ is the number of standard deviations from zero, the expected value of $z$.  Figure~\ref{fig:PID} shows the average $dE/dx$ of the measured charged particles plotted as a function of their momentum per charge ($p/q$) in Au+Au collisions at RHIC. The curves are the theoretical expectation from the Bethe-Bloch formula. The TPC can identify pions, kaons and (anti)protons at low momentum as seen from the figure. For higher momentum, particle identification needs information from the TOF detector. The TOF detector measures the time ($t$) taken by a track to traverse the distance ($L$) from the primary vertex to TOF. Knowing both $t$ and $L$, one can calculate the velocity ($\beta$) of each track. Figure~\ref{fig:PID2} shows the time-of-flight $1/\beta$ information as a function of momentum per charge ($p/q$). The particle species are well separated and the pion, kaon, and proton separation is good, even for momenta above 1 GeV/$c$. Similar PID results on Pb + Pb collisions at the LHC are presented in many publications, such as Refs.~\cite{Adam:2015vda, Acharya:2017bso}.

\section{Antimatter Nuclei in High-Energy Nucleus-Nucleus Collisions}

\subsection{Characteristics of Heavy-Ion Accelerators}
The discoveries of antinuclei produced at proton accelerators, as reviewed in Sec.~\ref{pA}, reached a plateau in the 1970s, and no new species have been found in $pp$ or in $pA$ collisions since the work of Vishnevsky {\it et al.} in 1974 \cite{Vishnevsky:1974ks}. Meanwhile, the first clear evidence for antiprotons in the field of cosmic ray experiments was reported by Golden {\it et al.} in 1979 \cite{Golden:1979bw}, whereas even up to the present day, no heavier antinucleus has been observed in a cosmic ray event \cite{Fuke:2005it, Sasaki:2008zzb, Abe:2012tz, Aguilar:2016kjl}. In recent decades, the search for new and heavier antinuclei has consequently migrated to the field of nucleus-nucleus collisions. Nucleus-nucleus collisions at high energy create suitable conditions for the production of antinuclei, because a large amount of energy is deposited into a volume that is more extended than in the case of elementary particle collisions. On the other hand, in contrast to the Big Bang, high-energy nuclear collisions do not involve gravitational attraction and allow the antimatter to escape, and avoid being destroyed by annihilation.

To quantitatively study the fundamental properties of antimatter, the rate of production of antinuclei in nuclear collisions over a range of energies, up to the highest available energy, is an important measurement. The relevant facilities include the Alternating Gradient Synchrotron (AGS) at Brookhaven National Laboratory, the Super Proton Synchrotron (SPS) at CERN, the Relativistic Heavy-Ion Collider (RHIC) at Brookhaven, and the Large Hadron Collider (LHC) at CERN. The AGS and SPS had prominent roles in antinucleus research during the 1950s, 60s and 70s using proton-nucleus collisions (see Section \ref{pA}). In the 1980s, these two fixed-target machines were upgraded by adding the capability to accelerate heavy ions \cite{Barton:1983ch, Barton:1987mk, Gutbrod:2016}. In the case of the AGS (SPS), $^{197}$Au ($^{208}$Pb) beams up to 11.5$A$ GeV (158$A$ GeV) were available, corresponding to a center-of-mass energy per nucleon pair of $\sqrt{s_{NN}} = 4.8$ (17) GeV in gold on gold (lead on lead) collisions. Beginning in 2000, a dedicated heavy-ion collider, RHIC \cite{Harrison:2003sb}, began operation with Au + Au collisions at up to $\sqrt{s_{NN}} = 200$ GeV, and less frequently, with $^{238}$U + $^{238}$U collisions at $\sqrt{s_{NN}} = 193$ GeV. With the advent of the LHC heavy-ion program in 2010, Pb + Pb events at 2.76 TeV (later up to $\sqrt{s_{NN}} = 5$ TeV) came within reach. These major facilities paved the way for detailed and comprehensive studies of the then-known antinuclei, as well as searches for new, heavier species.  

In heavy-ion collisions, the physics conditions are quite different from those in $pp$ interactions. Such collisions at ultrarelativistic energies produce a hot and dense phase of matter containing approximately equal numbers of quarks and antiquarks. This phase, called quark-gluon plasma (QGP) \cite{Adams:2005dq, Adcox:2004mh, Arsene:2004fa, Back:2004je} persists for only a few times $10^{-23}$ seconds and exhibits fluid properties with exceptionally low viscosity \cite{Muller:2006ee, Jacak:2010zz, Shuryak:2014zxa, Braun-Munzinger:2015hba,Luo:2017NST} and exceptionally high vorticity \cite{STAR:2017ckg,Hattori:2017NST}. Then the hot and dense plasma cools and undergoes a transition into a hadron gas, producing mesons, baryons, antibaryons, and occasionally, antinuclei.  %Nucleus-nucleus collisions at high energy create suitable conditions for the production of antinuclei, because a large amount of energy is deposited into a volume that is more extended than in the case of elementary particle collisions. On the other hand, in contrast to the Big Bang, high-energy nuclear collisions do not involve gravitational attraction, and thus it is easier for antimatter to escape before undergoing annihilation.

High-energy nuclear collisions are controlled experiments. Unlike the single event of the Big Bang, the ``little bang" of heavy-ion collisions can be repeated at will. We can switch between energies, colliding species, etc., in contrast to the ``passive" observation process of a cosmic ray experiment.   

In later sections of this paper, the possible production mechanisms for antinuclei are reviewed. We begin by reviewing the measured production characteristics of antinuclei in relativistic heavy-ion collisions at the four facilities discussed above, namely, the AGS, SPS, RHIC and LHC.   

\subsection{Yields of Antinuclei in Heavy-Ion Collisions at the AGS and SPS
}\label{Yields-HIC}

The first journal publication on measured production of antinuclei in nucleus-nucleus collisions dates from 1992, when Aoki {\it et al.} \cite{Aoki:1992mb} (AGS/E858 collaboration), reported results from the early phase of heavy-ion operations at the AGS using a $^{28}$Si beam at 14.6$A$ GeV.  This beam energy is just below the threshold for production of a $d\bar{d}$ pair in a binary nucleon-nucleon interaction.  Targets of Al, Cu and Au were investigated, and measurements of antideuterons were performed at zero degrees using a beamline focusing spectrometer that was qualitatively similar to those described in Sec.~\ref{pA}. The mechanism of direct antideuteron production via {\it NN} $\rightarrow$ {\it NNNN}$\bar{d}$ or {\it NN} $\rightarrow$ {\it NN}$d\bar{d}$, allowing for the effects of Fermi momentum, was reported to be inconsistent with the measurements \cite{Aoki:1992mb}. 

The other antideuteron production mechanism investigated by Aoki {\it et al.} was independent production of an antiproton and an antineutron in the same event, followed by antideuteron formation via statistical coalescence.  This mechanism has already been discussed in Sec.~\ref{pA} in the context of antideuteron formation in proton-nucleus collisions.  Furthermore, in extensive studies of kinematic regions dominated by participants from heavy-ion collisions at mostly lower energies, the formation of deuterons and heavier composites has been shown to be largely explained by statistical coalescence \cite{Gutbrod:1988gt, Lemaire:1980qw, Jacak:1985zz, Hayashi:1988en, Saito:1994tg, Abbott:1994np, Wang:1994rua}. Equation (1) in Sec.~\ref{pA} can be rewritten in terms of a coalescence parameter $B_A$ for a composite nucleus of mass number $A$ and charge $Ze$:
\begin{equation}
E_A \frac{d^3 N_A}{dp^3_A} = B_A \left( E_p \frac{d^3 N_p}{dp^3_p} \right)^Z
                                            \left( E_n \frac{d^3 N_n}{dp^3_n} \right)^{A-Z}, 
\label{eq:coalescence1}
\end{equation}
\begin{equation}
{\rm or~~} E_A \frac{d^3 N_A}{dp^3_A} \approx B_A \left( E_p \frac{d^3 N_p}{dp^3_p} \right)^A, \\\
\label{eq:coalescence2}
\end{equation}
where $E \frac{d^3 N}{dp^3}$ is the invariant momentum-space density, and $p_A,\,p_p$ and $p_n$ are the momenta of a nucleus, proton and neutron, respectively, assuming $p_A = Ap_p$.  The second expression above is a good approximation at high relativistic energies, where the difference between proton and neutron spectra can be safely neglected.  

Aoki {\it et al.} \cite{Aoki:1992mb} reported that while the coalescence approach describes deuteron production in their experiment, it overpredicts $\overline{B_2}$ for antideuterons by about an order of magnitude. In contrast, they also point out that a similar approach works well for antideuteron production in proton-nucleus collisions. However, it should be noted that the coalescence studies in the literature for composite nuclei \cite{Gutbrod:1988gt, Lemaire:1980qw, Jacak:1985zz, Hayashi:1988en, Saito:1994tg, Abbott:1994np, Wang:1994rua,Yan:2006} generally report agreement with the coalescence model insofar as they observe a constant coalescence parameter $B_A$ as a function of appropriate kinematic variables within their datasets, but either do not compare $B_A$ measurements across different experiments, or do not attribute differences in $B_A$ between different experiments to a breakdown of the coalescence picture.  In contrast, Aoki {\it et al.} \cite{Aoki:1992mb} did not have suitable data for investigating the constancy of $\overline{B_2}$ within their dataset, and instead tested the agreement of their antideuteron and antiproton yields with $B_2$ and $\overline{B_2}$ values from other experiments.  As they pointed out, several experiments have reported similar $B_2$ values, on the order of 1 to 2 $\times 10^{-2}$ GeV$^2/c^3$, but in general, $B_A$ values are expected to vary with the size of the particle source, which can obviously depend on various factors like $A_{\rm proj} + A_{\rm targ}$, centrality, and probably beam energy. 

Spatial correlation effects are ignored in Eqs. (\ref{eq:coalescence1}) and (\ref{eq:coalescence2}), and might not be negligible \cite{Mekjian:1977ei, Sato:1981ez, Csernai:1986qf}. (Anti)nuclei have a relatively small binding energy compared with the typical thermal energies in the participant fireball of a heavy-ion collision. A recent paper by Zhang and Ko~\cite{ZHANG2018191} considers how this relates to production of hypertritons with small binding energy and large size. On the other hand, a process where two nucleons form a deuteron, for example, requires the mediation of other matter in order to conserve energy and momentum.  These countervailing factors mean that (anti)nuclei have the potential to strongly constrain specific properties of the emitting source~\cite{Shah:2015oha, Zhu:2015voa}.  They imply a shell-like formation region, especially for antinuclei \cite{Mrowczynski:1989jd, Leupold:1993ms}.  Consequently, there has been theoretical speculation that $\overline{B_A}$ might be measurably smaller than $B_A$ \cite{Mrowczynski:1989jd, Leupold:1993ms, Bleicher:1995dw}.  An alternative picture based on thermalization favors no such difference by assuming that nucleons and antinucleons have the same temperature, flow and freezeout density distributions prior to forming (anti)clusters~\cite{Scheibl:1998tk}.  Depending on the beam energy and the rapidity region under study, production of nuclei has some possibility to be influenced by initial-state matter, whereas antinuclei are guaranteed to be unaffected by initial-state matter.  Overall, $B_A$ and $\overline{B_A}$ measurements are a valuable complement to the source size information (homogeneity lengths) provided by femtoscopy measurements \cite{Lisa:2005dd}. 
 
%%%%%%%%%Figure%%%%%%%%%%%%%%%%
\begin{figure}[!thb]
\centering
\centerline{\includegraphics[scale=0.30, bb=1200 -50 -230 750]{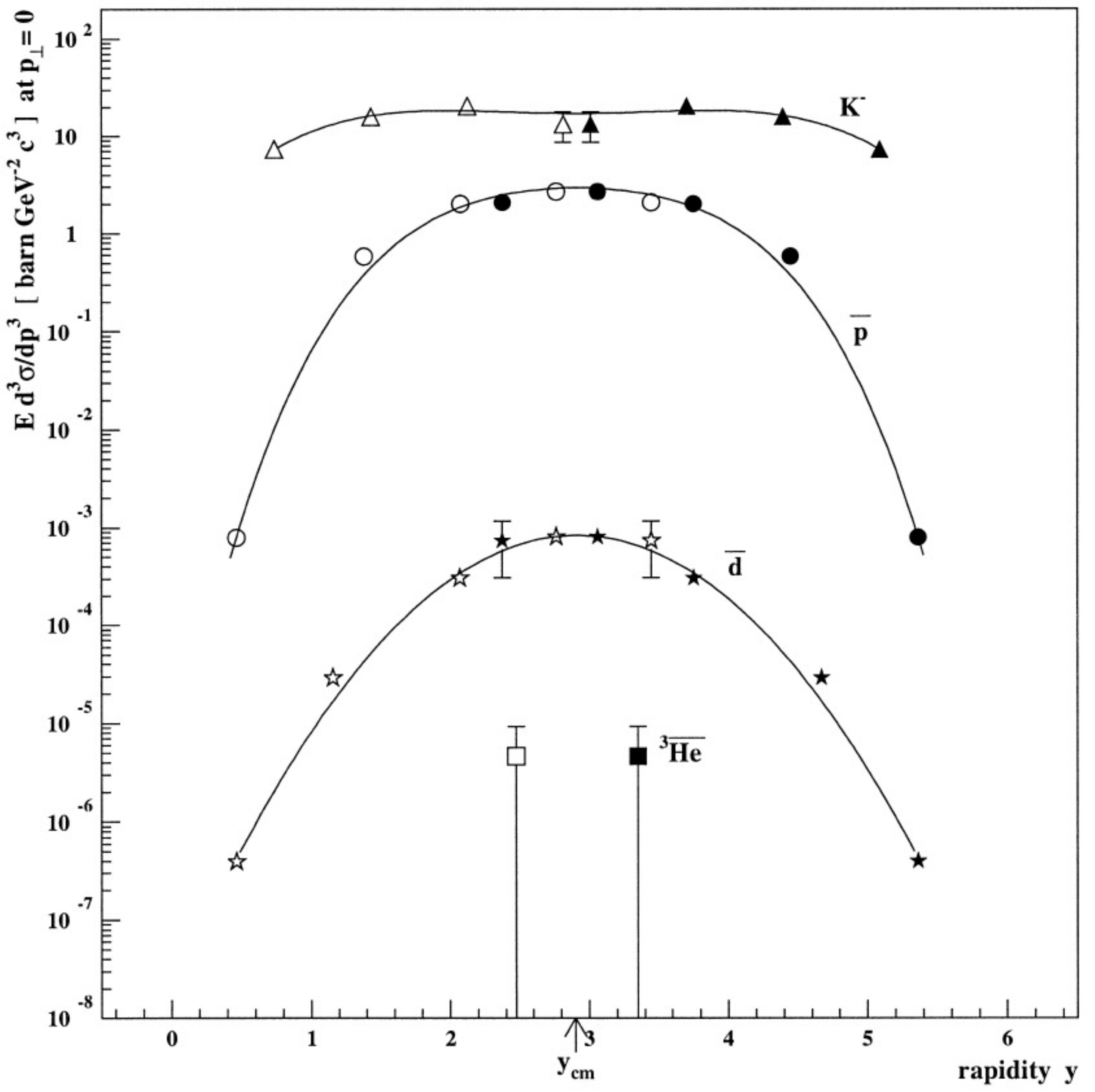}}
\caption{Invariant production cross sections at 0$^\circ$ versus rapidity for $K^-$, antiprotons, antideuterons, and $^3\overline{\rm He}$ in Pb + Pb collisions at $\sqrt{s_{NN}} = 17$ GeV (158$A$ GeV beam energy) with a minimum-bias trigger, measured by the NA52 collaboration \cite{Appelquist:1996qy, Ambrosini:1997bf}. The solid markers are the measurements, and the open markers are reflected about midrapidity.  The error bars are statistical uncertainties only.}
\label{fig:NA52}
\end{figure}
%%%%%%%%%%%%%%%%%%%%%%%%%%%%%%

The CERN NA52 collaboration reported on production of antideuterons in a heavy nucleus-nucleus system, 158$A$ GeV Pb + Pb ($\sqrt{s_{NN}} =17$ GeV) in 1996 and 1998 \cite{Appelquist:1996qy, Ambrosini:1997bf}. Measurements were performed at zero degrees, again using a beamline spectrometer.  Figure \ref{fig:NA52} illustrates that useful antideuteron statistics were obtained as a function of rapidity, with a standard deviation of 0.6 units about midrapidity. NA52 also reported one $^3\overline{\rm He}$ candidate, which is included on Fig.~\ref{fig:NA52}.  The inferred $\overline{B_2}$ from the 1996 paper of NA52 was $(1.5 \pm 0.4) \times 10^{-3}$ GeV$^2/c^3$, similar to the E858 value \cite{Aoki:1992mb} and about an order of magnitude smaller than typical values for deuterons at lower energies.  The authors of Ref. \cite{Appelquist:1996qy} attributed the difference to an increase in the antiproton source size at AGS and SPS energies \cite{Mekjian:1977ei, Sato:1981ez}. The 1998 NA52 paper \cite{Ambrosini:1997bf} utilized the antideuteron $\overline{B_2}$ to infer a source radius in the context of the model of Sato and Yazaki \cite{Sato:1981ez}.  Radii for particles and antiparticles were consistent (for $\bar{d},~ R_{\rm rms} = 6.8 \pm 0.8$ fm and for $d,~ R_{\rm rms} = 7.2 \pm 0.7$ fm), and larger than the projectile radius, suggesting that the source volume expands prior to freezeout of the antideuterons and deuterons. 

%%%%%%%%%Figure%%%%%%%%%%%%%%%%
\begin{figure}[!hbt]
\centering
\centerline{\includegraphics[scale=0.38, bb=1000 -50 -230 600]{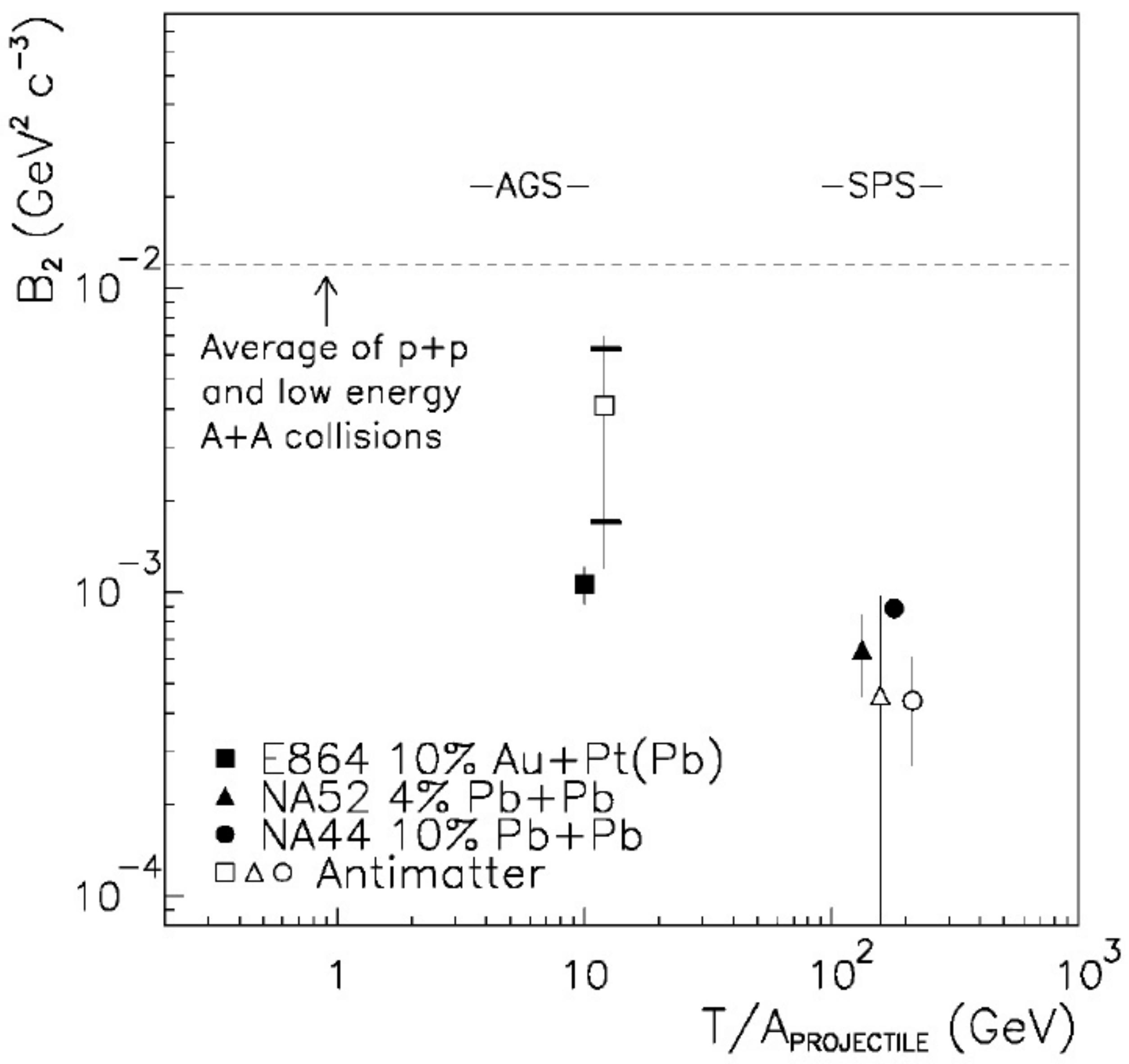}}
\caption{Coalescence parameters $B_2$ for antideuterons (open markers) and deuterons (solid markers) as a function of the projectile's total kinetic energy per mass number in central collisions of two heavy nuclei \cite{Armstrong:2000gd}. The measurements come from NA52 \cite{Appelquist:1996qy, Ambrosini:1997bf}, NA44 \cite{Bearden:2000we} and E864 \cite{Armstrong:2000gd, Armstrong:2000gz}.}
\label{fig:B2-E864}
\end{figure}
%%%%%%%%%%%%%%%%%%%%%%%%%%%%%%

The CERN NA44 collaboration reported results for antideuteron production in 0-10\% central collisions of a 158$A$ GeV Pb beam on a Pb target at the SPS \cite{Bearden:2000we}.  The NA44 spectrometer differed from that of other collaborations discussed above insofar as it selected secondaries emerging from the target at non-zero transverse momentum: the NA44 antideuteron acceptance was confined to $0.8 < p_T < 1.5$ GeV/$c$.  They conclude that source sizes based on deuteron and antideuteron coalescence are consistent with radii inferred from femtoscopy \cite{Lisa:2005dd} after applying appropriate corrections for various systematic effects, especially the assumption of a Gaussian source distribution in femtosopy analyses \cite{Bearden:2000we}.  The NA44 coalescence parameters are included in Fig.~\ref{fig:B2-E864}.  

In the year 2000, the Brookhaven E864 collaboration \cite{Armstrong:2000gd, Armstrong:2000gz} reported $\overline{B_2}$ and $B_2$ results for a 11.5$A$ GeV Au beam on a Pb or Pt target with a 0-10\% central trigger.  One of their papers mostly focuses on antideuterons \cite{Armstrong:2000gd} and includes the compilation plot reproduced in Fig.~\ref{fig:B2-E864}, showing $\overline{B_2}$ (antideuterons; open markers) and $B_2$ (deuterons; solid markers) as a function of the projectile's total kinetic energy per mass number in central collisions of two heavy nuclei; plotted points include data from NA52 \cite{Appelquist:1996qy, Ambrosini:1997bf}, NA44 \cite{Bearden:2000we} and E864 \cite{Armstrong:2000gd, Armstrong:2000gz}.  The E864 collaboration \cite{Armstrong:2000gd} concluded that these measurements favor the picture of Scheibl and Heinz \cite{Scheibl:1998tk} in which $\overline{B_2}$ and $B_2$ are about equal, and disfavor the alternative picture where $\overline{B_2}$ is smaller \cite{Mrowczynski:1989jd, Leupold:1993ms, Bleicher:1995dw}.

\subsection{Yields of Antinuclei in Heavy-Ion Collisions at RHIC and LHC
}\label{Yields-RHIC}

%%%%%%%%%Figure%%%%%%%%%%%%%%%%
\begin{figure}[!htb]
\centering
\centerline{\includegraphics[scale=0.30, bb=1200 -50 -230 750]{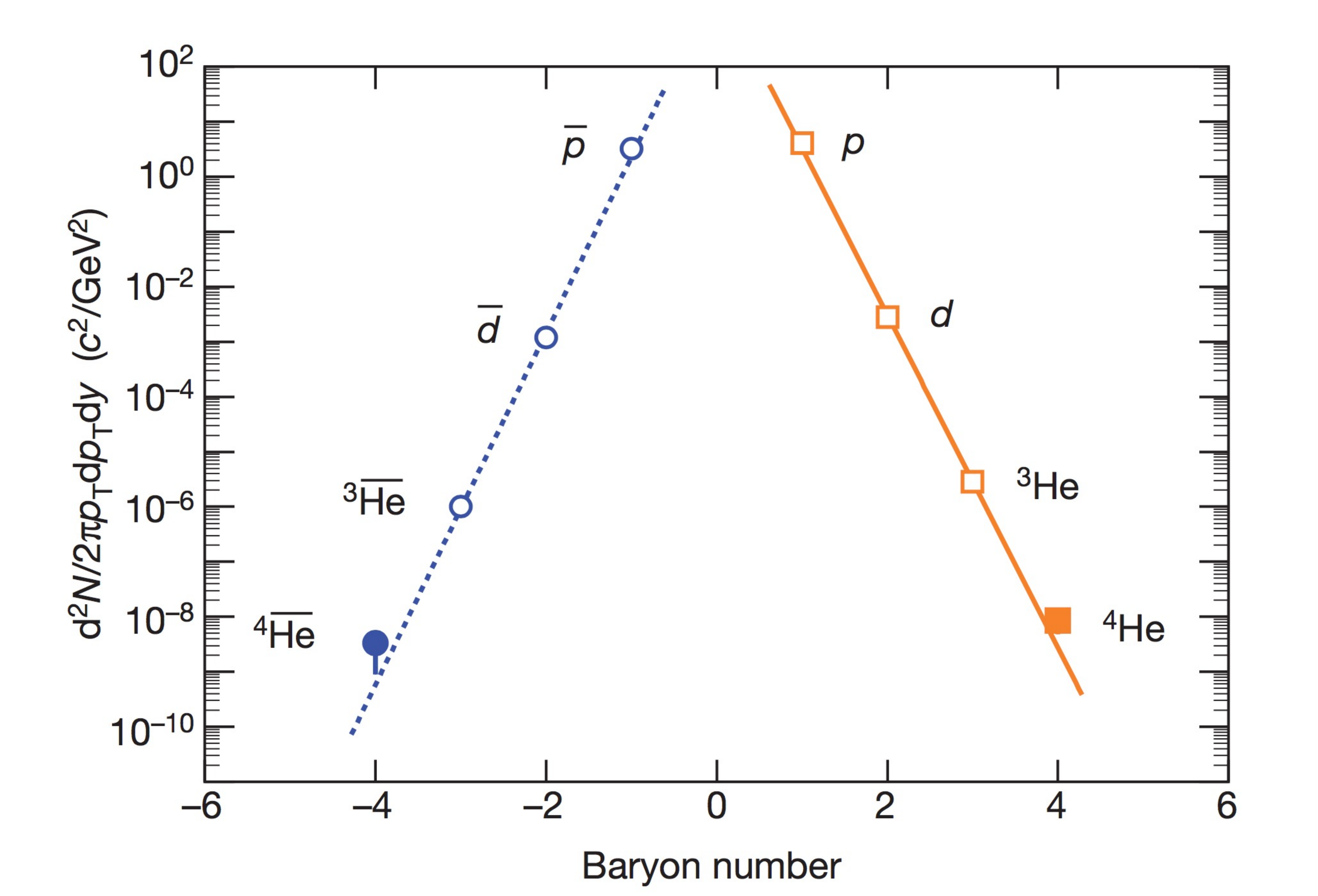}}
\caption{Differential invariant yields as a function of baryon number $B$. The yields were evaluated at $p_T/|B| = 0.875$ GeV/$c$ in central Au + Au collisions at $\sqrt{s_{NN}} = 200$ GeV.  The lines represent fits with an exponential formula $\propto$ $e^{-r|B|}$, where $r$ is the reduction factor (``penalty factor") for adding an additional (anti)nucleon~\cite{Agakishiev:2011ib}.}
\label{fig:RHIC_He4}
\end{figure}
%%%%%%%%%%%%%%%%%%%%%%%%%%%%%%

The BNL RHIC facility increases the available center-of-mass energy in nucleus-nucleus collisions by more than an order of magnitude over Pb + Pb collisions at the CERN SPS. Measurements of the antiproton-to-proton ratio at midrapidity~\cite{Adler:2001bp, Adams:2005dq} indicate that the central collision region at RHIC closely approaches the limit of zero net baryons.  Such a system with large multiplicity and small net-baryon density is well suited for the production of light antinuclei.  Measurements of antideuteron and $^3\overline{\rm He}$ production in Au + Au collisions at $\sqrt{s_{NN}} = 130$ GeV~\cite{Adler:2001uy} show that the production rates are much larger than in nucleus-nucleus collisions at lower energies.  A coalescence model analysis of the antideuteron and $^3\overline{\rm He}$ yields indicate that there is little or no increase in the antinucleon freezeout volume compared to collisions at CERN SPS energy~\cite{Adler:2001uy}.  The powerful capabilities of RHIC also led to the discovery of the $^4\overline{\rm He}$ antinuclei~\cite{Agakishiev:2011ib}, which are discussed in detail in Section~\ref{RHIC-AH3L-AHE4}.

The STAR collaboration at RHIC also carried out a study of antinucleus production up to $|A| =4$, extending the Protvino-based findings for proton-nucleus collisions presented in Fig.~\ref{fig:antiHe3}. Fig. \ref{fig:RHIC_He4} \cite{Agakishiev:2011ib} presents invariant momentum-space densities as a function of baryon number $B$ over the range -4 to +4. These yields were evaluated at $p_T/|B| = 0.875$ GeV/$c$ in central Au + Au collisions at $\sqrt{s_{NN}} = 200$ GeV.  The solid line and the dashed line are fits with an exponential formula $\propto e^{-r|B|}$ for nuclei and antinuclei, respectively, where $r$ is the production reduction factor, also known as the penalty factor.  For nucleons, $r =1.1^{+0.3}_{-0.2}\times 10^3$; for antinucleons, $r =1.6^{+1.0}_{-0.6}\times 10^3$. The reduction factor for adding an additional antinucleon at Protvino~\cite{Antipov:1970uc} is about a factor of 10 higher than at RHIC.  A calculation based on the thermodynamic model of Andronic {\it et al.} \cite{Andronic:2010qu} can predict ratios of the yields at RHIC: $\rm{^4He/^3He} = 3.1\times 10^{-3}$ and $\rm{^4\overline{He}/^3\overline{He}} = 2.4 \times 10^{-3}$, consistent with the measurement in Fig.~\ref{fig:RHIC_He4}. The reasoning outlined above can give a good prediction of the yield of heavier antinuclei. The yield of the stable antimatter nucleus next in line for discovery ($B=-6$) is predicted to be down by a factor of $2.6 \times 10^6$ compared to $\rm{^4\overline{He}}$, and is thus well beyond the reach of current accelerator technology.

%%%%%%%%Figure%%%%%%%%%%%%%%%%%%%%
\begin{figure}[!htb]
\centering
\centerline{\includegraphics[scale=0.45,bb=800 -50 -230 700]{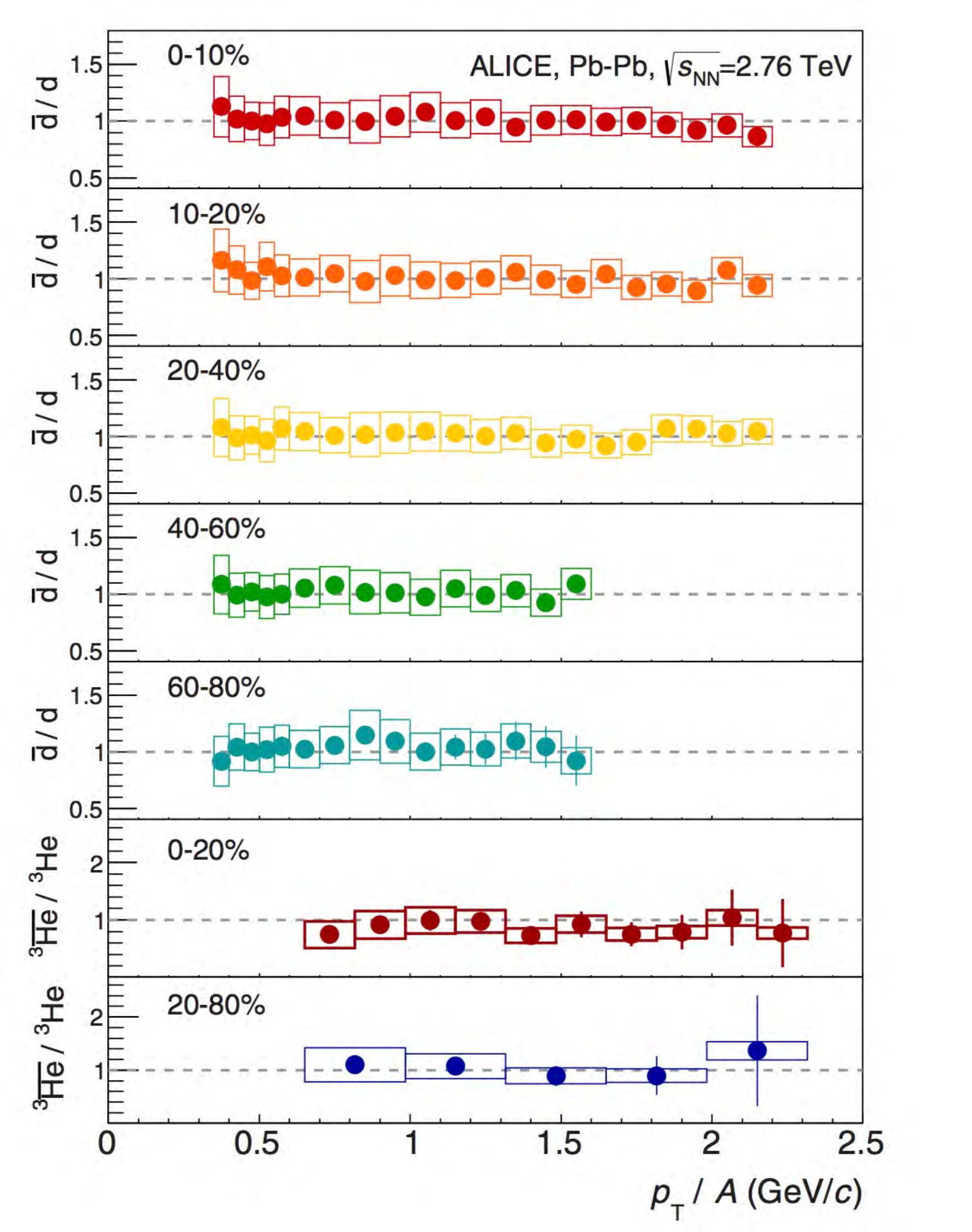}}
\caption{ Ratios of $\bar{d}/d = (\bar{p}/p)^2$ and $\rm{^3\overline{He}/^3He}$ versus $p_T$ per nucleon for various centrality classes in Pb + Pb collisions at $\sqrt{s_{NN}} = 2.76$ TeV~\cite{Adam:2015vda}. Error bars indicate statistical uncertainties, and boxes are systematic uncertainties.}
\label{fig:ALICE_yield}
\end{figure}
%%%%%%%%%%%%%%%%%%%%%%%%%%%%%%

The most energetic heavy-ion accelerator, the LHC at CERN, now provides up to 25 times the center-of-mass energy attainable at RHIC, while the LHC operated at almost 14 times RHIC energy during most of the time since operations began in 2010.  Antinucleus production relative to the coresponding nucleus at $\sqrt{s_{NN}} = 2.76$ TeV is indistinguishable from unity, as presented in Fig.~\ref{fig:ALICE_yield}~\cite{Adam:2015vda}.  This figure reproduces $\bar{d}/d$ and $\rm{^3\overline{He}/^3He}$ ratios for various centrality bins in Pb + Pb collisions at $\sqrt{s_{NN}} = 2.76$ TeV, as published by the ALICE collaboration.  All reported ratios are consistent with unity within errors, and exhibit a flat trend versus $p_T$ as well as versus collision centrality.  The ALICE collaboration concluded that the reduction factor is 307 $\pm$ 76 in 0-20\% central Pb + Pb collisions~\cite{Adam:2015vda}.  This penalty factor is about 5 times smaller than the corresponding factor at RHIC.  The ALICE experiment also measured  antinucleus production in $pp$ collisions at $\sqrt{s} = 7$ TeV~\cite{Acharya:2017fvb}, where the antideuteron rapidity density $dN/dy$ at midrapidity is as high as $10^{-4}$ and the $dN/dy$ for $^3\overline{\rm He}$ is $10^{-7}$. The reduction in the yield for each additional antinucleon under these conditions is thus about $10^3$. The ALICE paper~\cite{Acharya:2017fvb} also reports the first A=3 antinuclei observation in $pp$ collisions, where 6 $\rm{\overline{t}}$ and 10 $\rm{^3\overline{He}}$ were detected.

\subsection{Mechanisms for Production of Light Antinuclei}
 
%%\subsubsection{Proposed Mechanism for Production of Light Antimatter Nuclei: Vacuum Excitation}

The most successful and well tested models of antinucleus production are based on a thermal/statistical approach, or assume that antinucleons form antinuclei via statistical coalescence.  These established approaches are reviewed below in subsections \ref{therm} and \ref{coal}.  However, alternative production mechanisms have been discussed in the literature. Here we consider hypothesized collective antimatter production via vacuum excitation, advanced by Walter Greiner. Ref.~\cite{Greiner:2010zzb} offers the most up-to-date formulation of this idea, which has been developed over a period of time \cite{Greiner:1985ce}. 

The collective antimatter production mechanism has an analogy with spontaneous positron emission and the vacuum decay process in QED for strong fields~\cite{Muller:1984zz, Greiner:2003xg, Greiner:2010zzb,Greiner:1985ce}. In the framework of meson field theory, the energy spectrum of baryons has a peculiar structure, with an upper and a lower continuum, as in the familiar case of electrons. Of special interest in the case of the baryon spectrum is the potential well, built of the scalar and vector potential, which rises from the lower continuum~\cite{Auerbach:1986wrs}. It is known since the work of Dirac in the late 1920s that the negative energy states of the lower continuum have to be occupied, with the Pauli principle preventing decay. 

%%%%%%%%%%FIGURE%%%%%%%%%%%%%%%%%
\begin{figure}[!htb]
\centering
\includegraphics[scale=0.72,bb=420 -50 -230 150]{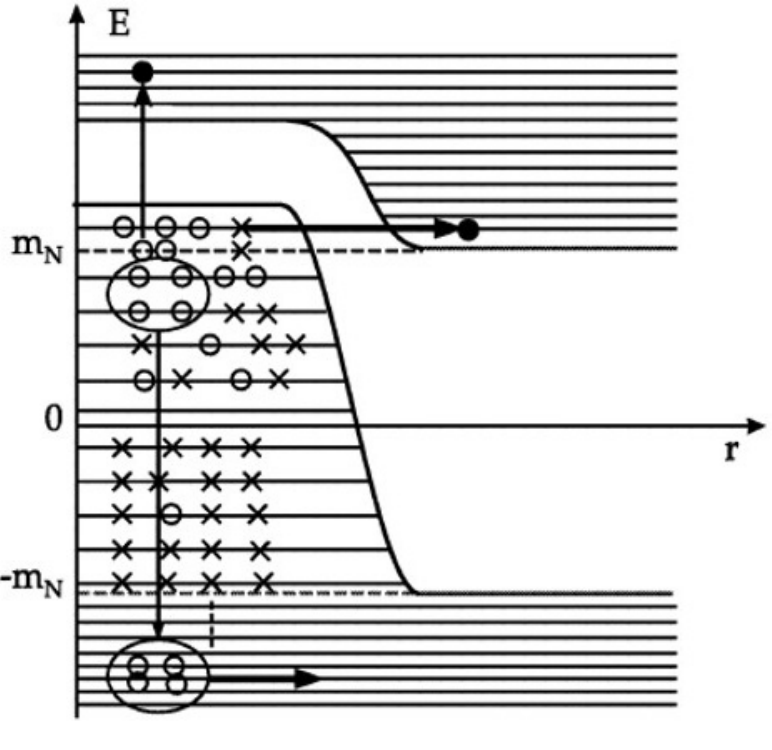}
\includegraphics[scale=0.72,bb=50 -50 -230 150]{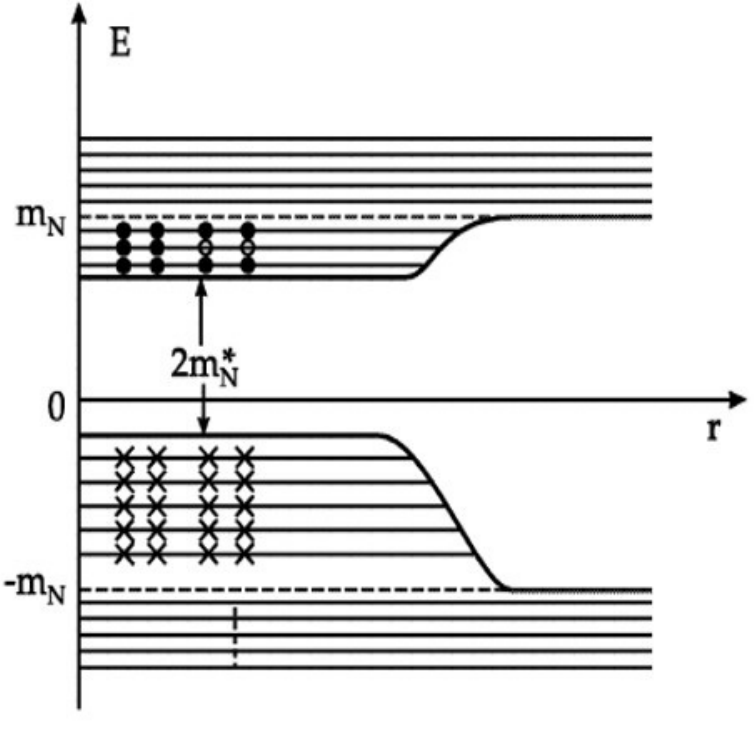}

\caption{Schematic potential well for nucleons and antinucleons at low (left) and high (right) nucleon densities. The horizontal arrow in the panel on the right denotes spontaneous creation of a baryon-antibaryon pair, while the antibaryons occupy bound states in the lower potential well. Such a situation, where the lower potential well reaches into the upper continuum, is called supercritical. Four of the bound hole states (bound antinucleons) are marked by a circle in the panel on the right, to illustrate possible formation of a ``quasi-antihelium" cluster. Greiner hypothesized that the energetic dynamics of the heavy-ion collision process might be capable of ejecting the quasi-antihelium into the lower continuum. If this mechanism were to occur under certain conditions, then the antinucleus production characteristics could deviate from what has been observed to date. This figure is reproduced from Refs.~\cite{Greiner:2010zzb}.}
\label{fig:potential_well}
\end{figure}
%%%%%%%%%%%%%%%%%%%%%%%%%%%%%%%%

It is important that the mechanism for the production of an antimatter cluster out of the highly correlated vacuum, as depicted in Greiner's conceptual diagram (Fig.~\ref{fig:potential_well}) does not proceed via phase space. The required coalescence of many particles in phase space suppresses the production of clusters, while it is favored by the direct production out of the highly correlated vacuum. In a certain sense, the highly correlated vacuum is a kind of ``cluster vacuum".

Greiner and others have pointed out that there are two problems associated with the coalescence picture for antinuclei, including the rather small rate, due to the small coalescence probability in phase space. The second problem is the reabsorption of antimatter fragments by the great amount of normal matter in the vicinity of the collision fireball; while it is argued that such creation and annihilation processes can be treated reasonably well using one of the nuclear transport models developed for describing collisions at high relativistic energies~\cite{Sorge:1989dy, Sorge:1990fw, Sorge:1992ej, Bass:1998ca, Bleicher:1999xi, Lin:2004en}, such calculations may underestimate the antinucleus production rate by many orders of magnitude if the hypothesized alternative mechanisms are possible.  Nonetheless, no experimental evidence has so far been found for collective antimatter production via vacuum excitation.   

\subsubsection{The Thermal-Statistical Approach}\label{therm}

Thermal models have been extensively used in describing the abundance of hadrons, including antinuclei, in heavy-ion collisions~\cite{Andronic:2010qu, Wheaton:2004qb, Torrieri:2004zz, Torrieri:2006xi, Petran:2013dva, Cleymans:2011pe}. Within the framework of this type of model, the system created in high-energy heavy-ion collisions is characterized by the chemical freeze-out temperature ($T_{\rm ch}$), kinetic freeze-out temperature ($T_{\rm kin}$), as well as the baryon, strangeness and charge chemical potentials, $\mu_B$, $\mu_S$ and $\mu_Q$, respectively. The thermal model \cite{BraunMunzinger:2003zd} represents a particle as an object emitted by the fireball with production rate 

\begin{equation}
n_i(T,\vec\mu)=\frac{\langle N_i \rangle}{V}=\frac{Tg_i}{2\pi^2}\sum_{k=1}^{\infty}\frac{1}{k^2}\lambda_{i}^{k}m_i^2K_2(\frac{km_i}{T}),\\\
\label{eq:thermal}
\end{equation}
where $\vec\mu = (\mu_B,\, \mu_S,\, \mu_Q)$, $g_i$ is the spin-isospin degeneracy factor, $K_2$ is the modified Bessel function, and fugacity is given by $\lambda_i = \exp(\frac{B_i\mu_B + S_i\mu_S + Q_i\mu_Q}{T})$. Antinuclei can contribute to the determination of the baryon chemical potential via thermal model fits of average production yields~\cite{Acharya:2017bso}.

%%%%%%%%%%%%FIGURE%%%%%%%%%%%%%%%%
\begin{figure}[!htb]
\centering
\centerline{\includegraphics[scale=0.98,bb=560 -50 -230 260]{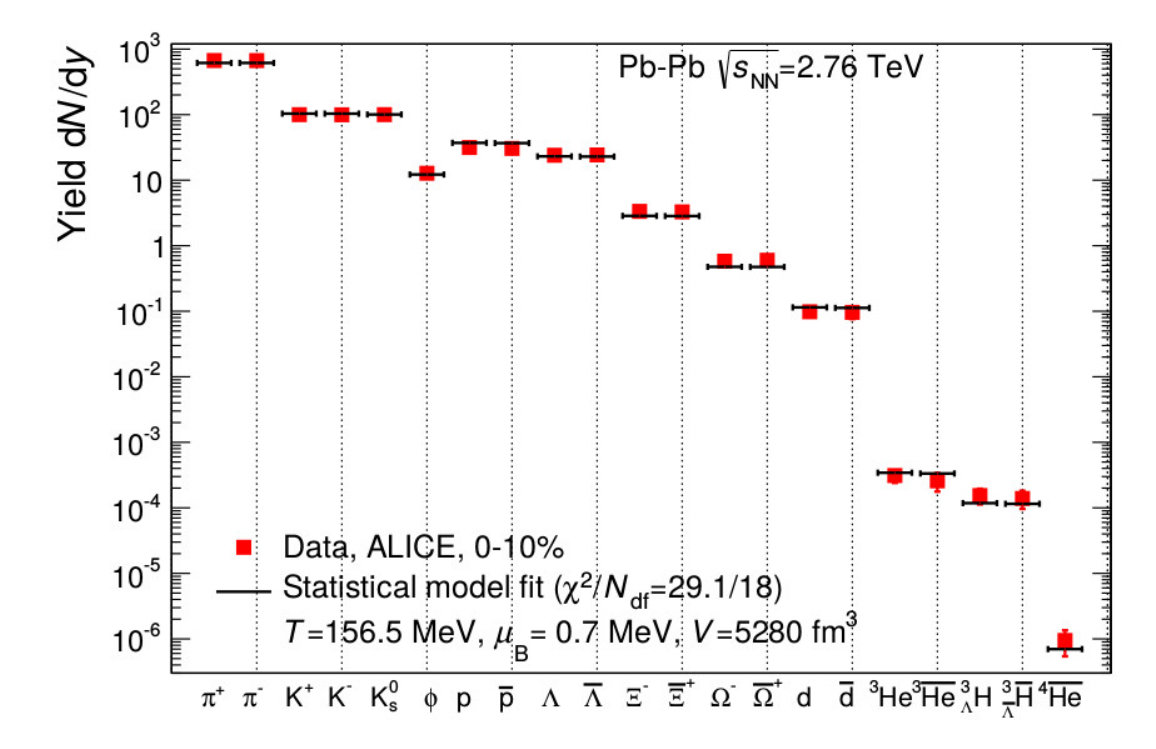}}
\caption{Thermal model fits to hadron multiplicities in central (0-10\%) Pb + Pb collisions at the LHC. This figure is reproduced from Ref.~\cite{Andronic:2016nof}.}
\label{fig:thermal-model2}
\end{figure}
%%%%%%%%%%%%%%%%%%%%%%%%%%%%%%%%

Fig.~\ref{fig:thermal-model2} shows thermal fits to the hadron yields in central Pb + Pb collisions at $\sqrt{s_{\rm NN}}$ = 2.76 TeV in comparisons with data. The fits with $T_{\rm chem}$ = 156 MeV, $\mu_B$ = 0.7 MeV and $V$ = 5280 fm$^3$ describe the data very well over 9 orders of magnitude~\cite{Andronic:2016nof}. The chemical freeze-out temperature becomes 154 $\pm$ 4 MeV if only data for nuclei (deuterons, \he, $\rm{^4He}$ and $\rm{^4\overline{He}}$) are considered in the fit \cite{Andronic:2010qu, Cleymans:2011pe}. It is impressive that a simple model with a small number of parameters describes the production of a large variety of hadrons over a wide range of energy with such good accuracy. Another important observation comes from the $d/p$ and \he/$p$ ratios, which are constant as a function of centrality~\cite{Adam:2015vda}. It is expected from a thermal statistical interpretation that $T_{\rm chem}$ and $\mu_B$ do not vary with centrality in high-energy nuclear collisions.

\subsubsection{The Coalescence Model}\label{coal}

%%%%%%%%%%%%FIGURE%%%%%%%%%%%%%%%%
\begin{figure}[!htb]
\centering
\centerline{\includegraphics[scale=0.35,bb=1300 -50 -230 650]{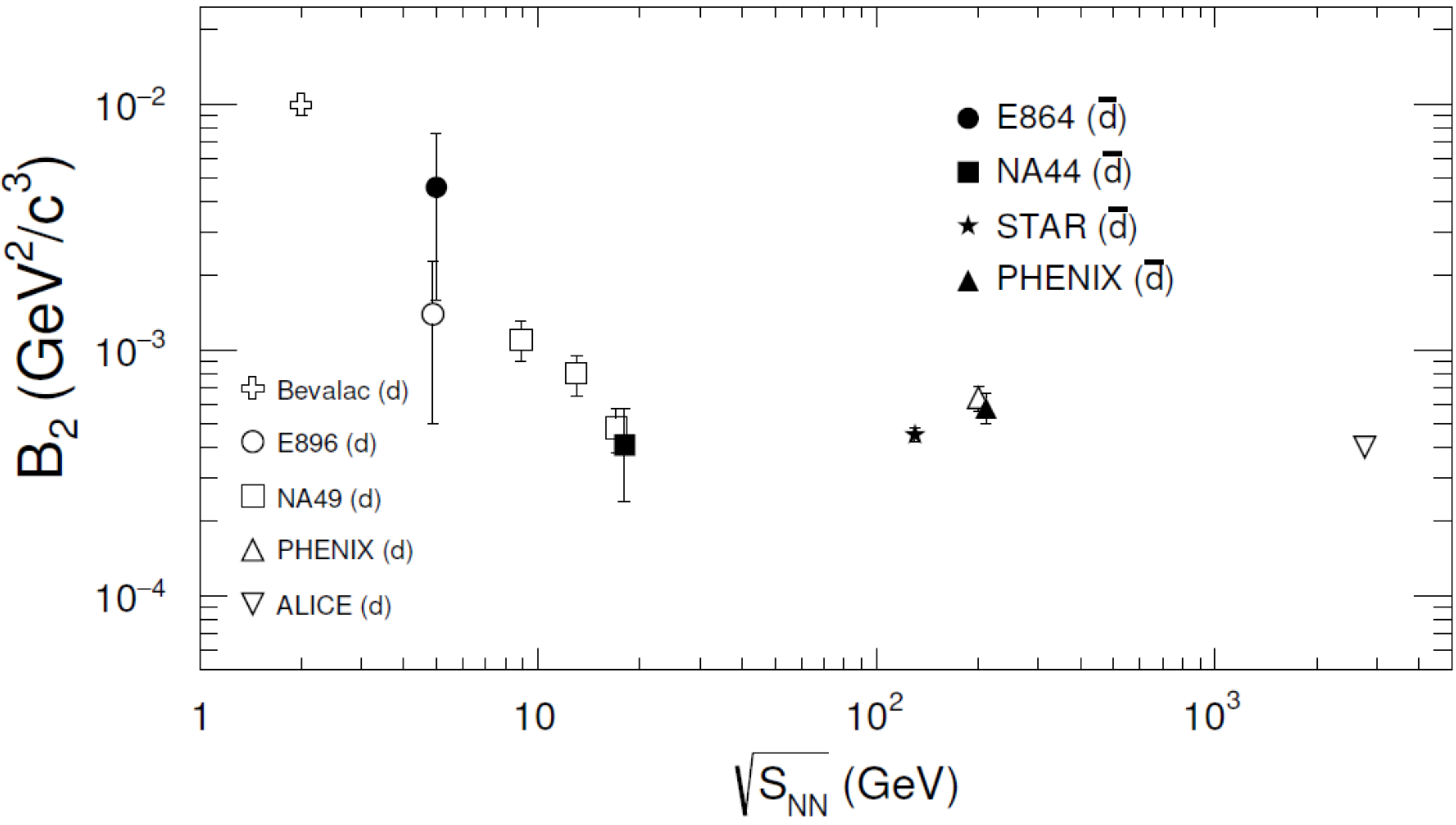}}
\caption{A compilation of coalescence parameters, $B_2$, for deuterons and antideuterons in central heavy-ion collisions~\cite{Wang:1994rua,Albergo:2002gi,Armstrong:2000gd,Anticic:2004yj,Bearden:2000we,Adler:2001uy,Adler:2004uy,Adam:2015vda}.}
\label{fig:B2}
\end{figure}
%%%%%%%%%%%%%%%%%%%%%%%%%%%%%%%%
The physics framework of the coalescence model for description of the production of light nuclei and antinuclei has been discussed in Section~\ref{pA} in the context of proton-nucleus collisions, and in Sections~\ref{Yields-HIC} and \ref{Yields-RHIC} in the context of relativistic heavy-ion collisions. Coalescence models based on Eqs.~(\ref{eq:coalescence1}) and (\ref{eq:coalescence2}) generally offer a reasonable description of the production of light nuclei and antinuclei at RHIC and LHC~\cite{Shah:2015oha, Zhu:2015voa, Sun:2015ulc}.

Figure~\ref{fig:B2} shows the coalescence parameter $B_2$ for deuterons and antideuterons in central Au+Au or Pb+Pb or Au+Pt(Pb) collisions, where the results at PHENIX and ALICE are evaluated at $p_T=1.3$ GeV/$c$ and 0-20\% centrality~\cite{Adler:2004uy,Adam:2015vda}. Figure~\ref{fig:B2} demonstrates that $B_2$ decreases rapidly from Bevalac energies through AGS and SPS energies, and then saturates at RHIC and LHC~\cite{Adler:2004uy, Adam:2015vda}. The value $B_2 \sim 4 \times 10^{-4}$ GeV$^2/c^3$ at $\sqrt{s_{\rm NN}}$ = 2.76 TeV~\cite{Adam:2015vda} is only marginally lower than the observation $B_2 \sim (6 \pm 1) \times 10^{-4}$ GeV$^2/c^3$ at RHIC~\cite{Adler:2004uy}.

%%%%%%%%%%%%FIGURE%%%%%%%%%%%%%%%%
\begin{figure}[!htb]
\centering
\centerline{\includegraphics[scale=0.5,bb=960 -50 -230 500]{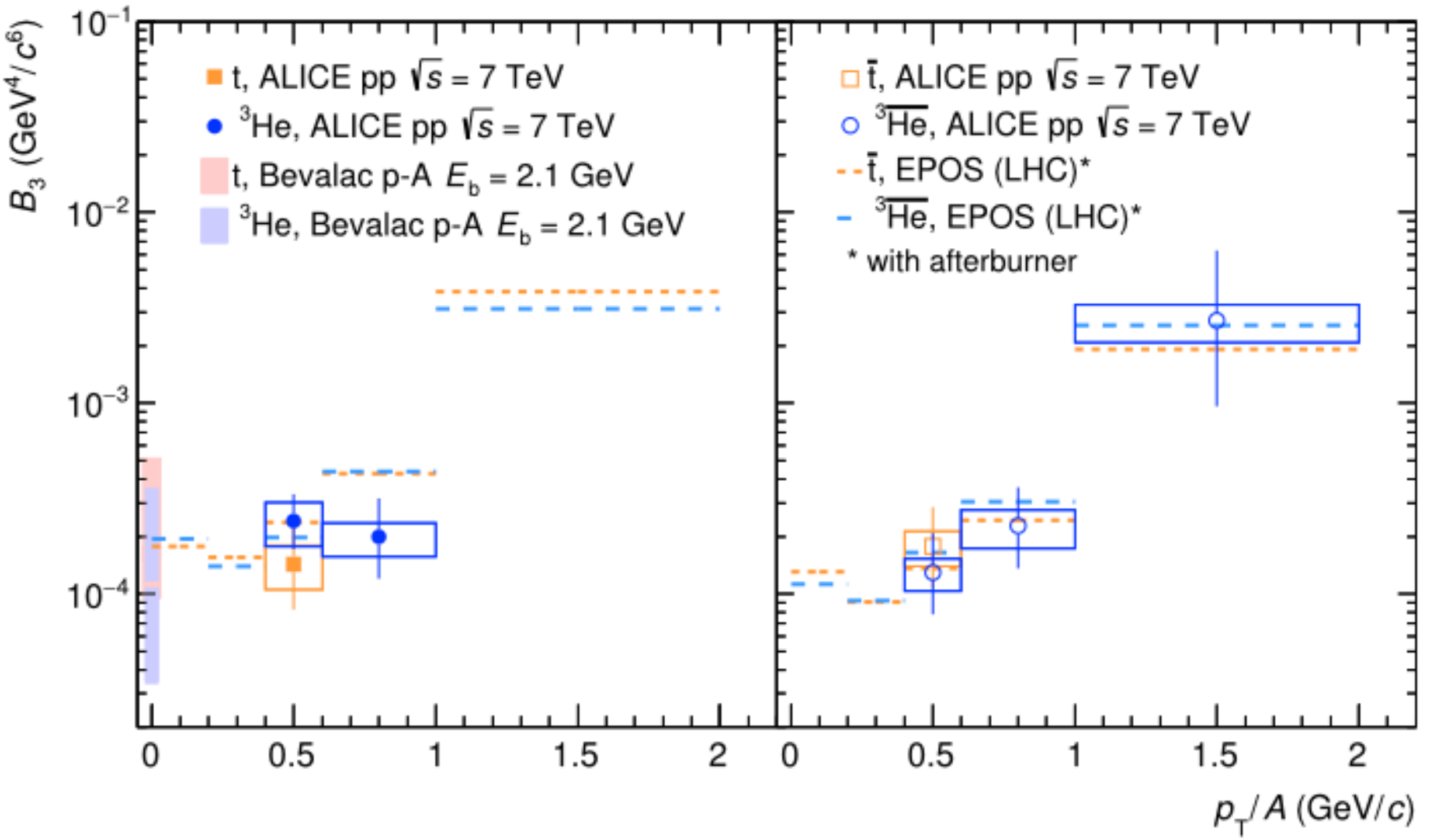}}
\caption{ Coalescence parameter $B_3$ for tritons and \he~(left panel), and for the corresponding antinuclei (right panel), in inelastic $pp$ collisions at $\sqrt{s} = 7$ TeV~\cite{Acharya:2017fvb}.}
\label{fig:B3}
\end{figure}
%%%%%%%%%%%%%%%%%%%%%%%%%%%%%%%%

Figure~\ref{fig:B3} presents coalescence parameters $\overline{B_3}$ for $A = -3$ and $B_3$ for $A = 3$ clusters produced in $pp$ collisions at $\sqrt{s} = 7$ TeV, measured by the ALICE collaboration~\cite{Acharya:2017fvb}. The observed $p_T$ dependence can be described by a QCD-inspired event generator with a coalescence-based afterburner~\cite{Acharya:2017fvb}. However, in heavy systems such as Au+Au and Pb+Pb, the prediction of the coalescence model that $B_n$ is independent of $p_T$ fails to reproduce early measurements at the AGS~\cite{Albergo:2002gi} as well as a recent measurement at the LHC~\cite{Acharya:2017dmc}, where significant variation of $B_2$ with $p_T$ for deuterons and antideuterons is reported. Overall, thermal models and coalescence models have each demonstrated some success in reproducing measured yields of nuclei and antinuclei, but with caveats as already noted \cite{Cleymans:2011pe, Steinheimer:2012tb, Xue:2012gx}.

\begin{figure}[!htb]
\centering
\centerline{\includegraphics[scale=0.80,bb=600 -50 -230 270]{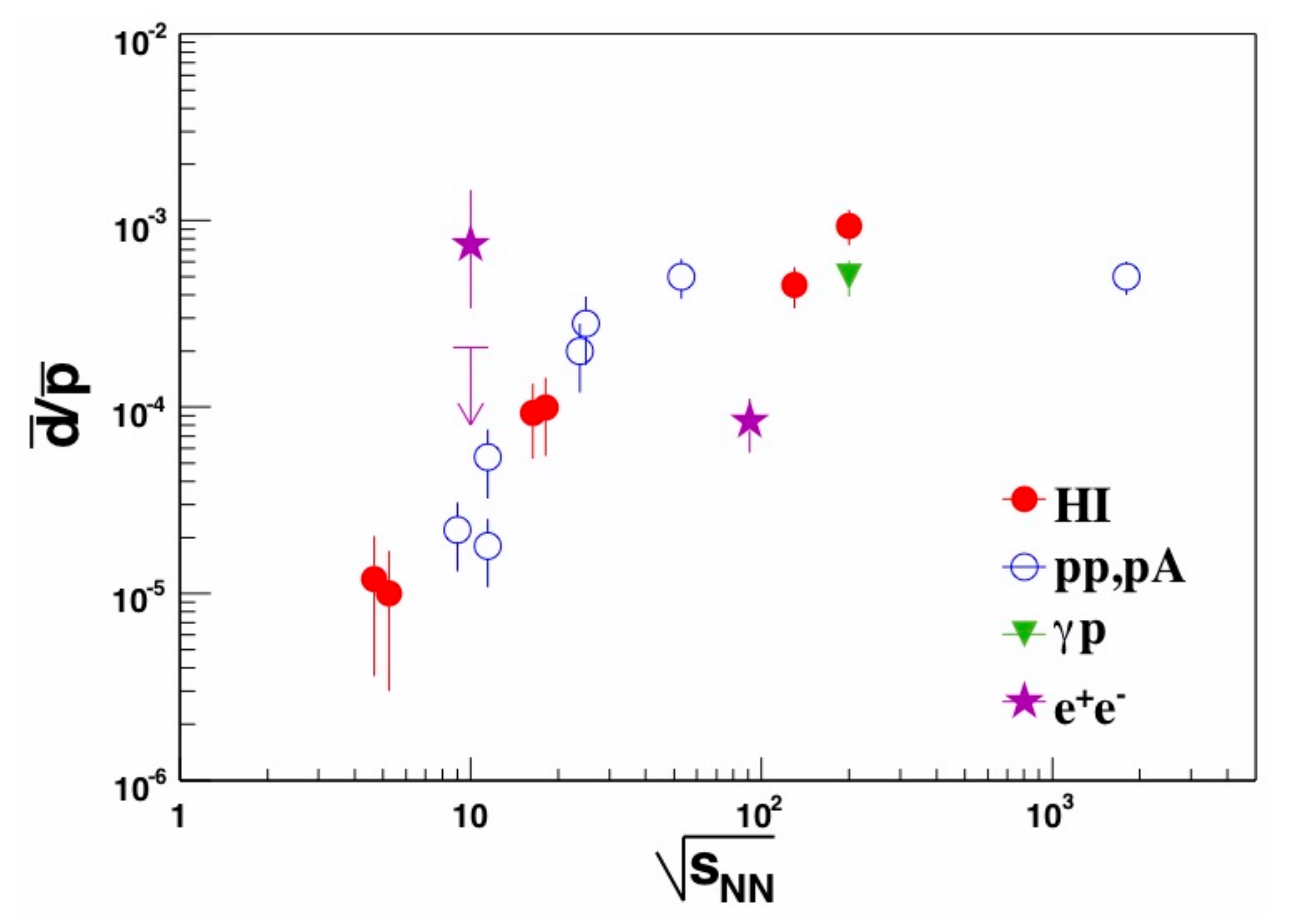}}
\caption{ A compilation by Liu and Xu \cite{Liu:2006my} of the antideuteron to antiproton ratio as a function of the center-of-mass energy of the colliding system. The datasets considered include collisions of heavy ions, $pA$, $pp$, $\bar{p}p$, $\gamma p$ and $e^+ e^-$~\cite{Binon:1969qz, Appel:1974fs, Aoki:1992mb, Ambrosini:1997bf, Armstrong:2000gd, Bearden:2000we, Armstrong:2000gz, Adler:2001uy, Adler:2004uy, Armstrong:1999xw, Alexopoulos:2000jk, Albrecht:1989ag, Aktas:2004pq, Schael:2006fd}. }
\label{fig:antid-antip-ratio}
\end{figure}

The observed ratio $\bar{d}/\bar{p}$ can be taken as a measure of the antibaryon phase-space density at kinetic freeze-out where coalescence also freezes. Figure \ref{fig:antid-antip-ratio} presents a compilation by Liu and Xu \cite{Liu:2006my, Steinheimer:2013lza} of the antideuteron to antiproton ratio as a function of the center-of-mass energy of a wide variety of colliding systems: $AA$, $pA$, $pp$, $\bar{p}p$, $\gamma p$ and $e^+ e^-$.  A univeral curve is observed in all cases except $e^+ e^-$, where there are reasons to expect a different pattern \cite{Liu:2006my}.  

%%%%%%%%%%FIGURE%%%%%%%%%%%%%%%%
%%%%%%%%%%%%%%%%%%%%%%%%%%%%%%%%

From Fig.~\ref{fig:antid-antip-ratio}, it is evident that the $\bar{d}/\bar{p}$ ratio increases from $10^{-5}$ at low energy to $10^{-3}$ at high energy. Each additional antinucleon added to an antinucleus decreases its production rate by that same penalty factor. At a center of mass energy of 100 GeV and above, the penalty factor is relatively flat, at slightly below $10^{-3}$. It is noteworthy that this effective measure of antibaryon density shows no difference among $pp$, $pA$ and $AA$ collisions. In heavy-ion collisions, more antiprotons are produced in each event than in $pp$ collisions. However, if more $pp$ collisions are accumulated such that $pp$ and $AA$ are compared with equivalent statistics, the same amount of coalesced antimatter is produced in $pp$ and $AA$ collisions. Accordingly, there are two determining factors related to the discovery at RHIC of the \hala~and the $^4\overline{\rm He}$: the energy to generate high antibaryon density for production of antinuclei, and high-luminosity heavy-ion collisions for efficient data collection and particle identification.  It is evident that the coefficient $B_n$ on its own does not provide a complete picture of the coalescence behavior of nucleons and antinucleons.  
  
\section{Recent Observation of Antinuclei in Heavy-Ion Collisions
}\label{RHIC-AH3L-AHE4}

\subsection{Observation of Antihypernuclei in Heavy-Ion Collisions}
Hypernuclei are nuclei in which at least one nucleon is replaced by a hyperon (a baryon that contains one or more strange quarks). The discovery of the first hyperons dates back to 1947, when Rochester and Butler~\cite{Rochester:1947} observed tracks produced by cosmic rays in a Wilson cloud chamber, which they interpreted as the spontaneous decay of a new type of elementary particle, one neutral and one charged. Three years later, Seriff {\it et al.}~\cite{Seriff:1950} obtained similar results in cloud-chamber observations made at a maximum altitude of 10,000 feet, and from a study of 11,000 cloud-chamber photographs, they verified many further examples of the observations of Rochester and Butler. In the same year, Hopper and Biswas~\cite{Hopper:1950} carried out an independent experiment using nuclear emulsion, where daughter tracks from a neutral parent were identified as a proton and pion. Furthermore, they concluded that the mass of the parent was 2370 $\pm$ 60 times the electron mass, assuming that the daughter products consisted solely of the observed proton and pion. It was recognized that the new particles reported by Rochester and Bulter \cite{Rochester:1947} were kaons, while the observations reported in Refs.~\cite{Seriff:1950} and~\cite{Hopper:1950} included $\Lambda$ hyperons. Over the next few years, many more hyperons were discovered, including the $\Sigma$ and the $\Xi$. 

In 1952, the first proton accelerator with energy in the GeV range, the Brookhaven Cosmotron, began operating. Soon it was possible to produce strange particles in the laboratory, whereas before this time, the only source of such particles had been cosmic rays. These advances led to the discovery of the $\Omega$~\cite{Barnes:1964}, the baryon with the highest strangeness content. Hyperons and their antiparticles enrich the physics of elementary particles, and their direct connection to this review is the hypernucleus and antihypernucleus, which are discussed further in this section.

The first hypernucleus was observed in a cosmic ray experiment by Danysz and Pniewski in 1953~\cite{Danysz:1953}. The lightest hypernucleus is the hypertriton (\hla) containing a proton, a neutron and a $\Lambda$ hyperon. The \hala~is likewise composed of an antiproton, an antineutron and a $\overline\Lambda$. 

\begin{figure}[htbp]
\centering
\centerline{\includegraphics[scale=0.68,bb=780 -30 -230 280]{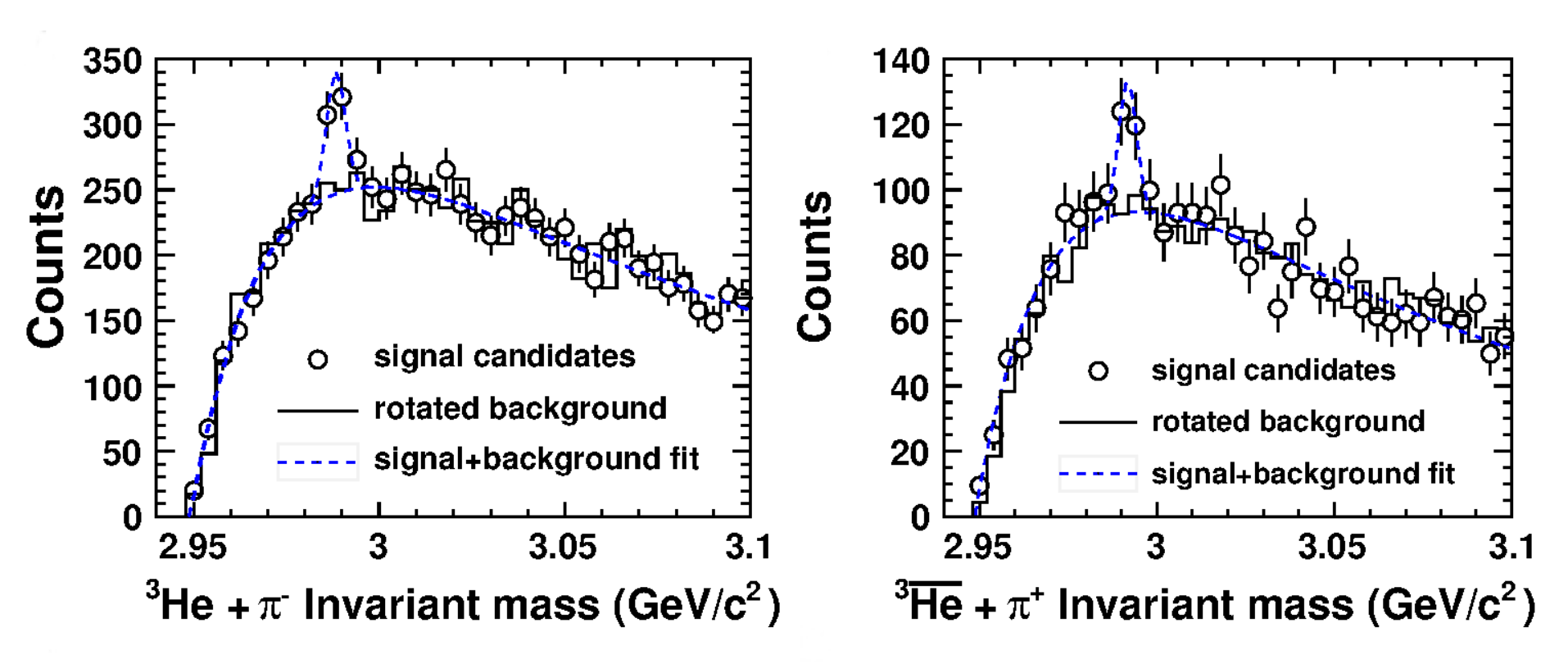}}
\caption{ The invariant mass distribution for the decay \hla$\rightarrow$\he+$\pi^{-}$ (left) and \hala$\rightarrow$\ahe+$\pi^{+}$ (right), in Au + Au collisions measured by the STAR collaboration~\cite{Abelev:2010rv}.}
\label{fig:antiH3L}
\end{figure}

\begin{figure}[htbp]
\centering
\centerline{\includegraphics[scale=0.76,bb=700 -30 -230 180]{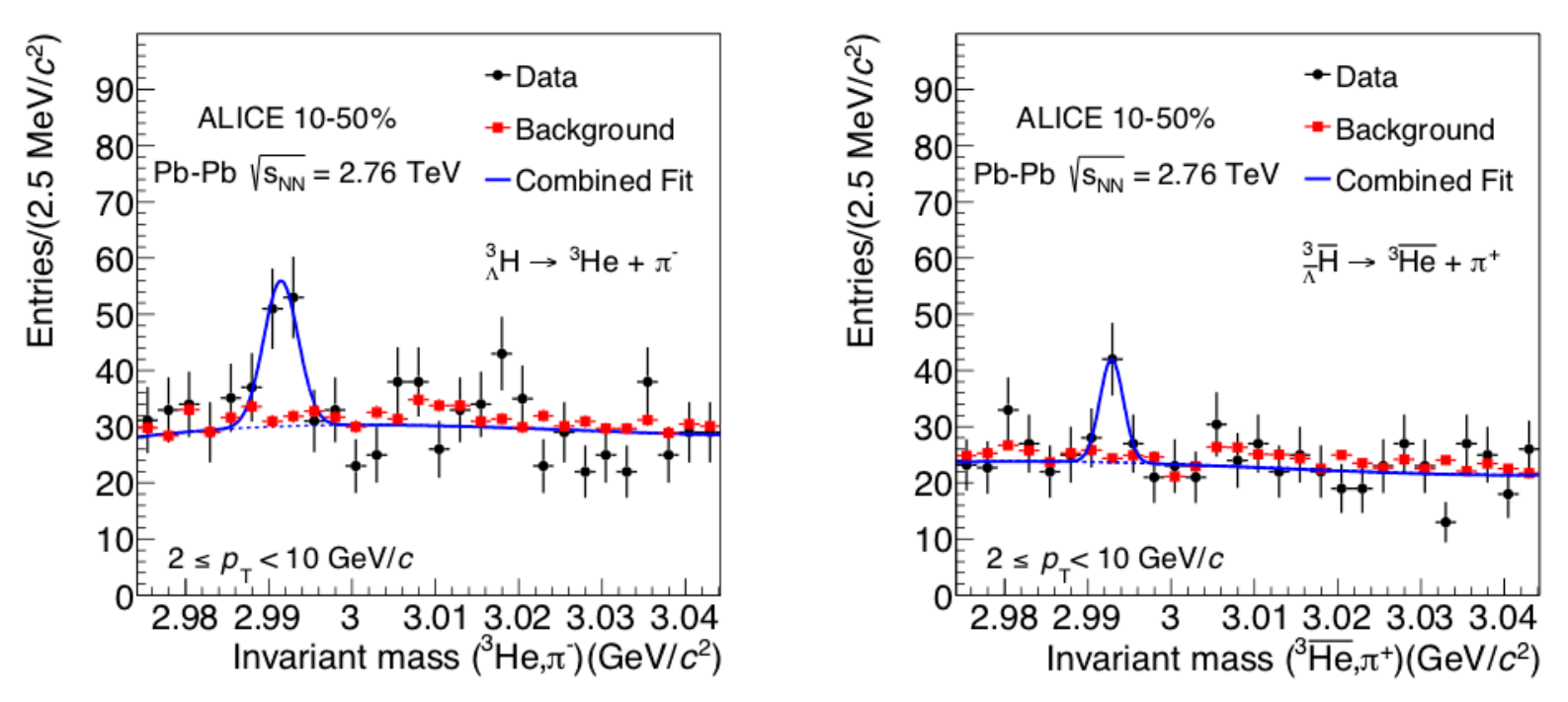}}
\caption{ The invariant mass distribution for the decay \hla$\rightarrow$\he+$\pi^{-}$ (left) and \hala$\rightarrow$\ahe+$\pi^{+}$ (right), in Pb + Pb collisions measured by the ALICE collaboration \cite{Adam:2015yta}.}
\label{fig:antiH3L2}
\end{figure}

The antihypertriton (\hala) is the next heaviest antinucleus after \ahe~and the antitriton. Although the \ahe~and $\bar{t}$ were discovered at proton accelerators in the 1970s, the detection of \hala~remained a challenge due to its short lifetime~\cite{Alberico:2001jb}. In 2010, the first observation of an antimatter hypernucleus was reported by the STAR experiment at RHIC~\cite{Chen:2009ku, Chen:2010zzc, Abelev:2010rv}. In STAR, 70 $\pm$ 17 \hala~candidates were observed in an analysis of the invariant mass distribution for \ahe~+~$\pi^+$ at a distance of a few centimeters away from the primary collision vertex. The data sample consisted of 9.1 $\times$ $10^7$ minimum-bias trigger events, along with 2.2 $\times$ $10^7$ central trigger events, corresponding to Au + Au collisions at $\sqrt{s_{NN}}$ = 200 GeV collected in 2004 and 2007. A total of 2168 \ahe~candidates were identified essentially free of background~\cite{Abelev:2010rv}. Figure~\ref{fig:antiH3L} shows the invariant mass distributions for the \ahe~+~$\pi^+$ decay channels, with mass $m(^3_{\bar{\Lambda}} \overline{\rm H}$) = 2.991 $\pm 0.001 \,(\mathrm{stat.}) \pm 0.002 \,(\mathrm{syst.})$ GeV/$c^{2}$. The STAR collaboration also observed 157 $\pm$ 30 \hla~candidates in the same datasets using the \he~+~$\pi^-$ channel (see Fig.~\ref{fig:antiH3L}), with a mass consistent with that of the \hala. In 2015, the production of \hala~and \hla~in Pb + Pb collisions at $\sqrt{s_{NN}}$ = 2.76 TeV was also measured by the ALICE experiment~\cite{Adam:2015yta}, as shown in Fig.~\ref{fig:antiH3L2}.  

%%%%%%%%%%%%%%%%%%%%%%%%%%%%%%%%%

%%%%%%%%%%%%%%%%%%%%%%%%%%%%%%%%%%%%%%%%%

Antihypernuclei can be produced in heavy-ion collisions through coalescence at the late stage of collision evolution. In heavy-ion collisions, the coalescence mechanism has been successful in explaining production of light nuclei, and also production of antinuclei like $\bar{d}$ and \ahe~\cite{Aoki:1992mb, Gutbrod:1988gt, Lemaire:1980qw, Jacak:1985zz, Hayashi:1988en, Saito:1994tg, Abbott:1994np, Wang:1994rua, Appelquist:1996qy, Ambrosini:1997bf, Armstrong:2000gd, Bearden:2000we, Armstrong:2000gz, BraunMunzinger:2007zz}. In this approach, the production ratio of $^3_{\bar{\Lambda}} \overline{\rm H}$ to $^3_{\Lambda} \rm H$ should be proportional to $(\bar\Lambda/\Lambda)(\bar{p}/p)(\bar{n}/n)$. This product of ratios from STAR data is 0.45 $\pm$ 0.08 (stat.) $\pm$ 0.10 (syst.)~\cite{Abelev:2010rv}, consistent with the observed number of 0.49 $\pm$ 0.18 (stat.) $\pm$ 0.07 (syst.) for $^3_{\bar{\Lambda}} \overline{\rm H}$/$^3_{\Lambda} \rm H$. Equilibration among strange quarks and light quarks is one of the proposed signatures of QGP formation~\cite{Koch:1986ud}, which would result in high (anti)hypernucleus yields. Both the STAR and ALICE collaborations have measured these ratios, which are summarized in Table~\ref{table_ratio}. The STAR results are higher than those of ALICE, but still consistent within uncertainties. One should note that by assuming a branching ratio of 25\%~\cite{Kamada:1997rv}, the STAR \ahe~and \he~yields in the ratio calculation have subtracted the feed-down contribution from \hala~and \hla~decay.

\begin{table}[h!]
\centering \caption{ Particle ratios from Au + Au collisions at 200 GeV and Pb + Pb collisions at 2.76 TeV. } \label{table_ratio}
%\begin{ruledtabular}
\begin{tabular}
{cc}\\ \hline \hline Particle type & Ratio\\\hline
Au+Au collisions:\\
\hala/\ahe &  0.89 $\pm$ 0.28 $\pm$ 0.13 \\
\hla/\he   &  0.82 $\pm$ 0.16 $\pm$ 0.12 \\
\\
Pb+Pb collisions at 0-10\% centrality:\\
\hala/\ahe & 0.42 $\pm$ 0.10 $\pm$ 0.13 \\
\hla/\he   & 0.47 $\pm$ 0.10 $\pm$ 0.13 \\
\hline\hline
\end{tabular}
%\end{ruledtabular}
\end{table}

\begin{figure}[!hbt]
\centering
\centerline{\includegraphics[scale=1.2, bb=460 -30 -230 240]{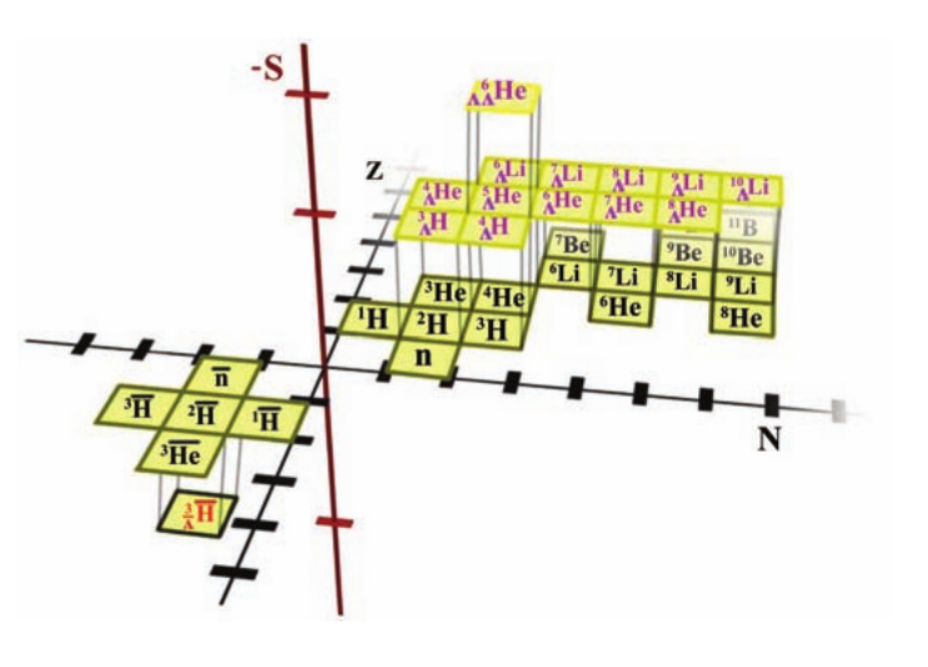}}
\caption{ The chart of the nuclides in the space of number of protons, $Z$, and number of neutrons, $N$, showing the extension into the strangeness ($S$) dimension~\cite{Abelev:2010rv}. Hypernuclei lie at positive ($N,Z$) above the plane and antihypernuclei in the negative ($N,Z$) region. }
\label{fig:Nucl_chart}
\end{figure}

The ability to produce antihypernuclei in relativistic heavy-ion collisions extends the conventional two-dimensional chart of the nuclides into a third dimension, based on the strangeness quantum number as shown in Fig.~\ref{fig:Nucl_chart}. With nonzero strangeness, new ideas related to the structure of nuclear matter can be explored~\cite{Heinz:1985pm, Greiner:1988pc}.

%%%%%%%%%%%%%%%%%%%%%%%%%%%%%%%%%
%%%%%%%%%%%%%%%%%%%%%%%%%%%%%%%%%%%%%%%%% 

Hypernuclei serve as an excellent laboratory to study hyperon-nucleon ($YN$) and hyperon-hyperon ($YY$) interactions, which are important for nuclear physics and nuclear astrophysics. The $YN$ interaction plays a crucial role in understanding the structure of neutron stars~\cite{Lattimer:2006xb}. Additional degrees of freedom, such as strangeness, lower the maximum mass of neutron stars~\cite{Burgio:2011wt, Lonardoni:2014bwa}.  This makes it extremely difficult to describe recent observations~\cite{Demorest:2010bx, Antoniadis:2013pzd} of neutron stars with masses above twice the solar mass.  Theoretical difficulties include the fact that realistic relativistic equations containing hyperons become soft at high densities; this is known as the ``hyperon puzzle". Possible solutions include deconfinement to quark matter, or alternative $YN$ and $YY$ couplings \cite{Dexheimer:2008ax, Weissenborn:2011ut, Steiner:2012rk, Lopes:2013cpa, Gusakov:2014ota, Gomes:2014aka, Dutra:2014qga, Bizarro:2015wxa}. 

On the other hand, our knowledge of the $YN$ interaction, measured in hypernuclei experiments, is hampered by limited statistics. Along with the binding energy of a hypernucleus, the strength of the $YN$ interactions affects hypernuclear lifetime~\cite{Rayet:1966, Kamada:1997rv}. Therefore, a precise determination of the lifetime of hypernuclei can provide direct information on the $YN$ interaction strength~\cite{Kamada:1997rv, Juric:1973zq}.

\begin{figure}[!hbt]
\centering
\centerline{\includegraphics[scale=0.65,bb=800 -30 -230 350]{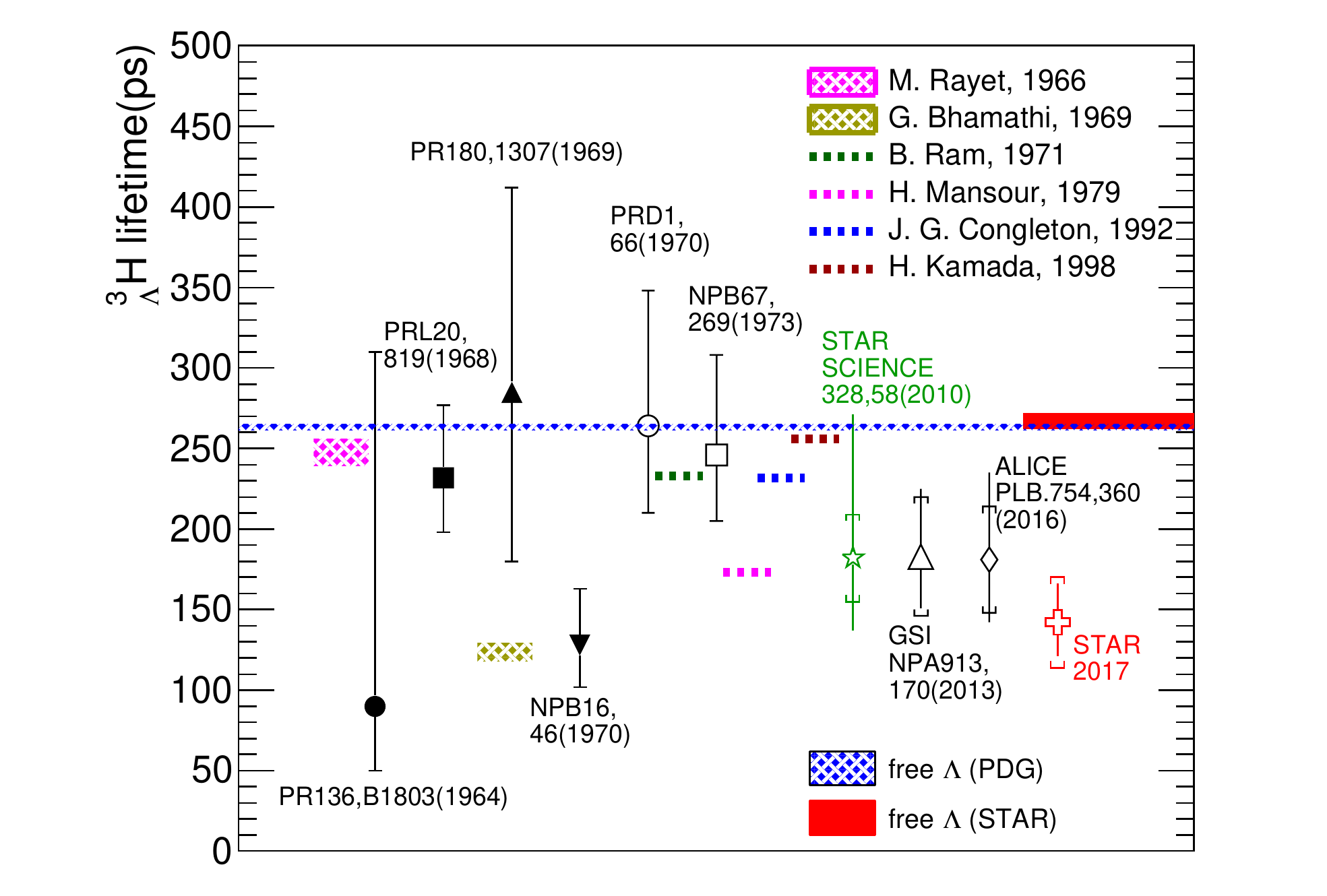}}
\caption{ The \hla~lifetime measurements from Refs.~\cite{Prem:1964, Phillips:1969uy, Bohm:1970se, Keyes:1970ck, Keyes:1974ev, Abelev:2010rv, Rappold:2013fic, Adam:2015yta,Adamczyk:2017buv}. The figure is reproduced from Ref.~\cite{Adamczyk:2017buv}.}
\label{fig:H3L_lifetime}
\end{figure}

The STAR experiment at RHIC, the HypHI experiment \cite{Saito:2016ulw} at the GSI Helmholtz Centre for Heavy Ion Research and the ALICE experiment at LHC have updated the lifetime of the hypertriton relatively recently~\cite{Abelev:2010rv, Rappold:2013fic, Adam:2015yta}. A 2014 reanalysis of the HypHI data and the worldwide measurements using a Bayesian analysis estimated the lifetime of the \hla~to be approximately $217^{+19}_{-16}$ ps and they specify an upper limit of 250 ps at the 95$\%$ confidence level~\cite{Rappold:2014jqa}. After including measurements from the ALICE experiment, the world average of the $^{3}_{\Lambda}{\rm H}$ lifetime shifts down to $215^{+18}_{-16}$ ps~\cite{Adam:2015yta}. The free $\Lambda$ lifetime is 263 $\pm$ 2 ps~\cite{Olive:2016xmw} and the world average of the $^3_\Lambda$H lifetime is shorter than that of the free $\Lambda$ by a factor of 0.82$^{+0.07}_{-0.06}$. The data are summarized in Fig.~\ref{fig:H3L_lifetime}.

%%%%%%%%%%%%%%%%%%%%%%%%%%%%%%%%%
%%%%%%%%%%%%%%%%%%%%%%%%%%%%%%%%%%%%%%%%%

A measurement of the \hla~lifetime that is shorter than the free $\Lambda$ lifetime challenges the current theoretical understanding of the \hla~as being comprised of a weakly-bound deuteron core and a $\Lambda$. This picture motivates the assumption that the \hla~lifetime is close to that of the free $\Lambda$. All currently available \hla~lifetime predictions are based on the assumption that the $\Lambda$ binding energy is very small, although this binding energy is poorly measured~\cite{Juric:1973zq, Keyes:1970ck}. The current measurements of the \hla~lifetime from heavy-ion experiments provide a new reference point for fine-tuning theoretical models~\cite{Rayet:1966, Kamada:1997rv, Bhamathi:1969ny, Mansour:1979, Kolesnikov:1988uy, Congleton:1992kk} and for advancing our understanding of the lightest hypernucleus. 
 
The precision of the \hla~lifetime is improved by measurements from the STAR collaboration based on data collected in 2010 and 2011~\cite{Adamczyk:2017buv} [see Fig.~\ref{fig:H3L_lifetime}], leading to a new lifetime estimate that is 50\% shorter than that of the free $\Lambda$, and contribute to a new worldwide average of $211^{+18}_{-16}$ ps. Those results, in combination with previous measurements, clearly motivate further study of \hla~\cite{DAVIS20053,Dalitz:2005mc,Gal:2016boi, Agnello:2016jlx, Liu:2017rjm}.

The STAR experiment will collect large datasets for Au + Au collisions during the upcoming Beam Energy Scan Phase-II program during 2019-2021~\cite{STAR:SN0598}. Similarly, improved measurements from the HypHI and ALICE experiments are also expected.  Looking ahead to the next decade, the Facility for Antiproton and Ion Research (FAIR) \cite{Senger:2017oqn} at Darmstadt is projected to begin operation, and will lead to measurements with further improvements in statistics~\cite{Steinheimer:2011hoa, Steinheimer:2013lza}. A precise experimental determination of the binding energy of \hla~and \hala~would help us to understand the structure of the lightest hypernucleus in detail. 

\subsection{Observation of Antimatter Helium-4
}\label{anti-He4}

One year after the observation of \hala, the first $^4\overline{\rm He}$ nuclei were observed by the STAR collaboration in Au + Au collisions at $\sqrt{s_{\rm NN}} = 200$ and 62.4 GeV.  A sample of one billion collision events was collected in 2007 and 2010~\cite{Agakishiev:2011ib, Xue:2011ej}. The STAR results combine energy loss ($\langle dE/dx \rangle$) in the TPC (see Fig.~\ref{fig:PID}) and the time of travel for tracks measured by the barrel TOF detector. Using the barrel TOF, the mass of particles can be calculated via $m^2 = p^{2}(t^2/L^2 -1)$, where $t$ and $L$ are the time of flight and path length of tracks, respectively. Meanwhile, the online high level trigger (HLT) was employed to perform preferential selection of collisions which contained tracks with electric charge $Ze =\pm 2e$ for fast offline analysis. In the STAR HLT analysis, a cut on the minimum distance from the collision vertex (Distance of Closest Approach, DCA) less than 3 cm for negative tracks, and 0.5 cm for positive tracks, was used to reject background. Then $n\sigma_{dE/dx}$, which quantifies the number of standard deviations of the observed track relative to the expected mean energy loss for that particle type (see. Eq.~\ref{Equ.PID_TPC2} in Sec.~\ref{Dect}), was applied as a further selection criterion to avoid misidentification of $^3\overline{\rm He}$ ($^3$He) and $^4\overline{\rm He}$ ($^4$He) at higher momenta. 

%%%%%%%%%%%%%%%%%%%%%%%%%%%%%%%%
\begin{figure}[!hbt]
\centering
\centerline{\includegraphics[scale=0.6,bb=720 -50 -230 520]{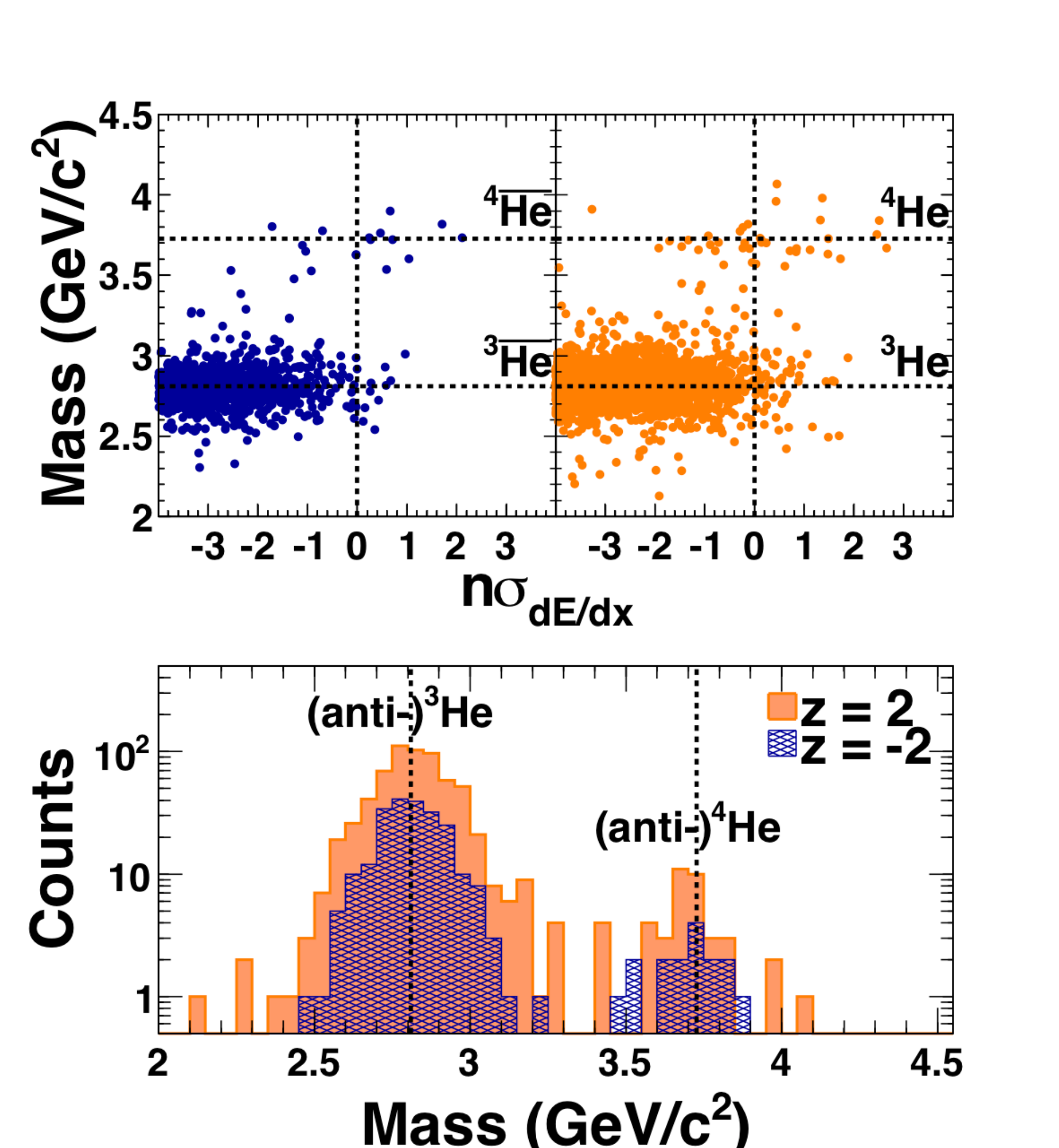}}
\caption{ The upper panel shows the distribution of n${\mathrm{\sigma_{dE/dx}}}$ versus mass in Au + Au collisions at $\sqrt{s_{\rm NN}} = 200$ GeV. The lower panel shows the projection of the same entries onto the mass axis~\cite{Agakishiev:2011ib}. }
\label{fig:antiHe4_tofpid}
\end{figure}
%%%%%%%%%%%%%%%%%%%%%%%%%%%%%%%%

Figure~\ref{fig:antiHe4_tofpid} shows the particle identification using the combination $n\sigma_{dE/dx}$ and mass measurement based on the TOF detector. Clusters of $^4\overline{\rm He}$ and $^4$He are located at the expected positions, and are clearly separated from $^3\overline{\rm He}$ and $^3$He, as well as from $^3$H and $^3\overline{\rm H}$.  By counting the $^4\overline{\rm He}$ signal within the cut windows $-2 < n\sigma_{dE/dx} < 3$ and 3.36 $<$ mass $<$ 4.04 GeV/$c^{2}$ in the upper panel of Fig.~\ref{fig:antiHe4_tofpid}, sixteen $^4\overline{\rm He}$ candidates were identified. Together with two $^4\overline{\rm He}$ candidates detected by the TPC alone in the year 2007, which are also presented in Fig.~\ref{fig:antiHe4_tofpid}, eighteen $^4\overline{\rm He}$ candidates were observed by the STAR collaboration, after a search over more than half a trillion tracks.  This constituted the rarest signal observed up to that time in any heavy-ion experiment.

\begin{figure}[!htb]
\centering
\centerline{\includegraphics[scale=0.6, bb=450 -50 100 400]{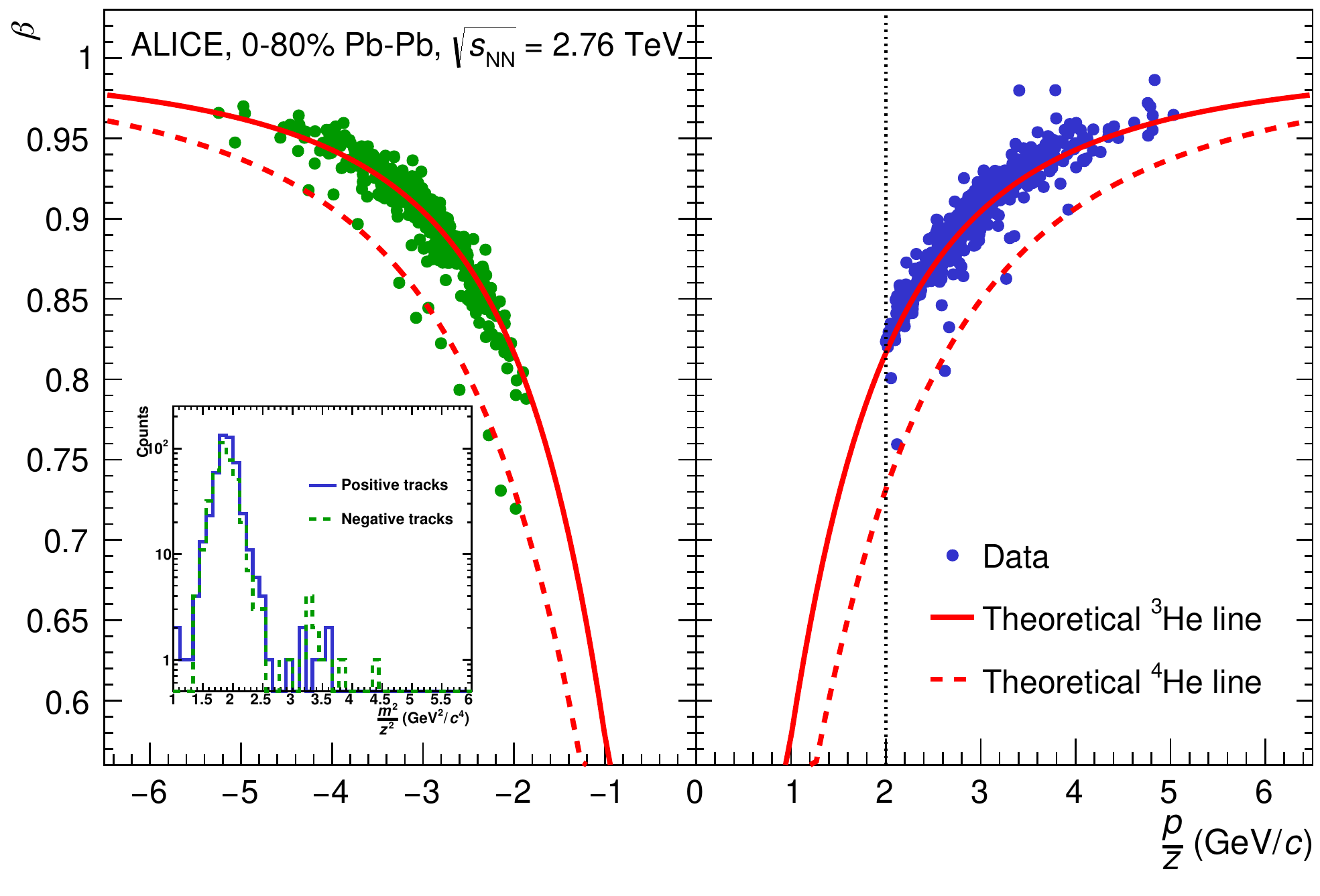}}
\caption{ Velocity, $\beta$, versus momentum per charge, measured using the ALICE TOF detector. A selection band of -1.5 to 3$\sigma$ around the mean of the TPC specific energy-loss distribution is applied in this figure~\cite{Acharya:2017bso}. }
\label{fig:antiHe4_tofpid_ALICE}
\end{figure}

In 2011, the ALICE collaboration reported at a conference~\cite{Sharma:2011ya} on their observation of four $^4\overline{\rm He}$ candidates in Pb + Pb collisions at $\sqrt{s_{\rm NN}} = 2.76$ TeV, while their final publication on this topic was published in 2018~\cite{Acharya:2017bso}. Fig.~\ref{fig:antiHe4_tofpid_ALICE} presents the velocity distributions of charged particles in the ALICE detector. A clean $^4\overline{\rm He}$ signal is observed, although statistics are limited in the LHC run-1 dataset. A factor of three times more statistics has been gathered at $\sqrt{s_{\rm NN}} = 5.02$ TeV in Pb + Pb collisions and will soon lead to further improvements compared with the analyses presented in Ref.~\cite{Acharya:2017bso}. 

%%%%%%%%%%%%%%%%%%%%%%%%%%%%%%%%%%%%%
%%%%%%%%%%%%%%%%%%%%%%%%%%%%%%%%%%%%%%

The $^4\overline{\rm He}$ is the heaviest antimatter nucleus observed to date, and its detection in heavy-ion collisions provides a point of reference for possible future observations in cosmic radiation, where hunting for antimatter and dark matter in the Universe is a science goal of very high interest and importance. The theoretical work of Blum {\it et al}.~\cite{Blum:2017,Blum:2017iwq} provides new insights on this topic. They took data on antinuclei for $pA$ and $AA$ collisions and combined it with the HBT scaling pointed out in Ref.~\cite{Scheibl:1998tk} to correctly predict $^3\overline{\rm He}$ production in $pp$ collisions at LHC energies. Their model predicts that the $^3\overline{\rm He}$ yield is 1-2 orders of magnitude higher than most earlier estimates~\cite{Blum:2017}. This in turn suggests that secondary production of antinuclei in the cosmos may occur at a measurable rate. We consider this point in more detail in Sec.~\ref{CR2}.

Considering the large penalty factor for producing antinuclei in heavy-ion collisions, in conjunction with the fact that there are no stable nuclides with $A=5$, it is likely that $^4\overline{\rm He}$ will remain the heaviest stable antimatter nucleus observed for the foreseeable future. Nevertheless, there is a theoretical calculation based on phase-space coalescence in the context of a special freezeout configuration associated with $^5\overline{\rm Li}$ and $^5$Li, whereby Sun and Chen argue that $^5\overline{\rm Li}$ might be feasible to observe at RHIC via the $^4\overline{\rm He} + \bar{p}$ decay channel~\cite{Sun:2015jta}.

\section{Properties of Antimatter Nuclei}

\subsection{Mass Difference of Antinuclei at the LHC
}\label{ALICE-mass}
As introduced in Sec. \ref{sym}, very precise measurements to search for a possible mass difference between the proton and antiproton have been carried out ­~\cite{Ulmer:2015jra, Gabrielse:1999kc, Hori:2011zz}, and large future improvements in such tests are expected \cite{Smorra:2015cuj}. The extension of {\it CPT} tests to nuclei and their antimatter partners is of particular interest, and opens the possibility to search for possible asymmetry associated with the nuclear binding mass in nuclei and antinuclei~\cite{Akindinov:2015hva}.  When explaining the importance of {\it CPT} tests involving the masses of nuclei and their antinuclei, Akindinov {\it et al.} emphasize their significance for understanding the color-confining transition from quark-gluon plasma to hadron gas~\cite{Akindinov:2015hva}.
 
%%%%%%%%%%%%%%%%%%%%%%%%%%%%%%
 \begin{figure}[!htb]
 \centering
\centerline{\includegraphics[scale=0.38, bb=1060 -50 -230 600]{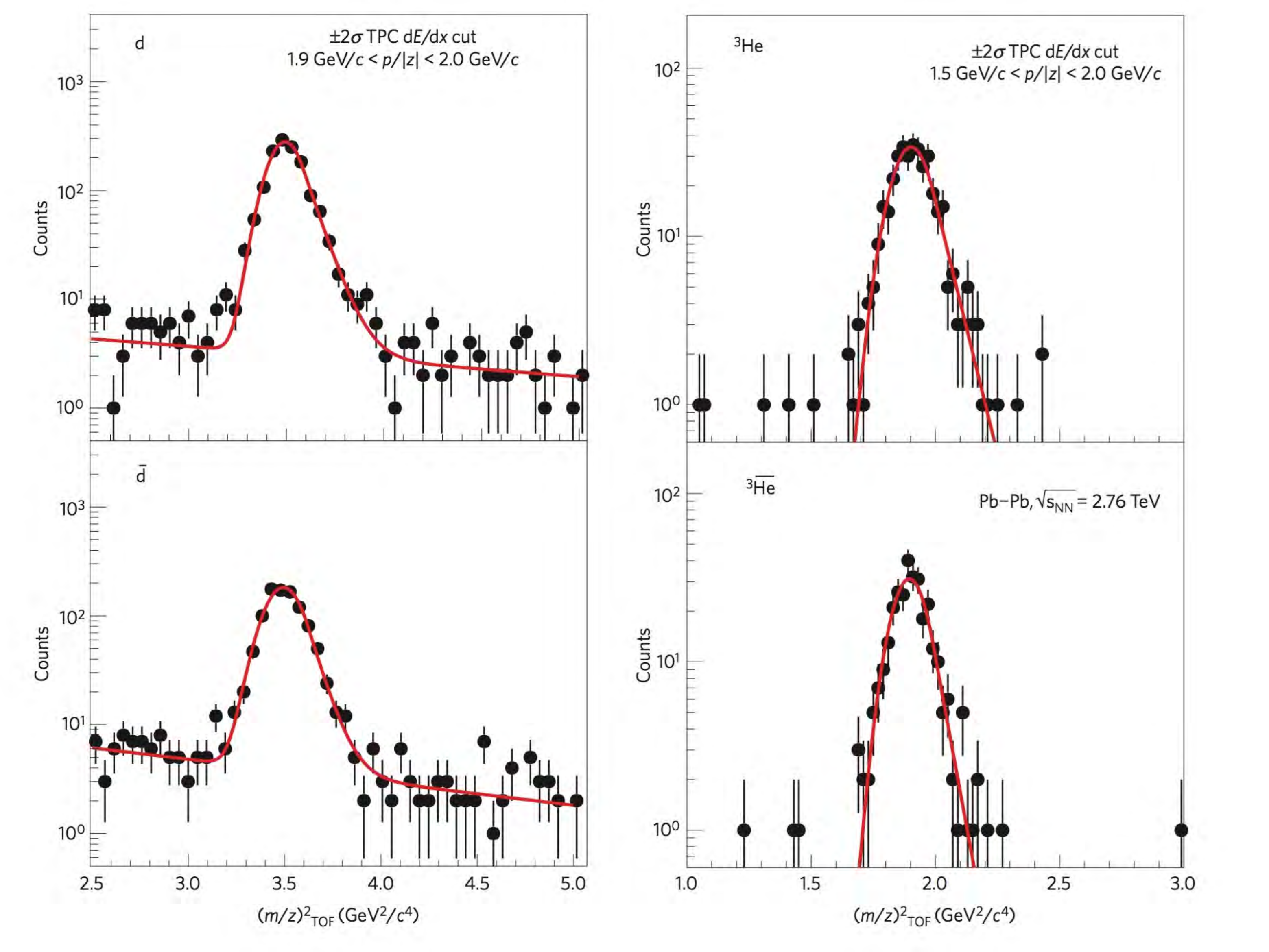}} 
\caption{Distributions of squared mass over charge, in selected rigidity intervals, for $d,\, \bar{d},\, ^3$He and $^3\overline{\rm He}$, measured by the ALICE experiment in Pb + Pb collisions at $\sqrt{s_{\rm NN}} = 2.76$ TeV \cite{Adam:2015pna}.}
\label{fig:Alice_massdiff}
\end{figure}
%%%%%%%%%%%%%%%%%%%%%%%%%%%%%%%%%%%
The ALICE collaboration analyzed the difference between squared ratios of mass over charge based on time-of-flight measurements, $(m/z)^2_{\rm TOF}$, for $d$ and $\bar{d}$, and for \he~and \ahe, in Pb + Pb collisions at $\sqrt{s_{\rm NN}} = 2.76$ TeV~\cite{Adam:2015pna}. The quantity $(m/z)^2$ was chosen for this analysis because it is directly proportional to time-of-flight squared, and follows a well-understood distribution.  The high-precision tracking and identification capabilities of the ALICE detectors provide an accurate measurement of the mass differences between nuclei and antinuclei \cite{Abelev:2014ffa}. The $(m/z)^2$ distributions were fitted in narrow intervals of $p/|z|$ and pseudorapidity, using a Gaussian with a small exponential tail that reflects the time signal distribution of the ALICE TOF detector. Figure~\ref{fig:Alice_massdiff} shows examples of $(m/z)^2_{\rm TOF}$ distributions for \dbar~and \ahe~candidates in selected rigidity intervals, and hereafter, the notation $\mu$ refers to $m/z$.   

To reduce the systematic uncertainties related to tracking, spatial alignment and time calibration, the ALICE measurement used the mass differences rather than the absolute masses. Nevertheless, imperfections in the detector alignment and in the description of the magnetic field can lead to position-dependent systematic uncertainties. The measurement of momentum brings the largest uncertainties for the mass differences \cite{Abelev:2014ffa}. The uncertainties are independent of the mass and are the same for all particles of a given charge in a given momentum interval. The (anti-)deuteron and the (anti-)$^3$He masses were corrected by scaling factors based on masses compiled by the Particle Data Group \cite{Olive:2016xmw}. Specifically, the corrected mass is given by 
$\mu_{A(\bar{A})} = \mu^{\rm TOF}_{A(\bar{A})} \,(\mu^{\rm PDG}_{p(\bar{p})} / \mu^{\rm TOF}_{p(\bar{p})}).$ 
These correction factors depend on track rigidity and deviate from unity by at most 1\%~\cite{{Adam:2015pna}}. 

%%%%%%%%%%%%FIGURE%%%%%%%%%%%%%%%%
\begin{figure}[!htb]
\centering
\centerline{\includegraphics[scale=0.6, bb=800 -50 -230 460]{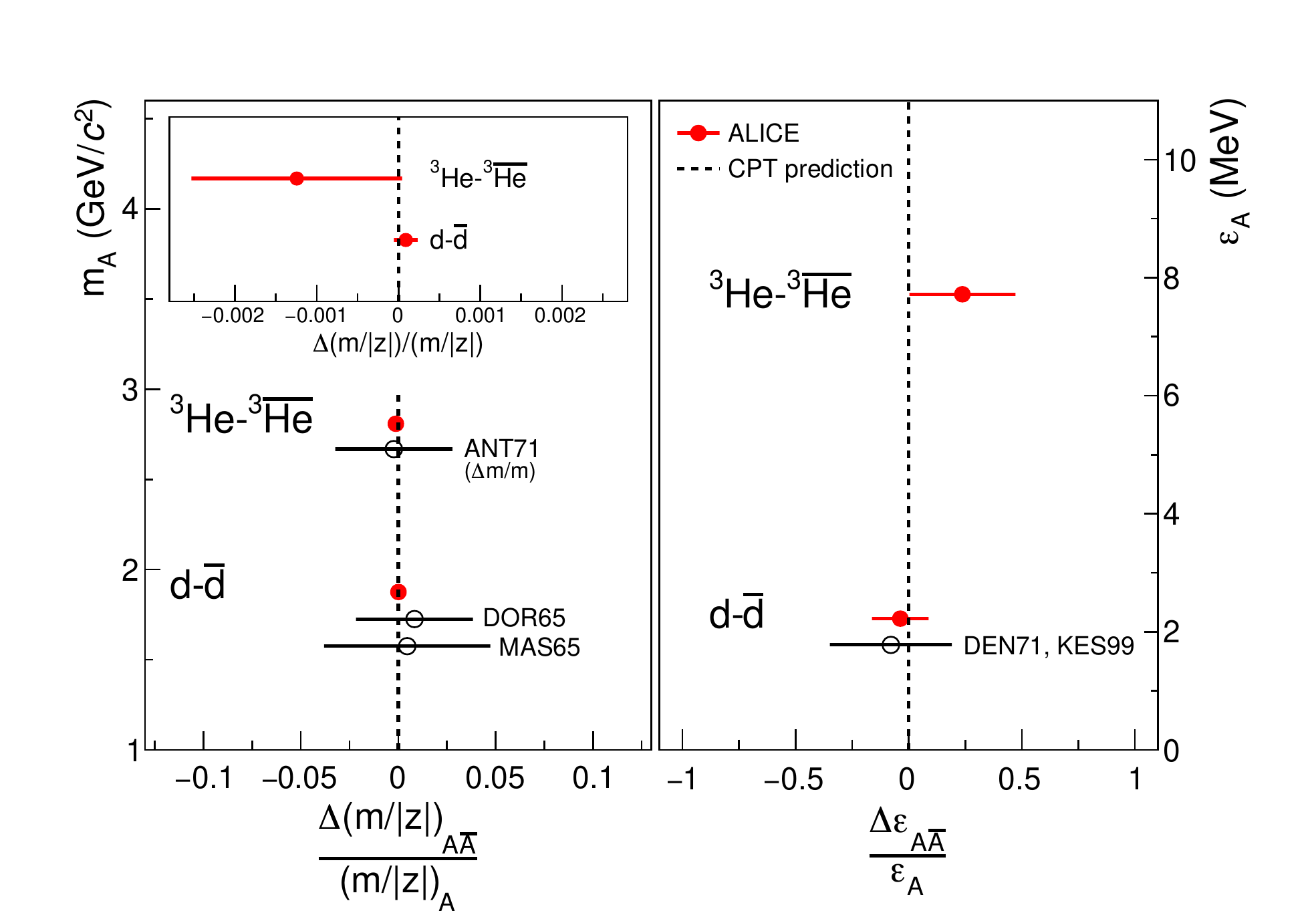}}
\caption{ Left panel: Nucleus mass versus measured ratios, as defined in Eqs. (\ref{mass_d_ratio}) and (\ref{mass_He3_ratio}). The solid red points correspond to the ALICE measurements \cite{Adam:2015pna}, while the open black circles, based on early measurements of the antideuteron \cite{Dorfan:1965zz, Massam1965} and anti-$^3$He \cite{Antipov:1970uc}, are seen to have much larger error bars. The vertical dashed line corresponding to zero on the horizontal axis marks the expection of {\it CPT} symmetry.  Right panel: A similar comparison for binding energy $\varepsilon_A$ versus $\Delta\varepsilon_{A\overline{A}}/ \varepsilon_A$. In this case, the older point of comparison is based on measurements published in Refs.~\cite{Denisov:1971im, Kessler:1999zz}. }
\label{fig:Alice_mass2}
\end{figure}
 %%%%%%%%%%%%%%%%%%%%%%%%%%%%%%%%

The corrected $\mu$ values at the peak positions lead to differences $\Delta\mu$ between nucleus and antinucleus, and these in turn were reported in Ref.~\cite{Adam:2015pna} as dimensionless ratios:  
\begin{equation}
\label{mass_d_ratio}
\frac{\Delta\mu_{d\bar{d}}}{\mu_d} = [0.9 \pm 0.5\,({\rm stat.}) \pm 1.4\,({\rm syst.})] \times 10^{-4}, 
\end{equation}
\begin{equation}
\label{mass_He3_ratio}
\frac{\Delta\mu_{\,\rm ^3He\,^3\overline{He}}}{\mu_{\,\rm ^3He}} = \rm{[-1.2 \pm 0.9\,(stat.) \pm 1.0\,(syst.)] \times 10^{-3}}, 
\end{equation}
where $\mu_d$ and $\mu_{\,\rm{^3He}}$ are the recommended fundamental physical constants from the CODATA group \cite{Mohr:2012tt}. The mass-over-charge differences are consistent with zero within errors, as expected from {\it CPT} symmetry. The left panel of Fig.~\ref{fig:Alice_mass2} presents these two ratios as solid red points.  

Assuming that $z_{\bar{d}} = -z_d$ and $z_{\,\rm ^3\overline{He}} = -z_{\,\rm ^3He}$ (as for the proton and antiproton~\cite{Hori:2011zz}), the binding energy differences between the two studied nucleus-antinucleus pairs were obtained by the ALICE collaboration using the mass-over-charge differences discussed above and the mass differences between the proton and antiproton~\cite{Gabrielse:1999kc, Hori:2011zz} and between the neutron and antineutron~\cite{Cresti:1986eg, DelAguila:1987ic}. Defining $\varepsilon_A = Zm_p + (A - Z)m_n - m_A$, where $m_p$ and $m_n$ are the proton and neutron masses from the PDG \cite{Olive:2016xmw}, and $m_A$ is the mass of the corresponding nucleus with atomic number $Z$ and mass number $A$ published by the CODATA group \cite{Mohr:2012tt}, the obtained binding energy differences are 
\begin{equation}
\label{binding_d_ratio}
\frac{\varepsilon_{d\bar{d}}}{\varepsilon_d} = \rm{-0.04 \pm 0.05\,(stat.) \pm 0.12\,(syst.)},
\end{equation}
\begin{equation}
\label{binding_He3_ratio}
\frac{\varepsilon_{\rm{^3He\,^3\overline{He}}}}{\varepsilon_{\,\rm ^3He}} = \rm{0.24 \pm 0.16\,(stat.) \pm 0.18\,(syst.)}.
\end{equation}
The ALICE collaboration plotted these results in the right panel of Fig.~\ref{fig:Alice_mass2}; they are the first-ever measurement of the binding energy difference for the case of $^3$He minus $\rm{^3\overline{He}}$, and they offer a significant improvement over the prior measurement of binding energy difference in the case of deuteron minus antideuteron.  As in the case of the mass differences for nuclei and their partner antinuclei, results are consistent with no measurable deviation from {\it CPT} symmetry.

\subsection{Antiproton-Antiproton Interaction
}\label{pbar-pbar}

Section~\ref{ALICE-mass} was devoted in part to the binding energy difference between $\bar{d}$ and $d$, and between \ahe~and \he, and this form of analysis verifies that the attractive forces which bind the antinucleons inside antinuclei are indistinguishable, within errors, from the corresponding forces in nuclei, as required by {\it CPT} symmetry. A different window into this physics can be accessed by directly studying the strong interaction between two antiprotons. This section describes measurements by the STAR collaboration of the momentum correlation function for antiproton-antiproton pairs produced in Au + Au collisions at $\sqrt{s_{\rm NN}}$ = 200 GeV \cite{Adamczyk:2015hza}. The observation of antideuterons and heavier antinuclei demonstrates the existence of an attractive interaction among antinucleons, but the 2015 study by the STAR collaboration goes beyond that observation by determining the scattering length ($f_0$, related to elastic cross sections) and the effective range ($d_0$) for the strong interaction between antiproton pairs \cite{Adamczyk:2015hza}.  This analysis was made possible for the first time by the large event samples and high rate of antiproton production in heavy-ion collisions at RHIC, and this form of characterization of the interaction is a fundamental ingredient for understanding the structure of more complex antinuclei.

The technique used for probing the antiproton-antiproton interaction involves momentum correlation, and it resembles the space-time correlation technique used in Hanbury-Brown and Twiss (HBT) intensity interferometry~\cite{HanburyBrown:1954amm, Lisa:2005dd, Ma:2015b, Ma:2017c, Yang:2016NST}. In an experiment, the two-particle correlation function is defined as $A(k^*)/B(k^*)$, where $A(k^*)$ is the distribution of the relative momentum ($k^*$) measured for the correlated pairs from the same event, while $B(k^*)$ is the same for non-correlated pairs, where each member of a pair must come from a different event (event-mixing technique). The measured correlation strength can be reduced by particle ID impurities in the sample. This effect can be corrected by 
\begin{equation}
\label{eq_1}
{\rm CF_{corrected}}(k^*) = \frac{{\rm CF_{measured}}(k^*)-1}{{\rm PairPurity}(k^*)}+1,
\end{equation}
where ${\rm PairPurity}(k^*)$ is the pair purity for the two particles, and $\mathrm{CF_{measured}}(k^*)$ and $\mathrm{CF_{corrected}}(k^*)$ are, respectively, the measured and corrected correlation functions.

In this type of study, it is important to simultaneously analyze both $pp$ and $\bar{p}\bar{p}$ pairs.  Inside the (anti)proton sample, there are secondary (anti)protons that come from weak decays of already-correlated primary particles. Thus, the measured correlation function is contaminated by residual correlations. The dominant contaminations are from $p\Lambda$ and $\Lambda\Lambda$ correlations or their antiparticle partners. The residual contamination in the correlation functions are taken into account by simultaneously fitting the data with the primary correlation function and the residual correlation function. Taking the two-proton correlation function as an example~\cite{Zbroszczyk:2008jja},  
\begin{align}
C_{\rm inclusive}(k^*) = 1 &+ x_{pp}[C_{pp}(k^*;R_{pp})-1] \nonumber \\ 
&+ x_{p\Lambda}[C_{p\Lambda}(k^*;R_{p\Lambda})-1] + x_{\Lambda\Lambda}[C_{\Lambda\Lambda}(k^*)-1], 
\label{eq:Inclusive}
\end{align}
where $C_{\rm inclusive}(k^*)$ is the inclusive CF, and $C_{pp}(k^*;R_{pp})$ is the true proton-proton CF, which can be described by the Lednick\'{y} and Lyuboshitz analytical model~\cite{Lednicky:1981su}; other quantities in Eq. (\ref{eq:Inclusive}) are defined below.  $C_{p\Lambda}(k^*;R_{p\Lambda})$ is the $p\Lambda$ CF taken from a theoretical calculation~\cite{Lednicky:1981su} that includes all final-state interactions and has been successful in explaining experimental data~\cite{Adams:2005ws}. $C_{\Lambda\Lambda}(k^*)$ is from an experimental measurement by the STAR collaboration with a purity correction~\cite{Adamczyk:2014vca}. $R_{pp}$ and $R_{p\Lambda}$ are the invariant Gaussian radii~\cite{Adams:2005ws} for the proton-proton correlation and the proton-$\Lambda$ correlation, respectively. In the fits performed in Ref.~\cite{Adamczyk:2015hza}, they are assumed to be the same. The quantities $x_{pp}$, $x_{p\Lambda}$ and $x_{\Lambda\Lambda}$ represent the relative contributions from pairs with both daughters from the primary collision, pairs with one daughter from the primary collision and the other from a $\Lambda$ decay, and pairs with both daughters from a $\Lambda$ decay, respectively; they are obtained from the THERMINATOR2 model~\cite{Chojnacki:2011hb}. Finally, 
$$C_{p\Lambda}(k^*_{pp})={\displaystyle \int} C_{p\Lambda}(k^*_{p\Lambda}) T(k^*_{p\Lambda},k^*_{pp}) dk^*_{p\Lambda},$$ 
where $T(k^*_{p\Lambda},k^*_{pp})$ is a matrix generated by THERMINATOR2 model to transform $k^*_{p\Lambda}$ to $k^*_{pp}$~\cite{Zbroszczyk:2008jja}.

%%%%%%%%%%FIGURE%%%%%%%%%%%%%%%%%
\begin{figure}[!htb]
\centering
\centerline{\includegraphics[scale=0.60,bb=700 -50 -230 660]{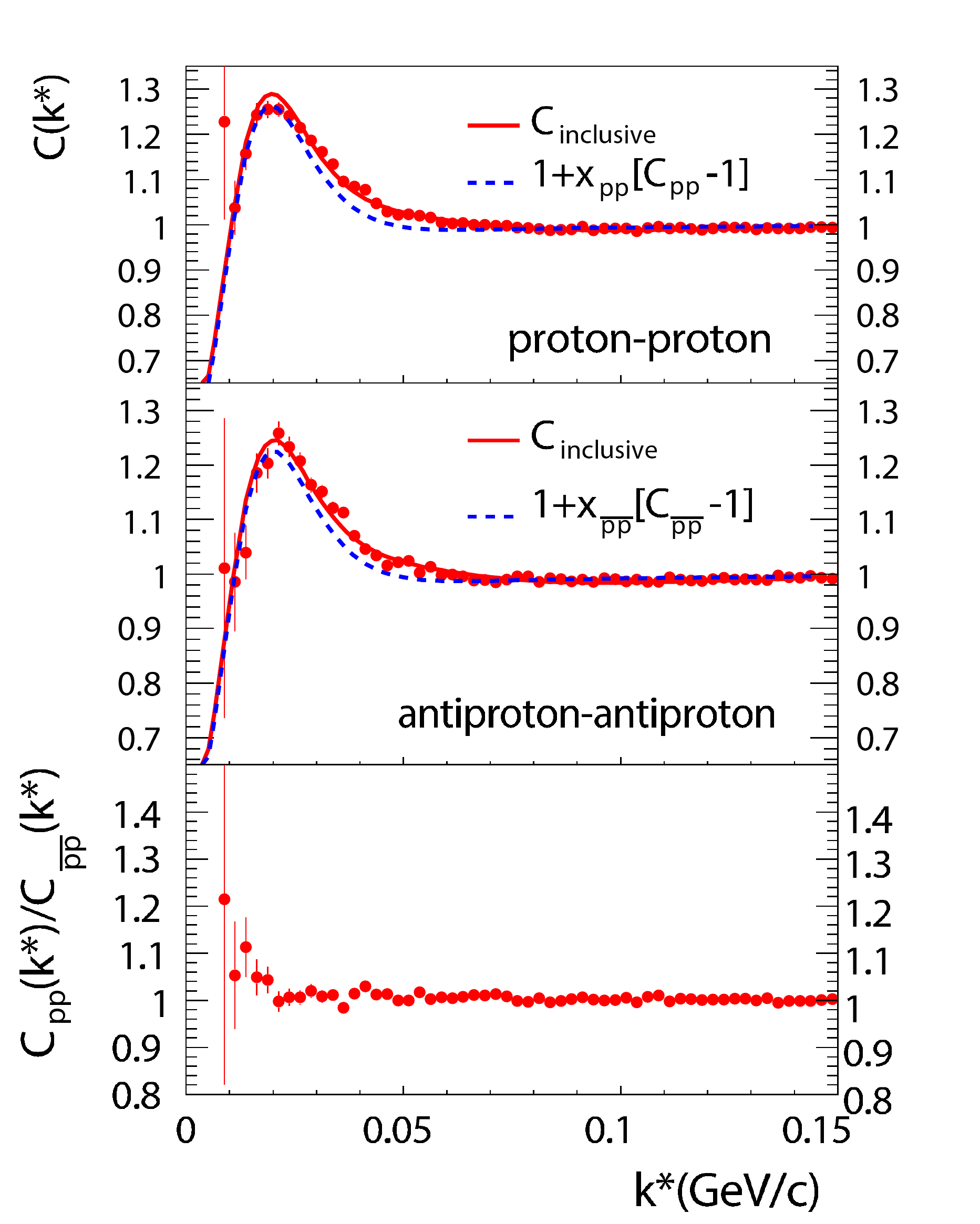}}
\caption{Panel (a) presents the proton-proton correlation, panel (b) the antiproton-antiproton correlation, and panel (c) the ratio of (a) to (b). Errors are statistical only. The fits to the data using Eq.~(\ref{eq:Inclusive}), $C_{\rm inclusive}(k^*)$, are plotted as solid lines, and the term $1 + x_{pp}[C_{pp}(k^*;R_{pp})-1]$ is shown as dashed lines. The $\chi^2/$NDF of the fit is 1.66 and 1.61 for cases (a) and (b), respectively. The figure is edited from the published version in~\cite{Adamczyk:2015hza}.}
\label{fig_3}
\end{figure} 
%%%%%%%%%%%%%%%%%%%%%%%%%%%%%%%

In the year-2015 measurement by the STAR Collaboration~\cite{Adamczyk:2015hza}, they fixed the scattering length $f_0 = 7.82$ fm and effective range $d_0 = 2.78$ fm when fitting the parton-proton correlation function, as they are well-determined from proton-proton elastic scattering \cite{Mathelitsch:1984hq}. Only the radius $R_{pp}$ is treated as a free parameter. When fitting the antiproton-antiproton correlation function, the radii $R_{\bar{p}\bar{p}}$, $f_0$ and $d_0$ are treated as free parameters, since there is no measurement of $f_0$ or $d_0$ for an antiproton pair. 

Fig.~\ref{fig_3} presents the purity-corrected CF for proton-proton pairs, Fig.~\ref{fig_3}(a), and for antiproton-antiproton pairs, Fig.~\ref{fig_3}(b), for 30-80\% centrality Au + Au collisions at $\sqrt{s_{\rm NN}} = 200$ GeV.  The red solid lines, $C_{\rm inclusive}$, are the fits to the data, and the dashed blue lines are the CFs for the $pp$ contribution. The proton-proton CF exhibits a maximum at $k^* \sim 0.02$ GeV which is caused by the attractive $S$-wave interaction between the two protons, and is consistent with previous measurements. The antiproton-antiproton CF has an exactly similar structure, which indicates that the interaction between two antiprotons is also attractive. 

For the proton-proton CF, $R_{pp} = 2.75  \pm 0.01$ fm; $\chi^2/$NDF = 1.66. For the antiproton-antiproton CF, $R_{\bar{p}\bar{p}} = 2.80 \pm 0.02$ fm; $f_0 =7.41 \pm 0.19 \,(\mathrm{stat.}) \pm 0.36 \,(\mathrm{syst.})$ fm; $d_0 =2.14 \pm 0.27 \,(\mathrm{stat.}) \pm 1.34 \,(\mathrm{syst.})$ fm; $\chi^2/$NDF = 1.61. Figure~\ref{fig_3}(c) shows the ratio of the inclusive CF for proton-proton pairs to that for antiproton-antiproton pairs. Throughout the studied $k^*$ region, this ratio is indistinguishable from unity within errors, as expected if proton-proton and antiproton-antiproton pairs have the same strong interaction. 

\begin{table}[]
\centering
\caption{The scattering length ($f_0$) and effective range ($d_0$) for nucleon-nucleon and antiproton-antiproton interactions.}
\label{f0table}
\begin{tabular}{cccc}
\\
\hline\hline
        & Proton-proton            & Antiproton-antiproton            \\ \hline
$f_0$ (fm) & $7.82\pm0.01$             & $7.41 \pm 0.19\,({\rm stat}) \pm 0.36\,({\rm sys})$     \\
$d_0$ (fm) & $2.78\pm0.01$             & $2.14 \pm 0.27\,({\rm stat}) \pm 1.34\,({\rm sys})$     \\ \hline
        & Proton-neutron (triplet) & Neutron-neutron                   &                          \\
$f_0$ (fm) & $-5.43\pm0.02$           & $16.70\pm0.38$                &                          \\
$d_0$ (fm) & $1.71\pm0.05$            & $2.78\pm0.13$                &                          \\ \hline
        & Proton-neutron (singlet)                  &                          \\
$f_0$ (fm) & $23.72\pm0.02$                        &                          \\
$d_0$ (fm) & $2.66\pm0.06$                        &                          \\ 
\hline\hline
\end{tabular}
\end{table}

Table \ref{f0table} presents the strong interaction parameters $f_0$ (scattering length) and $d_0$ (effective range) for antiproton-antiproton interactions, as well as prior measurements for nucleon-nucleon interactions \cite{Mathelitsch:1984hq, Slaus89}. The $f_0$ and $d_0$ for the antiproton-antiproton interaction is consistent with the same for the proton-proton interaction within errors.  This measurement can provide the parameterization input for describing the interaction among cold-trapped gases of antimatter ions, as in an ultra-cold environment, where $S$-wave scattering dominates, and effective-range theory shows that $f_0$ and $d_0$ can suffice to describe elastic collisions \cite{Mott:1965, Rom:2006}. This result also provides a quantitative verification of matter-antimatter symmetry with respect to the forces that are responsible for the binding of nuclei and antinuclei, and complements the alternative approach reviewed in Section~\ref{ALICE-mass}.

\section{Implications for Space-Based Experiments and for Other Fields}

\subsection{Space-Based Cosmic Ray Baryon Measurements
}\label{CR}

By convention, the term ``cosmic rays" refers to the non-photonic component of the radiation found in space. Primary cosmic rays are composed of 99\% hadrons and 1\% electrons; the hadronic component is $\sim 90$\% protons, $\sim 9$\% alpha particles, with heavier nuclei making up the remainder. Positrons and antiprotons are also present in cosmic rays ~\cite{Schlickeiser2002, Stanev2010}.  At one time, it was speculated that the latter component had a possible connection to the baryogenesis puzzle (see Sec.~\ref{sym}), but now it is recognized that the primary flux interacts with interstellar matter as it traverses the cosmos, and that such interactions are sufficient to explain the observation of stable antiparticles, namely positrons and antiprotons, among cosmic rays~\cite{Schlickeiser2002, Stanev2010}.  

With the advent of space-based detectors, the study of these positrons and antiprotons has made rapid progress. For example, the recent direct detection by the DAMPE (DArk Matter Particle Explorer) collaboration of a spectral break at $E \sim 0.9$ TeV for cosmic ray electrons and positrons is an unexpected development and has been a source of excitement in the field~\cite{Ambrosi:2017wek}. To date, there has been no observation of any antinucleus other than the antiproton. Such a discovery would have a profound impact on our understanding of the Universe; for example, any observation of a $|Z| > 2$ antinucleus would suggest the existence of antistellar nucleosynthesis in antimatter domains, while the detection of antihelium would point towards the existence of residual antimatter from Big-Bang nucleosynthesis~\cite{Schlickeiser2002, Stanev2010, Adriani:2014pza}. However, a recent study suggests that secondary production of antinuclei in cosmic ray collisions could be larger that previously assumed~\cite{Blum:2017,Blum:2017iwq}. The flux could be as high as one $^3\overline{\rm He}$ event per 5-year exposure of current active experiments.

%%%%%%%%%%FIGURE%%%%%%%%%%%%%%%%%
\begin{figure}[!htb]
\centering
\centerline{\includegraphics[scale=1.40, bb=460 -30 -230 180]{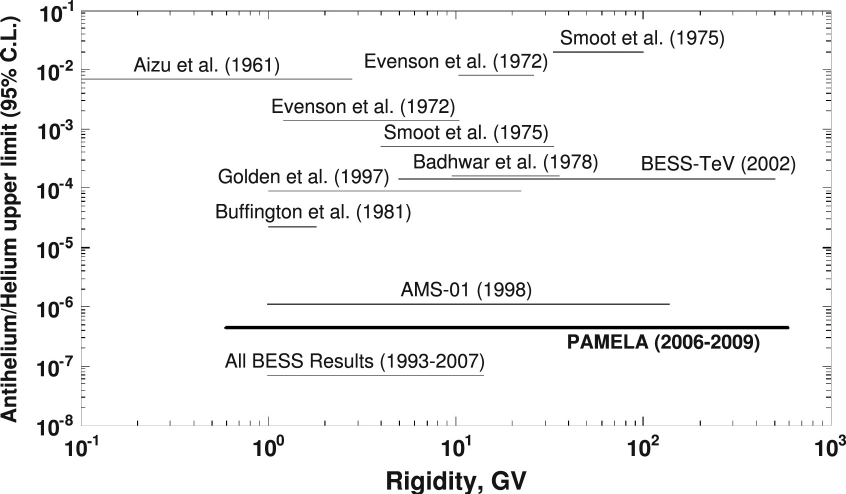}}
\caption{ A compilation of data on the upper limit of $\overline{\mathrm{He}}$/He in cosmic rays~\cite{Aizu:1961zz, Smoot:1975bv, Badhwar:1978hz, Buffington:1981zz, Golden:1997iz, Alcaraz:2000ss, Adriani:2014pza, Sasaki:2008zzb, Abe:2012tz}. This plot is reproduced from Ref.~\cite{Adriani:2014pza}. }
\label{fig:PAMELA_AntiHe}
\end{figure}
%%%%%%%%%%%%%%%%%%%%%%%%%%%%%%%

Given the above perspective, searching for any antinucleus signal in the cosmos beyond antiprotons is one of the major motivations for space detectors such as the current-generation Alpha Magnetic Spectrometer (AMS-02)~\cite{Aguilar:2013qda} and the Payload for Antimatter Matter Exploration and Light nuclei Astrophysics (PAMELA)~\cite{Adriani:2008zr}. The AMS-02 detector was launched via the Space Shuttle and installed on the International Space Station in May 2011; one of its unique characteristics is its long-duration mission, on the order of 20 years.  The PAMELA detector was launched into Earth orbit in June 2006.  Neither experiment has detected any antihelium signal to date. Fig.~\ref{fig:PAMELA_AntiHe} presents the world data on the upper limit of the $\overline{\mathrm{He}}$/He ratio. Data from the PAMELA experiment spans the largest energy range~\cite{Adriani:2014pza}. In particular, no antihelium events with negative rigidities were found among 6.3 million PAMELA events with charge $|Z| \geq 2$ selected in the rigidity range from 0.6 to 600 GeV/$c$ per charge in units of the electron charge~\cite{Adriani:2014pza}. 

As discussed in Sec.~\ref{RHIC-AH3L-AHE4}, the measurements of $\mathrm{^4\overline{He}}$ in high-energy heavy-ion collisions can help to establish a baseline for the expected rate of $\mathrm{^4\overline{He}}$ in space from secondaries produced in primary cosmic ray collisions. Our information about this baseline comes from the large penalty factor to add an extra antinucleon to an antinucleus ($10^3$ at the LHC, and a higher penalty factor at lower energies --- see Sec.~\ref{RHIC-AH3L-AHE4}), and the low rate of forming $\mathrm{^4\overline{He}}$ in central heavy-ion collisions, for example, a probability of $1.10 \times 10^{-7}$ per central Au + Au collision at \sNN~= 200 GeV~\cite{Xue:2012gx}.  Accordingly, it is clear that the current sensitivity of any of the cosmic ray detectors in operation today is well below the rate estimated from accelerator-based measurements.

\subsection{Space-Based Cosmic Ray Electron-Positron Measurements}\label{CR2}

To date, the most novel and compelling results from the current generation of space-based cosmic-ray experiments have come from positron measurements, notably the unexpected positron excess. The implications of recent antimatter research in the field of relativistic heavy-ion collisions are most directly felt in cosmic-ray physics in the area of searches for antibaryons and antinuclei in space, as set out in Sec.~\ref{CR} above.  However, positron observations also have implications for the primordial bulk antimatter hypothesis.  Furthermore, there are many aspects of overlap between the baryon and electron measurements in space, and interdisciplinary implications connected to heavy-ion physics necessarily spill over those boundaries.  Many positrons originate in our own galaxy, and recent observations appear to reduce or possibly eliminate the need for explanations involving exotica such as dark matter \cite{Weidenspointner2018}.  

%%%%%%%%%%FIGURE%%%%%%%%%%%%%%%%%%
\begin{figure}[!htb]
\centerline{\includegraphics[scale=0.30,bb=1360 -50 -230 980]{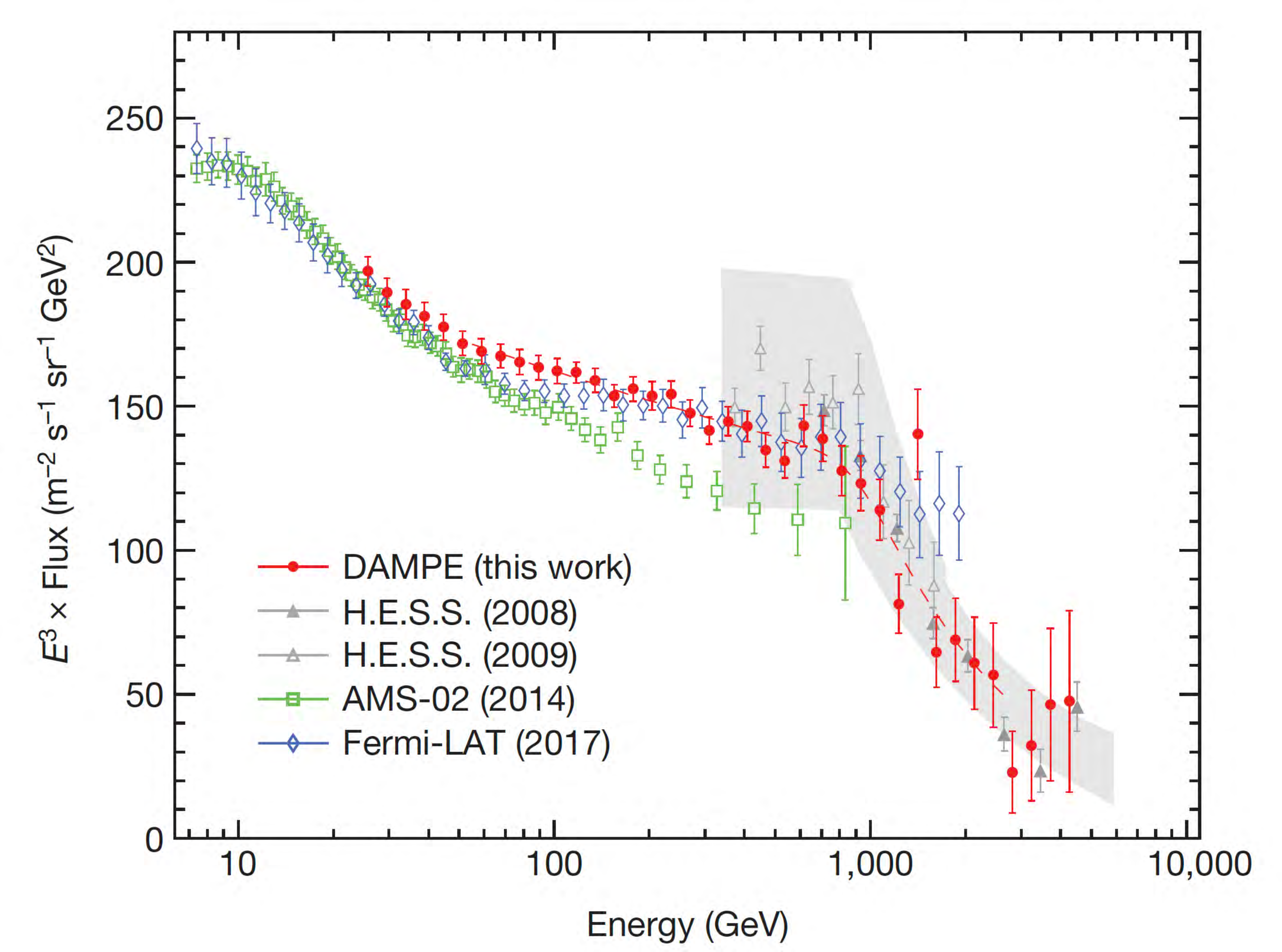}}
\caption{ The cosmic-ray electron and positron spectrum (multiplied by $E^3$) measured by the DAMPE collaboration ~\cite{Ambrosi:2017wek}. Data from AMS-02~\cite{Aguilar:2014fea}, Fermi-LAT~\cite{Abdollahi:2017nat}, as well as the ground-based experiment of the H.E.S.S. Collaboration~\cite{Aharonian:2008aa} are plotted for comparison. The red dashed line represents a smoothly-broken power-law model that best fits the DAMPE data. This plot is reproduced from Ref.~\cite{Ambrosi:2017wek}. }
\label{fig:DAMPE}
\end{figure}
%%%%%%%%%%%%%%%%%%%%%%%%%%%%%%%%%

In the first report from the AMS-02 collaboration~\cite{Aguilar:2013qda}, over 25 billion cosmic ray events have been analyzed. An anomalous positron fraction (the ratio of the positron flux to the combined flux of positrons and electrons) has been reported, where the positron fraction increases steadily from 10 to $\sim$250 GeV \cite{Aguilar:2013qda}. This observation cannot be explained by considering only secondary production of positrons~\cite{Serpico:2011wg, Delahaye:2008ua, Moskalenko:1997gh}. Conclusions about the behavior above 250 GeV requires more statistics than have been accumulated to date.

The latest result on cosmic ray electron and positron fraction comes from the DAMPE collaboration~\cite{Ambrosi:2017wek}, and has connections to the positron anomaly. Figure~\ref{fig:DAMPE} shows the first data from DAMPE, resulting from approximately 530 days of operation. Unprecedentedly high energy resolution and low background are evident. The data in the energy range 55 GeV to 2.63 TeV fit much better to a smoothly-broken power-law model than to a single power-law model~\cite{Ambrosi:2017wek}. The precise measurement of the cosmic-ray electron and positron spectrum by DAMPE can narrow-down the parameter space of models, including those related to the positron anomaly~\cite{Ambrosi:2017wek}. The parameters include, for example, the spectral cutoff energy of the electrons accelerated by nearby pulsars or supernova remnants, or the rest mass and the annihilation cross section of a dark-matter particle. Together with data from the cosmic microwave background or $\gamma$-rays, these improved constraints on the model parameters obtained by DAMPE may clarify the connection between the positron anomaly and the annihilation or decay of dark matter~\cite{Ambrosi:2017wek}.  The reader is referred to Ref. \cite{Blum:2017iwq} for a comprehensive review.

\subsection{Muonic Antiatoms}

Muonic atoms are hydrogen-like systems made of a hadron and a muon. In the past, muonic atom systems were studied mostly by examining the products of particle decays. For example, the $\pi \mu$ system in $K_L^0$ decay \cite{Coombes:1976hi, Aronson:1982bz}. Recently, measurements of the Lamb shift \cite{Lamb1947} in the context of muonic atoms have been used to study proton structure \cite{Pohl:2010zza, Antognini:1900ns}.  In heavy-ion collisions, when the high-temperature and high-density matter freezes out, it produces a mix of particle species that are close in phase space, which is an ideal condition for the production of muonic atoms. Unlike systems that are created during freeze-out and are bound by the strong force, muonic atoms are created {\it after} freeze-out and are bound by the Coulomb force. If it is close to hadrons in phase space, a muon can be captured by a charged hadron to form a $K \mu$, $p \mu$ or $\pi \mu$ system~\cite{Baym:1993ae}. Such systems are perfect tools to access the thermal emission of muons from a QCD system. Only thermal muons, or muons from short-lived resonances, are able to form muonic systems. They may also allow access to primordial single-lepton spectra~\cite{Kapusta:1998fh}.

\begin{figure}[!htb]
\centering
\centerline{\includegraphics[scale=0.55,bb=560 -50 -230 320]{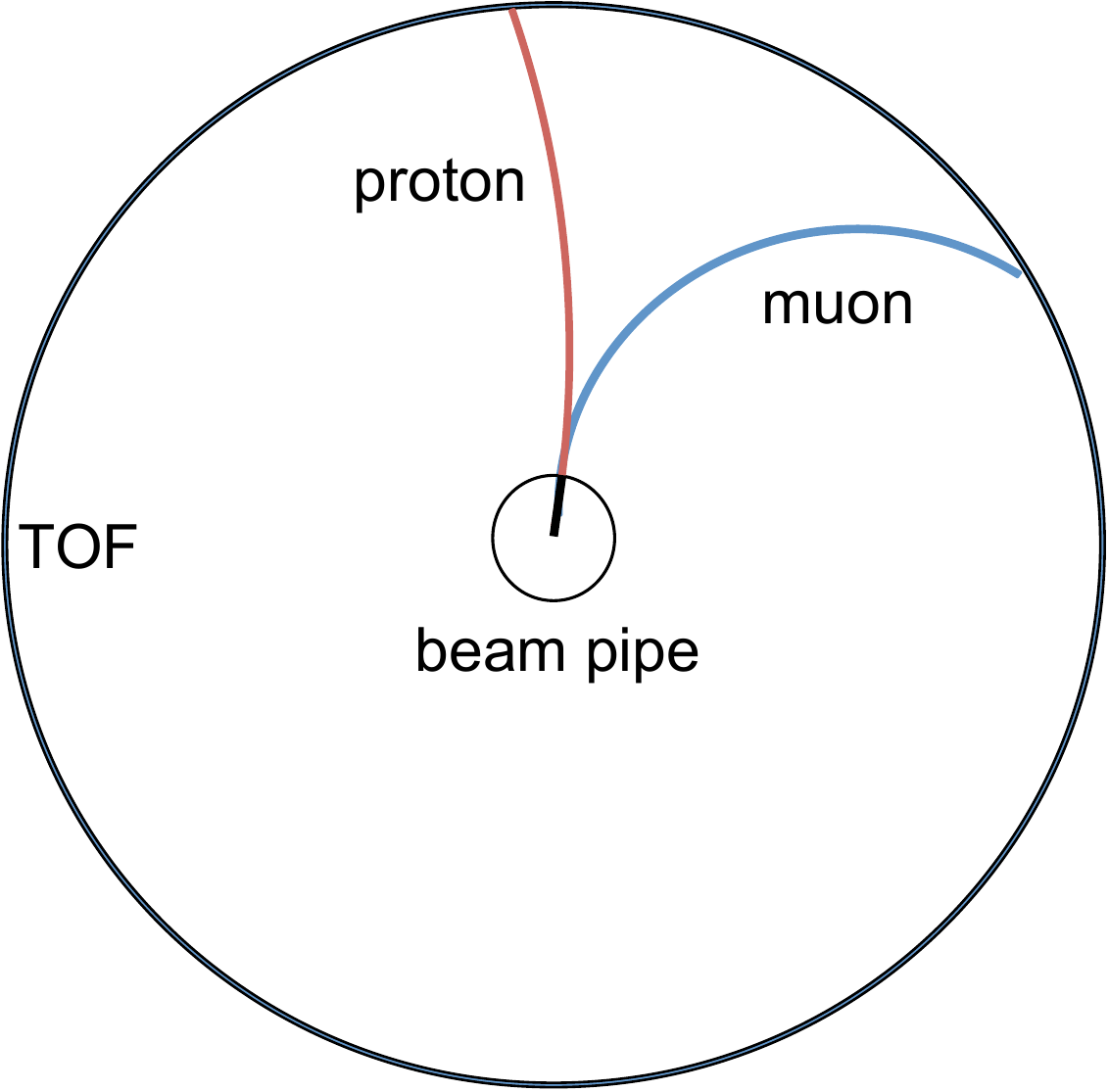}}
\caption{A schematic representation, in the plane perpendicular to the beam direction, of a muonic atom being dissociated at the STAR beam pipe, then transiting into the STAR Time Projection Chamber as a proton and a muon.}
\label{fig:muonAtomDissociation}
\end{figure}

Neither the antimatter muonic hydrogen, nor the hyper-muonic atom, $K\mu$, have yet been discovered. Since the installation of the STAR TOF detector, this RHIC experiment has been in a good position to search for such systems. When a muonic atom is formed in a heavy-ion collision at STAR, it travels relatively slowly until it encounters the beam pipe, where it becomes dissociated into a free muon and a free hadron. The free muon and hadron continue their motion, depositing energy along their trajectories by ionizing the gas inside the TPC, after which they reach the TOF detector where they produce a timing signal. Fig.~\ref{fig:muonAtomDissociation} shows a schematic illustration of such a process.\\

%%%%%%%%%%%%%% Figure %%%%%%%%%%%%%%%%%%%
%%%%%%%%%%%%%%%%%%%%%%%%%%%%%%%%%%%%%%%%%

A special trigger has been implemented at STAR to select particles with late arrival time at the TOF detector. For each hadron-muon pair, the invariant mass of a possible muonic atom can be reconstructed using the daughter momenta and masses. The following quantity has been studied~\cite{Xin:2015noa} as a function of the difference between the possible invariant mass of an atom and the mass sum for the daughters, $\delta_m \equiv (M_{\rm inv}-M_{\mu}-M_{\rm hadron})$:
\begin{eqnarray*}
\Theta(\delta_m) = {\rm \frac{UL \times LS}{ME^2} -1},
\end{eqnarray*}
where UL and LS stand for the yield of unlike-sign and like-sign pairs, respectively, and ME stands for the mixed-event background. The purpose of taking the product UL $\times$ LS is to cancel the Coulomb effect, since at low relative momentum between pairs, the Coulomb effect will enhance the UL yield and suppress the LS yield.  However, the product of the two is not sensitive to such an effect.  A sharp peak with a positive value at zero in the $\Theta(\delta_m)$ distribution would indicate that hadron-muon pairs have been formed. The STAR collaboration has shown~\cite{Xin:2015noa} such peaks for $p\mu^-$, $\bar{p}\mu^+$, $K^-\mu^+$, and $K^+\mu^-$ pairs, providing evidence for the existence of muonic atoms with these configurations.

The $K\mu$ system has been further investigated via a study of emission asymmetry between two daughters. This technique has been used to study emission sequence in heavy-ion collisions~\cite{Lednicky:1995vk, Voloshin:1997jh}. The effect of Coulomb and strong interactions are different for the case of two particles initially moving towards each other versus the case of moving apart. Pairs are divided into two groups, representing either the case where the kaons catch up with the muons, or the case where they move away from the muons. Two 2-particle correlation functions, $C_+$ and $C_-$, are defined. Each is constructed as the correlated yield normalized by the uncorrelated yield, similar to the correlation function used in the antiproton-antiproton study described earlier in this review. The subscripts $+$ and $-$ reflect the sign of $\vec{v} \cdot \vec{k}_K^*$, where $\vec{v}$ is the pair velocity and $\vec{k}_K^*$ is the kaon momentum vector in the pair's rest frame. If the average space-time emission points of kaons and muons coincide, e.g., there is emission from a single muonic atom, both correlation functions are identical and the ratio $C_+/C_-$ will be unity. Otherwise, the Coulomb correlation strength will be enhanced for the catching-up case compared to the moving-away case, and the ratio $C_+/C_-$ will deviate from unity. 

The STAR collaboration has calculated~\cite{Xin:2015noa} the $C_+/C_-$ ratio for both $K\mu$ and $K\pi$ systems. The ratio for $K\mu$ is indeed close to unity, while the ratio for the $K\pi$ system, which is not expected to form in heavy-ion collisions and serves as a consistency check, deviates from unity. This observation again supports the hypothesis that formation of $K\mu$ muonic atoms has occurred. As a side note, recently it was reported that $K\pi$ systems can be formed in platinum or nickel foil targets when bombarded by protons from the CERN PS~\cite{DIRAC:2016rpv}. This observation indicates that the kinematic range in heavy-ion collisions for kaons and pions are well separated, which prevent them from binding together.

Out of the four systems that the STAR collaboration has observed, if confirmed, the $\bar{p}\mu^+$ system would be the first observation of an antimatter muonic hydrogen atom. The $K\mu$ system would also be the first observation of hyper-muonic atoms. It is clearly desirable to have similar studies repeated at LHC energies.

\section{Outlook}  

Research during recent years in the field of relativistic nucleus-nucleus collisions has yielded much progress in the study of antimatter nuclei, helped in particular by opening up of new center-of-mass energy frontiers as well as constantly improving luminosity, detector acceptance and data acquisition performance at the Relativistic Heavy-Ion Collider \cite{Huang:2013xta, STAR:SN0598, STAR:SN0644} and at the Large Hadron Collider \cite{Martinengo:2017fuc, Abelevetal:2014cna, Abelevetal:2014dna}.  Some of these upgrades and improvements have already yielded physics results, as reviewed in previous sections and also in an early brief review~\cite{Ma:2013xn}, while others are still in progress.  Future experimental possibilities will also open up at new accelerator facilities, namely HIAF (High Intensity Heavy-Ion Accelerator Facility) in Huizhou, China \cite{Wu:2018, Li:2017}, NICA at JINR (Nuclotron-based Ion Collider fAcility at the Joint Institute for Nuclear Research) in Dubna \cite{Kekelidze:2017tgp}, FAIR (Facility for Antiproton and Ion Research) at Darmstadt \cite{Senger:2017oqn}, and J-PARC-HI (Japan - Proton Accelerator Research Complex - Heavy Ions) \cite{Sakaguchi:2017ggo}; while these new facilities will operate at much lower energies than RHIC and LHC, their performance in terms of beam intensity and event rates will allow novel aspects of lighter antinuclei to be explored.  

Looking ahead on the timescale of the coming decade or sooner, progress can be anticipated in many specific aspects of research into antinuclei.  In the domain of antihypernuclei and hypernuclei, where there has been relatively gradual progress since the 1960s, we can look forward to much improved data for both lifetime and binding energy of the antihypertriton and the hypertriton \cite{Adamczyk:2017buv}, with expected physics contributions from several facilities over a wide range of beam energies \cite{Steinheimer:2013lza, Steinheimer:2011hoa, STAR:SN0598, Senger:2017oqn}.  Greatly improved statistics at the LHC will enhance many antimatter measurements, such those reviewed in Sections \ref{Yields-RHIC}, \ref{therm} and \ref{coal}, as well as the {\it CPT} symmetry tests in Section~\ref{ALICE-mass}.  With regard to the antinuclear force derived from antiproton-antiproton interactions (Section~\ref{pbar-pbar}), the  scattering length $f_0$ is relatively well determined, but not the effective range $d_0$. Future improvements in this determination will require a more accurate calculation of the $p\Lambda$ correlation function and a better measurement of the $\Lambda\Lambda$ correlation function.

It is also appropriate here to summarize the prospects for future discovery of new species in the area of antinuclei.  In Ref.~\cite{Agakishiev:2011ib} and in Sec.~\ref{anti-He4}, it is explained that it is likely that $^4\overline{\rm He}$ will remain the heaviest stable antimatter nucleus observed for the foreseeable future. Nevertheless, there is a theoretical calculation based on phase-space coalescence in the context of a special freeze out configuration associated with $^5\overline{\rm Li}$ and $^5$Li, whereby it is argued that $^5\overline{\rm Li}$ might be feasible to observe at RHIC via the $\rm{^4\overline{He}} + \bar{p}$ decay channel~\cite{Sun:2015jta}.  A search for the anti-resonance decay $^4\overline{\rm Li} \rightarrow \rm{^3\overline{He}} + \bar{p}$ is also deserving of attention, but specific guidance on the likely rate for this antinucleus is lacking.  The possible abundance of ${\rm ^4_{\bar{\Lambda}}\overline{H}}$ is argued to be comparable to that of $^4\overline{\rm He}$ \cite{Sun:2015ulc}, but for the proposed decay channel ${\rm ^4_{\bar{\Lambda}}\overline{H}} \rightarrow {\rm ^4\overline{He}} + \pi^-$, it is challenging (at least at RHIC) to distinguish the signal from background due to the very low production rate of ${\rm ^4\overline{He}}$ and the large combinatorial background that arises from pairing with pions. The ALICE collaboration is better-positioned to pursue this search, because they are expected to have samples of about 5500 ${\rm ^4\overline{He}}$ and $^4$He in LHC run 3, which begins in 2021 \cite{Acharya:2017bso}. Note that at 2.76 TeV, the ${\rm ^4\overline{He}}$ to $^4$He ratio is consistent with unity \cite{Acharya:2017bso}, which puts them in an advantageous position for a successful search.  Finally, investigations of muonic atoms and antiatoms are expected to continue, with prospects for confirmed observations. 
  
\section*{Acknowledgements}
Discussions with Drs. Veronica Dexheimer, John Millener, Neha Shah, Song Zhang and Zhengqiao Zhang are gratefully acknowledged. The work of J.H.C. and Y.G.M. is supported in part by the National Natural Science Foundation of China under Nos. 11421505, 11775288, 11520101004 and 11220101005, by the Major State Basic Research Development Program (973) of China under Nos. 2014CB845401 and 2015CB856904, by the Key Research Program of Frontier Sciences of CAS under Grant No. QYZDJSSW-SLH002, and by the Strategic Priority Research Program of the Chinese Academy of Sciences under Grant No. XDB16. D.K. acknowledges support by the U.S. Department of Energy, Office of Science, under Grant DE-FG02-89ER40531. A.H.T. and Z.B.X. are supported in part by the U.S. Department of Energy, Office of Science, under Grant DE-SC-0012704.   

\pagebreak
\section*{References}
\bibliography{review-refs}

\begin{thebibliography}{100}
\expandafter\ifx\csname url\endcsname\relax
  \def\url#1{\texttt{#1}}\fi
\expandafter\ifx\csname urlprefix\endcsname\relax\def\urlprefix{URL }\fi
\expandafter\ifx\csname href\endcsname\relax
  \def\href#1#2{#2} \def\path#1{#1}\fi

\bibitem{Schuster:1898}
A.~Schuster, {Potential Matter.— A Holiday Dream}, Nature 58 (1898) 367.
\newblock \href {http://dx.doi.org/10.1038/058367a0}
  {\path{doi:10.1038/058367a0}}.

\bibitem{Dirac:1928hu}
P.~A.~M. Dirac, {The Quantum Theory of the Electron}, Proc. Roy. Soc. Lond.
  A117 (1928) 610--624.
\newblock \href {http://dx.doi.org/10.1098/rspa.1928.0023}
  {\path{doi:10.1098/rspa.1928.0023}}.

\bibitem{Dirac:1928ej}
P.~A.~M. Dirac, {The Quantum Theory of the Electron. 2.}, Proc. Roy. Soc. Lond.
  A118 (1928) 351.
\newblock \href {http://dx.doi.org/10.1098/rspa.1928.0056}
  {\path{doi:10.1098/rspa.1928.0056}}.

\bibitem{Dirac:1930ek}
P.~A.~M. Dirac, {A Theory of Electrons and Protons}, Proc. Roy. Soc. Lond. A126
  (1930) 360--365.
\newblock \href {http://dx.doi.org/10.1098/rspa.1930.0013}
  {\path{doi:10.1098/rspa.1930.0013}}.

\bibitem{Dirac:1930bga}
P.~A.~M. Dirac, {On the Annihilation of Electrons and Protons}, Proc. Cambridge
  Phil. Soc. 26 (1930) 361--375.
\newblock \href {http://dx.doi.org/10.1017/S0305004100016091}
  {\path{doi:10.1017/S0305004100016091}}.

\bibitem{Oppenheimer:1930}
J.~R. Oppenheimer, {On the Theory of Electrons and Protons}, Phys. Rev. 35
  (1930) 562--563.
\newblock \href {http://dx.doi.org/10.1103/PhysRev.35.562}
  {\path{doi:10.1103/PhysRev.35.562}}.

\bibitem{Weyl:1927vd}
H.~Weyl, {Quantenmechanik und Gruppentheorie}, Z. Phys. 46 (1927) 1--46.
\newblock \href {http://dx.doi.org/10.1007/BF02055756}
  {\path{doi:10.1007/BF02055756}}.

\bibitem{Weyl:1931}
H.~Weyl, {Gruppentheorie und Quantenmechanik, 1st ed., 1928; 2nd ed., 1931},
  Hirzel, Leipzig, 1931, p. 263, {Translated by H. P. Robertson, as {\it The
  Theory of Groups and Quantum Mechanics}, Methuen, London, 1931; Dover, New
  York, NY 1949}.

\bibitem{Dirac:1931kp}
P.~A.~M. Dirac, {Quantized Singularities in the Electromagnetic Field}, Proc.
  Roy. Soc. Lond. A133 (1931) 60--72.
\newblock \href {http://dx.doi.org/10.1098/rspa.1931.0130}
  {\path{doi:10.1098/rspa.1931.0130}}.

\bibitem{Dirac1983}
P.~A.~M. Dirac, {My Life as a Physicist}, Springer US, Boston, MA, 1983, pp.
  733--749.
\newblock \href {http://dx.doi.org/10.1007/978-1-4613-3655-6_15}
  {\path{doi:10.1007/978-1-4613-3655-6_15}}.

\bibitem{Zichichi:2009zza}
A.~Zichichi, {Why Antihydrogen and Antimatter Are Different}, CERN Cour. 49N4
  (2009) 15--17.

\bibitem{Anderson:1932zz}
C.~D. Anderson, {The Apparent Existence of Easily Deflectable Positives},
  Science 76 (1932) 238--239.
\newblock \href {http://dx.doi.org/10.1126/science.76.1967.238}
  {\path{doi:10.1126/science.76.1967.238}}.

\bibitem{Anderson:1933mb}
C.~D. Anderson, {The Positive Electron}, Phys. Rev. 43 (1933) 491--494.
\newblock \href {http://dx.doi.org/10.1103/PhysRev.43.491}
  {\path{doi:10.1103/PhysRev.43.491}}.

\bibitem{Blackett:1933}
P.~M.~S. Blackett, G.~P.~S. Occhialini, {Some Photographs of the Tracks of
  Penetrating Radiation}, Proc. Roy. Soc. Lond. A139 (1933) 699--727.
\newblock \href {http://dx.doi.org/10.1098/rspa.1933.0048}
  {\path{doi:10.1098/rspa.1933.0048}}.

\bibitem{Chadwick:1933}
J.~Chadwick, P.~M.~S. Blackett, G.~Occhialini, {New Evidence for the Positive
  Electron}, Nature 131 (1933) 473.
\newblock \href {http://dx.doi.org/10.1038/131473b0}
  {\path{doi:10.1038/131473b0}}.

\bibitem{Meitner:1933}
L.~Meitner, K.~Philipp, {Die bei Neutronenanregung auftretenden
  Electronenbahnen}, Naturwissenschaften 21 (1933) 286--287.
\newblock \href {http://dx.doi.org/10.1007/BF01490576}
  {\path{doi:10.1007/BF01490576}}.

\bibitem{Curie:1933a}
I.~Curie, F.~Joliot, {Contribution \`a l'\'etude des \'electrons positifs}, C.
  R. Acad. Sci. Paris 196 (1933) 1105--1107.

\bibitem{Curie:1933b}
I.~Curie, F.~Joliot, {Sur l'origine des \'electrons positifs}, C. R. Acad. Sci.
  Paris 196 (1933) 1581--1583.

\bibitem{Anderson:1933}
C.~D. Anderson, S.~H. Neddermeyer, Positrons from gamma-rays, Phys. Rev. 43
  (1933) 1034--1034.
\newblock \href {http://dx.doi.org/10.1103/PhysRev.43.1034}
  {\path{doi:10.1103/PhysRev.43.1034}}.

\bibitem{Anderson:1933zz}
C.~D. Anderson, {Cosmic-Ray Positive and Negative Electrons}, Phys. Rev. 44
  (1933) 406--416.
\newblock \href {http://dx.doi.org/10.1103/PhysRev.44.406}
  {\path{doi:10.1103/PhysRev.44.406}}.

\bibitem{Chadwick235}
J.~Chadwick, P.~M.~S. Blackett, G.~P.~S. Occhialini, Some experiments on the
  production of positive electrons, Proceedings of the Royal Society of London
  A: Mathematical, Physical and Engineering Sciences 144~(851) (1934) 235--249.
\newblock \href {http://dx.doi.org/10.1098/rspa.1934.0045}
  {\path{doi:10.1098/rspa.1934.0045}}.

\bibitem{Blackett:1933b}
P.~M.~S. Blackett, {The Positive Electron}, Nature 132 (1933) 917--919.
\newblock \href {http://dx.doi.org/10.1038/132917a0}
  {\path{doi:10.1038/132917a0}}.

\bibitem{Chamberlain:1955ns}
O.~Chamberlain, E.~Segr\`e, C.~Wiegand, T.~Ypsilantis, {Observation of
  Anti-protons}, Phys. Rev. 100 (1955) 947--950.
\newblock \href {http://dx.doi.org/10.1103/PhysRev.100.947}
  {\path{doi:10.1103/PhysRev.100.947}}.

\bibitem{Golden:1979bw}
R.~L. Golden, S.~Horan, B.~G. Mauger, G.~D. Badhwar, J.~L. Lacy, S.~A.
  Stephens, R.~R. Daniel, J.~E. Zipse, {Evidence for the Existence of Cosmic
  Ray Anti-protons}, Phys. Rev. Lett. 43 (1979) 1196--1199.
\newblock \href {http://dx.doi.org/10.1103/PhysRevLett.43.1196}
  {\path{doi:10.1103/PhysRevLett.43.1196}}.

\bibitem{Chamberlain:1956}
O.~Chamberlain, W.~W. Chupp, G.~Goldhaber, E.~Segr\`e, C.~Wiegand, E.~Amaldi,
  G.~Baroni, C.~Castagnoli, C.~Franzinetti, A.~Manfredini, {Antiproton Star
  Observed in Emulsion}, Phys. Rev. 101 (1956) 909--910.
\newblock \href {http://dx.doi.org/10.1103/PhysRev.101.909}
  {\path{doi:10.1103/PhysRev.101.909}}.

\bibitem{Dorfan:1965zz}
D.~E. Dorfan, J.~Eades, L.~M. Lederman, W.~Lee, C.~C. Ting, {Search for Massive
  Particle}, Phys. Rev. Lett. 14 (1965) 999--1003.
\newblock \href {http://dx.doi.org/10.1103/PhysRevLett.14.999}
  {\path{doi:10.1103/PhysRevLett.14.999}}.

\bibitem{Dorfan:1965uf}
D.~E. Dorfan, J.~Eades, L.~M. Lederman, W.~Lee, C.~C. Ting, {Observation of
  Antideuterons}, Phys. Rev. Lett. 14 (1965) 1003--1006.
\newblock \href {http://dx.doi.org/10.1103/PhysRevLett.14.1003}
  {\path{doi:10.1103/PhysRevLett.14.1003}}.

\bibitem{Massam1965}
T.~Massam, T.~Muller, B.~Righini, M.~Schneegans, A.~Zichichi, {Experimental
  Observation of Antideuteron Production}, Il Nuovo Cimento (1955-1965) 39~(1)
  (1965) 10--14.
\newblock \href {http://dx.doi.org/10.1007/BF02814251}
  {\path{doi:10.1007/BF02814251}}.

\bibitem{Brautti1965}
G.~Brautti, G.~Fidecaro, T.~Massam, M.~Morpurgo, T.~Muller, G.~Petrucci,
  B.~Rocco, P.~Schiavon, M.~Schneegans, A.~Zichichi, {A High-Intensity,
  Partially Separated, Beam of Antiprotons and K Mesons}, Il Nuovo Cimento
  (1955-1965) 38~(4) (1965) 1861--1874.
\newblock \href {http://dx.doi.org/10.1007/BF02750101}
  {\path{doi:10.1007/BF02750101}}.

\bibitem{Cocconi:1960zz}
V.~T. Cocconi, T.~Fazzini, G.~Fidecaro, M.~Legros, N.~H. Lipman, A.~W.
  Merrison, {Mass Analysis of the Secondary Particles Produced by the 25-GeV
  Proton Beam of the CERN Proton Synchrotron}, Phys. Rev. Lett. 5 (1960)
  19--21.
\newblock \href {http://dx.doi.org/10.1103/PhysRevLett.5.19}
  {\path{doi:10.1103/PhysRevLett.5.19}}.

\bibitem{Fitch:1962fq}
V.~L. Fitch, S.~L. Meyer, P.~A. Pirou\'e, {Particle Production at Large Angles
  by 30- and 33-BeV Protons Incident on Aluminum and Beryllium}, Phys. Rev.
  126~(5) (1962) 1849--1851.
\newblock \href {http://dx.doi.org/10.1103/PhysRev.126.1849}
  {\path{doi:10.1103/PhysRev.126.1849}}.

\bibitem{Schwarzschild:1963zz}
A.~Schwarzschild, C.~Zupan\v{c}i\v{c}, {Production of Tritons, Deuterons,
  Nucleons, and Mesons by 30-GeV Protons on Al, Be, and Fe Targets}, Phys. Rev.
  129 (1963) 854--862.
\newblock \href {http://dx.doi.org/10.1103/PhysRev.129.854}
  {\path{doi:10.1103/PhysRev.129.854}}.

\bibitem{Amaldi1963}
U.~Amaldi, T.~Fazzini, G.~Fidecaro, G.~Ghesqui{\`e}re, M.~Legros, H.~Steiner,
  {Deuteron and $\alpha$-Particle Production in Carbon and Polyethylene by 26.6
  GeV/$c$ Protons}, Il Nuovo Cimento (1955-1965) 29~(2) (1963) 476--486.
\newblock \href {http://dx.doi.org/10.1007/BF02750366}
  {\path{doi:10.1007/BF02750366}}.

\bibitem{Diddens1964}
A.~N. Diddens, W.~Galbraith, E.~Lillethun, G.~Manning, A.~G. Parham, A.~E.
  Taylor, T.~G. Walker, A.~M. Wetherell, {Particle Production in Proton-Proton
  Collisions at 19 and 24 GeV/$c$}, Il Nuovo Cimento (1955-1965) 31~(5) (1964)
  961--973.
\newblock \href {http://dx.doi.org/10.1007/BF02821668}
  {\path{doi:10.1007/BF02821668}}.

\bibitem{Blokhintsev:1958}
D.~I. Blokhintsev,
  \href{http://www.jetp.ac.ru/cgi-bin/dn/e_006_05_0995.pdf}{{On the
  Fluctuations of Nuclear Matter}}, Sov. Phys. JETP 6~(5) (1958) 995--999, [Zh.
  Eksp. Teor. Fiz. 33 (1957) 1295.].
\newline\urlprefix\url{http://www.jetp.ac.ru/cgi-bin/dn/e_006_05_0995.pdf}

\bibitem{Hagedorn:1960zz}
R.~Hagedorn, {Deuteron Production in High-Energy Collisions}, Phys. Rev. Lett.
  5 (1960) 276--277.
\newblock \href {http://dx.doi.org/10.1103/PhysRevLett.5.276}
  {\path{doi:10.1103/PhysRevLett.5.276}}.

\bibitem{Hagedorn1962}
R.~Hagedorn, Production of antideuterons, light nuclei and hyperfragments in
  high-energy pp-collisions, Il Nuovo Cimento (1955-1965) 25~(5) (1962)
  1017--1037.
\newblock \href {http://dx.doi.org/10.1007/BF02733726}
  {\path{doi:10.1007/BF02733726}}.

\bibitem{Butler:1961pr}
S.~T. Butler, C.~A. Pearson, {Deuterons from High-Energy Proton Bombardment of
  Matter}, Phys. Rev. Lett. 7 (1961) 69--71.
\newblock \href {http://dx.doi.org/10.1103/PhysRevLett.7.69}
  {\path{doi:10.1103/PhysRevLett.7.69}}.

\bibitem{BUTLER196277}
S.~T. Butler, C.~A. Pearson, {Deuterons from High-Energy Proton Bombardment of
  Matter}, Physics Letters 1~(3) (1962) 77 -- 81.
\newblock \href {http://dx.doi.org/10.1016/0031-9163(62)90274-3}
  {\path{doi:10.1016/0031-9163(62)90274-3}}.

\bibitem{Butler:1963pp}
S.~T. Butler, C.~A. Pearson, {Deuterons from High-Energy Proton Bombardment of
  Matter}, Phys. Rev. 129 (1963) 836--842.
\newblock \href {http://dx.doi.org/10.1103/PhysRev.129.836}
  {\path{doi:10.1103/PhysRev.129.836}}.

\bibitem{Binon:1969qz}
F.~G. Binon, et~al., {Production of Antideuterons by 43 GeV, 52 GeV and 70 GeV
  Protons}, Phys. Lett. 30B (1969) 510--513.
\newblock \href {http://dx.doi.org/10.1016/0370-2693(69)90186-5}
  {\path{doi:10.1016/0370-2693(69)90186-5}}.

\bibitem{Antipov:1971zs}
{\relax Yu}.~M. Antipov, et~al., {Production of Low Momentum Negative Particles
  by 70 GeV Protons}, Phys. Lett. 34B (1971) 164--166.
\newblock \href {http://dx.doi.org/10.1016/0370-2693(71)90697-6}
  {\path{doi:10.1016/0370-2693(71)90697-6}}.

\bibitem{Alper:1973my}
B.~Alper, et~al., {Large Angle Production of Stable Particles Heavier than the
  Proton and a Search for Quarks at the CERN Intersecting Storage Rings}, Phys.
  Lett. 46B (1973) 265--268.
\newblock \href {http://dx.doi.org/10.1016/0370-2693(73)90700-4}
  {\path{doi:10.1016/0370-2693(73)90700-4}}.

\bibitem{Appel:1974fs}
J.~A. Appel, M.~H. Bourquin, I.~Gaines, L.~M. Lederman, H.~P. Paar, J.~P.
  Repellin, D.~H. Saxon, J.~K. Yoh, B.~C. Brown, J.~M. Gaillard, {Heavy
  Particle Production in 300-GeV/$c$ Proton - Tungsten Collisions}, Phys. Rev.
  Lett. 32 (1974) 428--432.
\newblock \href {http://dx.doi.org/10.1103/PhysRevLett.32.428}
  {\path{doi:10.1103/PhysRevLett.32.428}}.

\bibitem{Antipov:1970uc}
{\relax Yu}.~M. Antipov, et~al., {Observation of Antihelium-3}, Yad. Fiz. 12
  (1970) 311--322.

\bibitem{Antipov:1971iq}
{\relax Yu}.~M. Antipov, et~al., {Observation of Antihelium-3}, Nucl. Phys. B31
  (1971) 235--252.
\newblock \href {http://dx.doi.org/10.1016/0550-3213(71)90228-8}
  {\path{doi:10.1016/0550-3213(71)90228-8}}.

\bibitem{Vishnevsky:1974ks}
N.~K. Vishnevsky, et~al., {Observation of Antitritium}, Yad. Fiz. 20 (1974)
  694--708.

\bibitem{Bozzoli:1978ud}
W.~Bozzoli, A.~Bussiere, G.~Giacomelli, E.~Lesquoy, R.~Meunier, L.~Moscoso,
  A.~Muller, F.~Rimondi, S.~Zylberajch, {Production of $d$, $t$, $^3$He,
  $\bar{d}$, $\bar{t}$ and $^3$$\overline{He}$ by 200-GeV Protons}, Nucl. Phys.
  B144 (1978) 317--328.
\newblock \href {http://dx.doi.org/10.1016/0550-3213(78)90373-5}
  {\path{doi:10.1016/0550-3213(78)90373-5}}.

\bibitem{Albrow:1975gr}
M.~G. Albrow, et~al., {Search for Stable Particles of Charge $\geq 1$ and Mass
  $\geq$ Deuteron Mass}, Nucl. Phys. B97 (1975) 189--200.
\newblock \href {http://dx.doi.org/10.1016/0550-3213(75)90030-9}
  {\path{doi:10.1016/0550-3213(75)90030-9}}.

\bibitem{Robinson2011}
M.~Robinson, {Symmetry and the Standard Model}, Springer-Verlag, New York, NY,
  2011.
\newblock \href {http://dx.doi.org/10.1007/978-1-4419-8267-4}
  {\path{doi:10.1007/978-1-4419-8267-4}}.

\bibitem{Costa2012}
G.~Costa, G.~Fogli, {Symmetries and Group Theory in Particle Physics: An
  Introduction to Space-Time and Internal Symmetries}, Springer-Verlag, Berlin,
  2012.
\newblock \href {http://dx.doi.org/10.1007/978-3-642-15482-9}
  {\path{doi:10.1007/978-3-642-15482-9}}.

\bibitem{Lee:1956qn}
T.~D. Lee, C.-N. Yang, {Question of Parity Conservation in Weak Interactions},
  Phys. Rev. 104 (1956) 254--258.
\newblock \href {http://dx.doi.org/10.1103/PhysRev.104.254}
  {\path{doi:10.1103/PhysRev.104.254}}.

\bibitem{Wu:1957my}
C.~S. Wu, E.~Ambler, R.~W. Hayward, D.~D. Hoppes, R.~P. Hudson, {Experimental
  Test of Parity Conservation in Beta Decay}, Phys. Rev. 105 (1957) 1413--1414.
\newblock \href {http://dx.doi.org/10.1103/PhysRev.105.1413}
  {\path{doi:10.1103/PhysRev.105.1413}}.

\bibitem{GellMann:1955jx}
M.~Gell-Mann, A.~Pais, {Behavior of Neutral Particles Under Charge
  Conjugation}, Phys. Rev. 97 (1955) 1387--1389.
\newblock \href {http://dx.doi.org/10.1103/PhysRev.97.1387}
  {\path{doi:10.1103/PhysRev.97.1387}}.

\bibitem{Christenson:1964fg}
J.~H. Christenson, J.~W. Cronin, V.~L. Fitch, R.~Turlay, {Evidence for the
  $2\pi$ Decay of the $K_2^0$ Meson}, Phys. Rev. Lett. 13 (1964) 138--140.
\newblock \href {http://dx.doi.org/10.1103/PhysRevLett.13.138}
  {\path{doi:10.1103/PhysRevLett.13.138}}.

\bibitem{Stueckelberg:1941rg}
E.~C.~G. Stueckelberg, {Remarks on the Creation of Pairs of Particles in the
  Theory of Relativity}, Helv. Phys. Acta 14 (1941) 588--594.

\bibitem{Feynman:1948ur}
R.~P. Feynman, {Space-time Approach to Nonrelativistic Quantum Mechanics}, Rev.
  Mod. Phys. 20 (1948) 367--387.
\newblock \href {http://dx.doi.org/10.1103/RevModPhys.20.367}
  {\path{doi:10.1103/RevModPhys.20.367}}.

\bibitem{Abouzaid:2010ny}
E.~Abouzaid, et~al., {Precise Measurements of Direct CP Violation, CPT
  Symmetry, and Other Parameters in the Neutral Kaon System}, Phys. Rev. D83
  (2011) 092001.
\newblock \href {http://dx.doi.org/10.1103/PhysRevD.83.092001}
  {\path{doi:10.1103/PhysRevD.83.092001}}.

\bibitem{Olive:2016xmw}
C.~Patrignani, et~al., {Review of Particle Physics}, Chin. Phys. C40~(10)
  (2016) 100001.
\newblock \href {http://dx.doi.org/10.1088/1674-1137/40/10/100001}
  {\path{doi:10.1088/1674-1137/40/10/100001}}.

\bibitem{Mittleman:1999it}
R.~K. Mittleman, I.~I. Ioannou, H.~G. Dehmelt, N.~Russell, {Bound on $CPT$ and
  Lorentz Symmetry with a Trapped Electron}, Phys. Rev. Lett. 83 (1999)
  2116--2119.
\newblock \href {http://dx.doi.org/10.1103/PhysRevLett.83.2116}
  {\path{doi:10.1103/PhysRevLett.83.2116}}.

\bibitem{Dehmelt:1999jh}
H.~Dehmelt, R.~Mittleman, R.~S. van Dyck, Jr., P.~Schwinberg, {Past
  Electron-Positron $g$ - 2 Experiments Yielded Sharpest Bound on $CPT$
  Violation for Point Particles}, Phys. Rev. Lett. 83 (1999) 4694--4696.
\newblock \href {http://dx.doi.org/10.1103/PhysRevLett.83.4694}
  {\path{doi:10.1103/PhysRevLett.83.4694}}.

\bibitem{Hanneke:2008tm}
D.~Hanneke, S.~Fogwell, G.~Gabrielse, {New Measurement of the Electron Magnetic
  Moment and the Fine Structure Constant}, Phys. Rev. Lett. 100 (2008) 120801.
\newblock \href {http://dx.doi.org/10.1103/PhysRevLett.100.120801}
  {\path{doi:10.1103/PhysRevLett.100.120801}}.

\bibitem{Bennett:2007aa}
G.~W. Bennett, et~al., {Search for Lorentz and $CPT$ Violation Effects in Muon
  Spin Precession}, Phys. Rev. Lett. 100 (2008) 091602.
\newblock \href {http://dx.doi.org/10.1103/PhysRevLett.100.091602}
  {\path{doi:10.1103/PhysRevLett.100.091602}}.

\bibitem{Grange:2015fou}
J.~Grange, et~al., {Muon ($g$-2) Technical Design Report}\href
  {http://arxiv.org/abs/1501.06858} {\path{arXiv:1501.06858}}.

\bibitem{Ulmer:2015jra}
S.~Ulmer, et~al., {High-Precision Comparison of the Antiproton-to-Proton
  Charge-to-Mass Ratio}, Nature 524~(7564) (2015) 196--199.
\newblock \href {http://dx.doi.org/10.1038/nature14861}
  {\path{doi:10.1038/nature14861}}.

\bibitem{Smorra:2015cuj}
C.~Smorra, et~al., {BASE - The Baryon Antibaryon Symmetry Experiment}, Eur.
  Phys. J. ST 224~(16) (2015) 3055--3108.
\newblock \href {http://dx.doi.org/10.1140/epjst/e2015-02607-4,
  10.1140/epjst/e2015-02336-2} {\path{doi:10.1140/epjst/e2015-02607-4,
  10.1140/epjst/e2015-02336-2}}.

\bibitem{Amole:2014vna}
C.~Amole, et~al., {The ALPHA Antihydrogen Trapping Apparatus}, Nucl. Instrum.
  Meth. A735 (2014) 319--340.
\newblock \href {http://dx.doi.org/10.1016/j.nima.2013.09.043}
  {\path{doi:10.1016/j.nima.2013.09.043}}.

\bibitem{Amole:2012zza}
C.~Amole, et~al., {Resonant Quantum Transitions in Trapped Antihydrogen Atoms},
  Nature 483 (2012) 439--443.
\newblock \href {http://dx.doi.org/10.1038/nature10942}
  {\path{doi:10.1038/nature10942}}.

\bibitem{Kuroda:2014yya}
N.~Kuroda, et~al., {A Source of Antihydrogen for In-Flight Hyperfine
  Spectroscopy}, Nature Commun. 5 (2014) 3089--3092.
\newblock \href {http://dx.doi.org/10.1038/ncomms4089}
  {\path{doi:10.1038/ncomms4089}}.

\bibitem{Schwinger:1951xk}
J.~S. Schwinger, {The Theory of Quantized Fields. 1.}, Phys. Rev. 82 (1951)
  914--927.
\newblock \href {http://dx.doi.org/10.1103/PhysRev.82.914}
  {\path{doi:10.1103/PhysRev.82.914}}.

\bibitem{Luders:1954zz}
G.~L{\"u}ders, {On the Equivalence of Invariance under Time Reversal and under
  Particle-Antiparticle Conjugation for Relativistic Field Theories}, Kong.
  Dan. Vid. Sel. Mat. Fys. Med. 28N5~(5) (1954) 1--17.

\bibitem{Lueders:1992dq}
G.~L{\"u}ders, {Proof of the TCP Theorem}, Annals Phys. 2 (1957) 1--15, [Annals
  Phys.281,1004(2000)].
\newblock \href {http://dx.doi.org/10.1016/0003-4916(57)90032-5}
  {\path{doi:10.1016/0003-4916(57)90032-5}}.

\bibitem{Luders:1957zz}
G.~L{\"u}ders, B.~Zumino, {Some Consequences of TCP-Invariance}, Phys. Rev. 106
  (1957) 385--386.
\newblock \href {http://dx.doi.org/10.1103/PhysRev.106.385}
  {\path{doi:10.1103/PhysRev.106.385}}.

\bibitem{Pauli:1955}
W.~Pauli, \href{https://lccn.loc.gov/56040984}{{Niels Bohr and the development
  of physics; essays dedicated to Niels Bohr on the occasion of his seventieth
  birthday}}, Pergamon Press, London, 1955.
\newline\urlprefix\url{https://lccn.loc.gov/56040984}

\bibitem{Bell:1996nh}
J.~S. Bell, {Time Reversal in Field Theory}, Proc. Roy. Soc. Lond. A231 (1955)
  479--495.
\newblock \href {http://dx.doi.org/10.1098/rspa.1955.0189}
  {\path{doi:10.1098/rspa.1955.0189}}.

\bibitem{Carosi:1990ms}
R.~Carosi, et~al., {A Measurement of the Phases of the {CP} Violating
  Amplitudes in $K^0 \to 2 \pi$ Decays and a Test of {CPT} Invariance}, Phys.
  Lett. B237 (1990) 303--312.
\newblock \href {http://dx.doi.org/10.1016/0370-2693(90)91448-K}
  {\path{doi:10.1016/0370-2693(90)91448-K}}.

\bibitem{AlaviHarati:2002ye}
A.~Alavi-Harati, et~al., {Measurements of Direct CP Violation, CPT symmetry,
  and Other Parameters in the Neutral Kaon System}, Phys. Rev. D67 (2003)
  012005, [Erratum: Phys. Rev. D70, 079904 (2004)].
\newblock \href {http://dx.doi.org/10.1103/PhysRevD.67.012005,
  10.1103/PhysRevD.70.079904} {\path{doi:10.1103/PhysRevD.67.012005,
  10.1103/PhysRevD.70.079904}}.

\bibitem{Kostelecky:2008ts}
V.~A. Kostelecky, N.~Russell, {Data Tables for Lorentz and CPT Violation}, Rev.
  Mod. Phys. 83 (2011) 11--31.
\newblock \href {http://dx.doi.org/10.1103/RevModPhys.83.11}
  {\path{doi:10.1103/RevModPhys.83.11}}.

\bibitem{Weinberg2008}
S.~Weinberg,
  \href{https://global.oup.com/academic/product/cosmology-9780198526827}{{Cosmology}},
  Oxford University Press, Oxford, UK, 2008.
\newline\urlprefix\url{https://global.oup.com/academic/product/cosmology-9780198526827}

\bibitem{Sozzi2012}
M.~Sozzi,
  \href{https://global.oup.com/academic/product/discrete-symmetries-and-cp-violation-9780199655427}{{Discrete
  Symmetries and CP Violation}}, Oxford University Press, Oxford, UK, 2012.
\newline\urlprefix\url{https://global.oup.com/academic/product/discrete-symmetries-and-cp-violation-9780199655427}

\bibitem{Canetti:2012zc}
L.~Canetti, M.~Drewes, M.~Shaposhnikov, {Matter and Antimatter in the
  Universe}, New J. Phys. 14 (2012) 095012.
\newblock \href {http://dx.doi.org/10.1088/1367-2630/14/9/095012}
  {\path{doi:10.1088/1367-2630/14/9/095012}}.

\bibitem{Schlickeiser2002}
R.~Schlickeiser, {Cosmic Ray Astrophysics}, Springer-Verlag, Berlin Heidelberg,
  2002.
\newblock \href {http://dx.doi.org/10.1007/978-3-662-04814-6}
  {\path{doi:10.1007/978-3-662-04814-6}}.

\bibitem{Stanev2010}
T.~Stanev, {High Energy Cosmic Rays, Second Edition}, Springer-Verlag, Berlin
  Heidelberg, 2010.
\newblock \href {http://dx.doi.org/10.1007/978-3-540-85148-6}
  {\path{doi:10.1007/978-3-540-85148-6}}.

\bibitem{Fukugita:2002hu}
M.~Fukugita, T.~Yanagida, {Resurrection of Grand Unified Theory Baryogenesis},
  Phys. Rev. Lett. 89 (2002) 131602.
\newblock \href {http://dx.doi.org/10.1103/PhysRevLett.89.131602}
  {\path{doi:10.1103/PhysRevLett.89.131602}}.

\bibitem{Huang:2016wwj}
W.-C. Huang, H.~P{\"a}s, S.~Zeissner, {Neutrino Assisted GUT Baryogenesis -
  Revisited}, Phys. Rev. D97~(5) (2018) 055040.
\newblock \href {http://dx.doi.org/10.1103/PhysRevD.97.055040}
  {\path{doi:10.1103/PhysRevD.97.055040}}.

\bibitem{Baron:2013eja}
J.~Baron, et~al., {Order of Magnitude Smaller Limit on the Electric Dipole
  Moment of the Electron}, Science 343 (2014) 269--272.
\newblock \href {http://dx.doi.org/10.1126/science.1248213}
  {\path{doi:10.1126/science.1248213}}.

\bibitem{Ackermann:2002ad}
K.~H. Ackermann, et~al., {STAR Detector Overview}, Nucl. Instrum. Meth. A499
  (2003) 624--632.
\newblock \href {http://dx.doi.org/10.1016/S0168-9002(02)01960-5}
  {\path{doi:10.1016/S0168-9002(02)01960-5}}.

\bibitem{Harrison:2003sb}
M.~Harrison, T.~Ludlam, S.~Ozaki, {RHIC Project Overview}, Nucl. Instrum. Meth.
  A499 (2003) 235--244.
\newblock \href {http://dx.doi.org/10.1016/S0168-9002(02)01937-X}
  {\path{doi:10.1016/S0168-9002(02)01937-X}}.

\bibitem{Anderson:2003ur}
M.~Anderson, et~al., {The STAR Time Projection Chamber: A Unique Tool for
  Studying High Multiplicity Events at RHIC}, Nucl. Instrum. Meth. A499 (2003)
  659--678.
\newblock \href {http://dx.doi.org/10.1016/S0168-9002(02)01964-2}
  {\path{doi:10.1016/S0168-9002(02)01964-2}}.

\bibitem{Bergsma:2002ac}
F.~Bergsma, et~al., {The STAR Detector Magnet Subsystem}, Nucl. Instrum. Meth.
  A499 (2003) 633--639.
\newblock \href {http://dx.doi.org/10.1016/S0168-9002(02)01961-7}
  {\path{doi:10.1016/S0168-9002(02)01961-7}}.

\bibitem{Kochenda:2002zz}
L.~Kochenda, S.~Kozlov, P.~Kravtsov, A.~Markov, M.~Strikhanov, B.~Stringfellow,
  V.~Trofimov, R.~Wells, H.~Wieman, {STAR TPC Gas System}, Nucl. Instrum. Meth.
  A499 (2003) 703--712.
\newblock \href {http://dx.doi.org/10.1016/S0168-9002(02)01967-8}
  {\path{doi:10.1016/S0168-9002(02)01967-8}}.

\bibitem{Beddo:2002zx}
M.~Beddo, et~al., {The STAR Barrel Electromagnetic Calorimeter}, Nucl. Instrum.
  Meth. A499 (2003) 725--739.
\newblock \href {http://dx.doi.org/10.1016/S0168-9002(02)01970-8}
  {\path{doi:10.1016/S0168-9002(02)01970-8}}.

\bibitem{CerronZeballos:1995iy}
E.~Cerron~Zeballos, I.~Crotty, D.~Hatzifotiadou, J.~Lamas~Valverde, S.~Neupane,
  M.~C.~S. Williams, A.~Zichichi, {A New Type of Resistive Plate Chamber: The
  Multigap RPC}, Nucl. Instrum. Meth. A374 (1996) 132--136.
\newblock \href {http://dx.doi.org/10.1016/0168-9002(96)00158-1}
  {\path{doi:10.1016/0168-9002(96)00158-1}}.

\bibitem{Geurts:2004}
F.~Geurts, et~al., {Performance of the Prototype MRPC Detector for STAR}, Nucl.
  Instrum. Meth. A533 (2004) 60--64.
\newblock \href {http://dx.doi.org/10.1016/j.nima.2004.07.001}
  {\path{doi:10.1016/j.nima.2004.07.001}}.

\bibitem{Ruan:2009ug}
L.~Ruan, et~al., {Perspectives of a Midrapidity Dimuon Program at RHIC: A Novel
  and Compact Muon Telescope Detector}, J. Phys. G36 (2009) 095001.
\newblock \href {http://dx.doi.org/10.1088/0954-3899/36/9/095001}
  {\path{doi:10.1088/0954-3899/36/9/095001}}.

\bibitem{Wang:2011ay}
Y.~Wang, et~al., {Performance of a New LMRPC Prototype for the STAR MTD
  System}, Nucl. Instrum. Meth. A640 (2011) 85--90.
\newblock \href {http://dx.doi.org/10.1016/j.nima.2011.03.012}
  {\path{doi:10.1016/j.nima.2011.03.012}}.

\bibitem{Xu:2016}
Y.~Xu, J.~Chen, Y.-G. Ma, A.~Tang, Z.~Xu, Y.~Zhu, Physics performance of the
  star zero degree calorimeter at relativistic heavy ion collider, Nucl. Sci.
  Tech. 27 (2016) 126.
\newblock \href {http://dx.doi.org/doi.org/10.1007/s41365-016-0129-z}
  {\path{doi:doi.org/10.1007/s41365-016-0129-z}}.

\bibitem{Yang:2014xta}
C.~Yang, et~al., {Calibration and Performance of the STAR Muon Telescope
  Detector Using Cosmic Rays}, Nucl. Instrum. Meth. A762 (2014) 1--6.
\newblock \href {http://dx.doi.org/10.1016/j.nima.2014.05.075}
  {\path{doi:10.1016/j.nima.2014.05.075}}.

\bibitem{Qiu:2014dha}
H.~Qiu, {STAR Heavy Flavor Tracker}, Nucl. Phys. A931 (2014) 1141--1146.
\newblock \href {http://dx.doi.org/10.1016/j.nuclphysa.2014.08.056}
  {\path{doi:10.1016/j.nuclphysa.2014.08.056}}.

\bibitem{Long:2017}
L.~Ma, X.~Dong, H.~Qiu, S.~Margetis, Y.-G. Ma, Alignment calibration and
  performance study of the star pxl detector, Nucl. Sci. Tech. 28 (2017) 25.
\newblock \href {http://dx.doi.org/10.1007/s41365-016-0177-4}
  {\path{doi:10.1007/s41365-016-0177-4}}.

\bibitem{Aamodt:2008zz}
K.~Aamodt, et~al., {The ALICE Experiment at the CERN LHC}, JINST 3 (2008)
  S08002.
\newblock \href {http://dx.doi.org/10.1088/1748-0221/3/08/S08002}
  {\path{doi:10.1088/1748-0221/3/08/S08002}}.

\bibitem{Abelev:2014ffa}
B.~B. Abelev, et~al., {Performance of the ALICE Experiment at the CERN LHC},
  Int. J. Mod. Phys. A29 (2014) 1430044.
\newblock \href {http://dx.doi.org/10.1142/S0217751X14300440}
  {\path{doi:10.1142/S0217751X14300440}}.

\bibitem{Aamodt:2010aa}
K.~Aamodt, et~al., {Alignment of the ALICE Inner Tracking System with
  Cosmic-Ray Tracks}, JINST 5 (2010) P03003.
\newblock \href {http://dx.doi.org/10.1088/1748-0221/5/03/P03003}
  {\path{doi:10.1088/1748-0221/5/03/P03003}}.

\bibitem{Carminati:2004fp}
P.~Cortese, et~al., {ALICE: Physics Performance Report, Volume I}, J. Phys. G30
  (2004) 1517--1763.
\newblock \href {http://dx.doi.org/10.1088/0954-3899/30/11/001}
  {\path{doi:10.1088/0954-3899/30/11/001}}.

\bibitem{Alme:2010ke}
J.~Alme, et~al., {The ALICE TPC, a Large 3-Dimensional Tracking Device with
  Fast Readout for Ultra-High Multiplicity Events}, Nucl. Instrum. Meth. A622
  (2010) 316--367.
\newblock \href {http://dx.doi.org/10.1016/j.nima.2010.04.042}
  {\path{doi:10.1016/j.nima.2010.04.042}}.

\bibitem{ALICE_TRD1}
{ALICE Collaboration, ALICE Transition Radiation Detector: Technical Design
  Report}, CERN-LHCC-2001-021, http://cds.cern.ch/record/519145.

\bibitem{ALICE_TOF1}
{ALICE Collaboration, ALICE Time-Of-Flight System (TOF): Technical Design
  Report}, CERN-LHCC-2000-012, http://cds.cern.ch/record/430132.

\bibitem{Alessandro:2006yt}
P.~Cortese, et~al., {ALICE: Physics Performance Report, Volume II}, J. Phys.
  G32 (2006) 1295--2040.
\newblock \href {http://dx.doi.org/10.1088/0954-3899/32/10/001}
  {\path{doi:10.1088/0954-3899/32/10/001}}.

\bibitem{Bichsel:2006cs}
H.~Bichsel, {A Method to Improve Tracking and Particle Identification in TPCs
  and Silicon Detectors}, Nucl. Instrum. Meth. A562 (2006) 154--197.
\newblock \href {http://dx.doi.org/10.1016/j.nima.2006.03.009}
  {\path{doi:10.1016/j.nima.2006.03.009}}.

\bibitem{Adam:2015vda}
J.~Adam, et~al., {Production of Light Nuclei and Antinuclei in $pp$ and Pb-Pb
  Collisions at Energies Available at the CERN Large Hadron Collider}, Phys.
  Rev. C93~(2) (2016) 024917.
\newblock \href {http://dx.doi.org/10.1103/PhysRevC.93.024917}
  {\path{doi:10.1103/PhysRevC.93.024917}}.

\bibitem{Acharya:2017bso}
S.~Acharya, et~al., {Production of $^{4}$He and $^{4}\overline{\textrm{He}}$ in
  Pb-Pb collisions at $\sqrt{s_{\mathrm{NN}}}$ = 2.76 TeV at the LHC}, Nucl.
  Phys. A971 (2018) 1--20.
\newblock \href {http://dx.doi.org/10.1016/j.nuclphysa.2017.12.004}
  {\path{doi:10.1016/j.nuclphysa.2017.12.004}}.

\bibitem{Fuke:2005it}
H.~Fuke, et~al., {Search for Cosmic-Ray Antideuterons}, Phys. Rev. Lett. 95
  (2005) 081101.
\newblock \href {http://dx.doi.org/10.1103/PhysRevLett.95.081101}
  {\path{doi:10.1103/PhysRevLett.95.081101}}.

\bibitem{Sasaki:2008zzb}
M.~Sasaki, et~al., {Search for Antihelium: Progress with BESS}, Adv. Space Res.
  42 (2008) 450--454.
\newblock \href {http://dx.doi.org/10.1016/j.asr.2007.09.012}
  {\path{doi:10.1016/j.asr.2007.09.012}}.

\bibitem{Abe:2012tz}
K.~Abe, et~al., {Search for Antihelium with the BESS-Polar Spectrometer}, Phys.
  Rev. Lett. 108 (2012) 131301.
\newblock \href {http://dx.doi.org/10.1103/PhysRevLett.108.131301}
  {\path{doi:10.1103/PhysRevLett.108.131301}}.

\bibitem{Aguilar:2016kjl}
M.~Aguilar, et~al., {Antiproton Flux, Antiproton-to-Proton Flux Ratio, and
  Properties of Elementary Particle Fluxes in Primary Cosmic Rays Measured with
  the Alpha Magnetic Spectrometer on the International Space Station}, Phys.
  Rev. Lett. 117~(9) (2016) 091103.
\newblock \href {http://dx.doi.org/10.1103/PhysRevLett.117.091103}
  {\path{doi:10.1103/PhysRevLett.117.091103}}.

\bibitem{Barton:1983ch}
M.~Q. Barton, {Acceleration of Heavy Ions in the AGS}, IEEE Trans. Nucl. Sci.
  30 (1983) 2019.
\newblock \href {http://dx.doi.org/10.1109/TNS.1983.4332702}
  {\path{doi:10.1109/TNS.1983.4332702}}.

\bibitem{Barton:1987mk}
D.~S. Barton, {Heavy Ion Program at BNL: AGS, RHIC}, Conf. Proc. C870316 (1987)
  804.

\bibitem{Gutbrod:2016}
H.~H. Gutbrod, {The Path to Heavy Ions at LHC and Beyond}, in: J.~Rafelski
  (Ed.), Melting Hadrons, Boiling Quarks - From Hagedorn Temperature to
  Ultra-Relativistic Heavy-Ion Collisions at CERN: With a Tribute to Rolf
  Hagedorn, 2016, pp. 97--106.
\newblock \href {http://dx.doi.org/10.1007/978-3-319-17545-4_13}
  {\path{doi:10.1007/978-3-319-17545-4_13}}.

\bibitem{Adams:2005dq}
J.~Adams, et~al., {Experimental and Theoretical Challenges in the Search for
  the Quark Gluon Plasma: The STAR Collaboration's Critical Assessment of the
  Evidence from RHIC Collisions}, Nucl. Phys. A757 (2005) 102--183.
\newblock \href {http://dx.doi.org/10.1016/j.nuclphysa.2005.03.085}
  {\path{doi:10.1016/j.nuclphysa.2005.03.085}}.

\bibitem{Adcox:2004mh}
K.~Adcox, et~al., {Formation of Dense Partonic Matter in Relativistic
  Nucleus-Nucleus Collisions at RHIC: Experimental Evaluation by the PHENIX
  Collaboration}, Nucl. Phys. A757 (2005) 184--283.
\newblock \href {http://dx.doi.org/10.1016/j.nuclphysa.2005.03.086}
  {\path{doi:10.1016/j.nuclphysa.2005.03.086}}.

\bibitem{Arsene:2004fa}
I.~Arsene, et~al., {Quark Gluon Plasma and Color Glass Condensate at RHIC? The
  Perspective from the BRAHMS Experiment}, Nucl. Phys. A757 (2005) 1--27.
\newblock \href {http://dx.doi.org/10.1016/j.nuclphysa.2005.02.130}
  {\path{doi:10.1016/j.nuclphysa.2005.02.130}}.

\bibitem{Back:2004je}
B.~B. Back, et~al., {The PHOBOS Perspective on Discoveries at RHIC}, Nucl.
  Phys. A757 (2005) 28--101.
\newblock \href {http://dx.doi.org/10.1016/j.nuclphysa.2005.03.084}
  {\path{doi:10.1016/j.nuclphysa.2005.03.084}}.

\bibitem{Muller:2006ee}
B.~Muller, J.~L. Nagle, {Results from the Relativistic Heavy Ion Collider},
  Ann. Rev. Nucl. Part. Sci. 56 (2006) 93--135.
\newblock \href {http://dx.doi.org/10.1146/annurev.nucl.56.080805.140556}
  {\path{doi:10.1146/annurev.nucl.56.080805.140556}}.

\bibitem{Jacak:2010zz}
B.~Jacak, P.~Steinberg, {Creating the Perfect Liquid in Heavy-Ion Collisions},
  Phys. Today 63N5 (2010) 39--43.
\newblock \href {http://dx.doi.org/10.1063/1.3431330}
  {\path{doi:10.1063/1.3431330}}.

\bibitem{Shuryak:2014zxa}
E.~Shuryak, {Strongly Coupled Quark-Gluon Plasma in Heavy Ion Collisions}, Rev.
  Mod. Phys. 89 (2017) 035001.
\newblock \href {http://dx.doi.org/10.1103/RevModPhys.89.035001}
  {\path{doi:10.1103/RevModPhys.89.035001}}.

\bibitem{Braun-Munzinger:2015hba}
P.~Braun-Munzinger, V.~Koch, T.~Schäfer, J.~Stachel, {Properties of Hot and
  Dense Matter from Relativistic Heavy-Ion Collisions}, Phys. Rept. 621 (2016)
  76--126.
\newblock \href {http://dx.doi.org/10.1016/j.physrep.2015.12.003}
  {\path{doi:10.1016/j.physrep.2015.12.003}}.

\bibitem{Luo:2017NST}
X.-F. Luo, N.~Xu, {Search for the QCD critical point with fluctuations of
  conserved quantities in relativistic heavy-ion collisions at RHIC: an
  overview}, Nucl. Sci. Tech. 28 (2017) 112.
\newblock \href {http://dx.doi.org/10.1007/s41365-017-0257-0}
  {\path{doi:10.1007/s41365-017-0257-0}}.

\bibitem{STAR:2017ckg}
L.~Adamczyk, et~al., {Global $\Lambda$ Hyperon Polarization in Nuclear
  Collisions: Evidence for the Most Vortical Fluid}, Nature 548 (2017) 62--65.
\newblock \href {http://dx.doi.org/10.1038/nature23004}
  {\path{doi:10.1038/nature23004}}.

\bibitem{Hattori:2017NST}
K.~Hattori, X.-G. Huang, Novel quantum phenomena induced by strong magnetic
  fields in heavy-ion collisions, Nucl. Sci. Tech. 28 (2017) 26.
\newblock \href {http://dx.doi.org/10.1107/s41365-016-0178-3}
  {\path{doi:10.1107/s41365-016-0178-3}}.

\bibitem{Aoki:1992mb}
M.~Aoki, et~al., {Measurements at 0$^\circ$ of Negatively Charged Particles and
  Antinuclei Produced in Collisions of 14.6$A$ GeV/$c$ Si on Al, Cu, and Au
  Targets}, Phys. Rev. Lett. 69 (1992) 2345--2348.
\newblock \href {http://dx.doi.org/10.1103/PhysRevLett.69.2345}
  {\path{doi:10.1103/PhysRevLett.69.2345}}.

\bibitem{Gutbrod:1988gt}
H.~H. Gutbrod, A.~Sandoval, P.~J. Johansen, A.~M. Poskanzer, J.~Gosset, W.~G.
  Meyer, G.~D. Westfall, R.~Stock, {Final State Interactions in the Production
  of Hydrogen and Helium Isotopes by Relativistic Heavy Ions on Uranium}, Phys.
  Rev. Lett. 37 (1976) 667--670.
\newblock \href {http://dx.doi.org/10.1103/PhysRevLett.37.667}
  {\path{doi:10.1103/PhysRevLett.37.667}}.

\bibitem{Lemaire:1980qw}
M.~C. Lemaire, S.~Nagamiya, S.~Schnetzer, H.~Steiner, I.~Tanihata, {Composite
  Particle Emission in High-Energy Heavy-Ion Collisions}, Phys. Lett. 85B
  (1979) 38--42.
\newblock \href {http://dx.doi.org/10.1016/0370-2693(79)90772-X}
  {\path{doi:10.1016/0370-2693(79)90772-X}}.

\bibitem{Jacak:1985zz}
B.~V. Jacak, D.~Fox, G.~D. Westfall, {Coalescence of Complex Fragments}, Phys.
  Rev. C31 (1985) 704--706.
\newblock \href {http://dx.doi.org/10.1103/PhysRevC.31.704}
  {\path{doi:10.1103/PhysRevC.31.704}}.

\bibitem{Hayashi:1988en}
S.~Hayashi, Y.~Y.~Miake, T.~Nagae, S.~Nagamiya, H.~Hamagaki, O.~Hashimoto,
  Y.~Shida, I.~Tanihata, K.~Kimura, O.~Yamakawa, T.~Kobayashi, X.~X. Bai,
  {Production of Pions and Light Fragments in 0.8$A$ GeV La + La Collisions},
  Phys. Rev. C38 (1988) 1229--1241.
\newblock \href {http://dx.doi.org/10.1103/PhysRevC.38.1229}
  {\path{doi:10.1103/PhysRevC.38.1229}}.

\bibitem{Saito:1994tg}
N.~Saito, et~al., {Composite Particle Production in Relativistic Au + Pt, Si +
  Pt, and p + Pt Collisions}, Phys. Rev. C49 (1994) 3211--3218.
\newblock \href {http://dx.doi.org/10.1103/PhysRevC.49.3211}
  {\path{doi:10.1103/PhysRevC.49.3211}}.

\bibitem{Abbott:1994np}
T.~Abbott, et~al., {Charged Hadron Distributions in Central and Peripheral Si +
  A Collisions at 14.6$A$ GeV/$c$}, Phys. Rev. C50 (1994) 1024--1047.
\newblock \href {http://dx.doi.org/10.1103/PhysRevC.50.1024}
  {\path{doi:10.1103/PhysRevC.50.1024}}.

\bibitem{Wang:1994rua}
S.~Wang, et~al., {Light Fragment Production and Power Law Behavior in Au + Au
  Collisions}, Phys. Rev. Lett. 74 (1995) 2646--2649.
\newblock \href {http://dx.doi.org/10.1103/PhysRevLett.74.2646}
  {\path{doi:10.1103/PhysRevLett.74.2646}}.

\bibitem{Yan:2006}
T.~Z. Yan, Y.~G. Ma, X.~Z. Cai, et~al., {Scaling of anisotropic flow and
  momentum-space densities for light particles in intermediate energy heavy ion
  collisions}, Phys. Lett. B 638 (2006) 50--54.
\newblock \href {http://dx.doi.org/10.1016/j.physletb.2006.05.018}
  {\path{doi:10.1016/j.physletb.2006.05.018}}.

\bibitem{Mekjian:1977ei}
A.~Mekjian, {Thermodynamic Model for Composite Particle Emission in
  Relativistic Heavy Ion Collisions}, Phys. Rev. Lett. 38 (1977) 640--643.
\newblock \href {http://dx.doi.org/10.1103/PhysRevLett.38.640}
  {\path{doi:10.1103/PhysRevLett.38.640}}.

\bibitem{Sato:1981ez}
H.~Sato, K.~Yazaki, {On the Coalescence Model for High-Energy Nuclear
  Reactions}, Phys. Lett. 98B (1981) 153--157.
\newblock \href {http://dx.doi.org/10.1016/0370-2693(81)90976-X}
  {\path{doi:10.1016/0370-2693(81)90976-X}}.

\bibitem{Csernai:1986qf}
L.~P. Csernai, J.~I. Kapusta, {Entropy and Cluster Production in Nuclear
  Collisions}, Phys. Rept. 131 (1986) 223--318.
\newblock \href {http://dx.doi.org/10.1016/0370-1573(86)90031-1}
  {\path{doi:10.1016/0370-1573(86)90031-1}}.

\bibitem{ZHANG2018191}
Z.~Zhang, C.~M. Ko, Hypertriton production in relativistic heavy ion
  collisions, Phys. Lett. B 780 (2018) 191 -- 195.
\newblock \href
  {http://dx.doi.org/https://doi.org/10.1016/j.physletb.2018.03.003}
  {\path{doi:https://doi.org/10.1016/j.physletb.2018.03.003}}.

\bibitem{Shah:2015oha}
N.~Shah, Y.~G. Ma, J.~H. Chen, S.~Zhang, {Production of Multistrange Hadrons,
  Light Nuclei and Hypertriton in Central Au + Au Collisions at
  $\sqrt{s_{NN}}=$ 11.5 and 200 GeV}, Phys. Lett. B754 (2016) 6--10.
\newblock \href {http://dx.doi.org/10.1016/j.physletb.2016.01.005}
  {\path{doi:10.1016/j.physletb.2016.01.005}}.

\bibitem{Zhu:2015voa}
L.~Zhu, C.~M. Ko, X.~Yin, {Light (Anti-)Nuclei Production and Flow in
  Relativistic Heavy-Ion Collisions}, Phys. Rev. C92~(6) (2015) 064911.
\newblock \href {http://dx.doi.org/10.1103/PhysRevC.92.064911}
  {\path{doi:10.1103/PhysRevC.92.064911}}.

\bibitem{Mrowczynski:1989jd}
S.~Mrowczynski, {Antideuteron Production and the Size of the Interaction Zone},
  Phys. Lett. B248 (1990) 459--463.
\newblock \href {http://dx.doi.org/10.1016/0370-2693(90)90322-W}
  {\path{doi:10.1016/0370-2693(90)90322-W}}.

\bibitem{Leupold:1993ms}
S.~Leupold, U.~W. Heinz, {Coalescence Model for Deuterons and Antideuterons in
  Relativistic Heavy-Ion Collisions}, Phys. Rev. C50 (1994) 1110--1128.
\newblock \href {http://dx.doi.org/10.1103/PhysRevC.50.1110}
  {\path{doi:10.1103/PhysRevC.50.1110}}.

\bibitem{Bleicher:1995dw}
M.~Bleicher, C.~Spieles, A.~Jahns, R.~Mattiello, H.~Sorge, H.~St{\"o}cker,
  W.~Greiner, {Phase Space Correlations of Antideuterons in Heavy-Ion
  Collisions}, Phys. Lett. B361 (1995) 10--13.
\newblock \href {http://dx.doi.org/10.1016/0370-2693(95)01159-N}
  {\path{doi:10.1016/0370-2693(95)01159-N}}.

\bibitem{Scheibl:1998tk}
R.~Scheibl, U.~W. Heinz, {Coalescence and Flow in Ultrarelativistic Heavy Ion
  Collisions}, Phys. Rev. C59 (1999) 1585--1602.
\newblock \href {http://dx.doi.org/10.1103/PhysRevC.59.1585}
  {\path{doi:10.1103/PhysRevC.59.1585}}.

\bibitem{Lisa:2005dd}
M.~A. Lisa, S.~Pratt, R.~Soltz, U.~Wiedemann, {Femtoscopy in Relativistic Heavy
  Ion Collisions}, Ann. Rev. Nucl. Part. Sci. 55 (2005) 357--402.
\newblock \href {http://dx.doi.org/10.1146/annurev.nucl.55.090704.151533}
  {\path{doi:10.1146/annurev.nucl.55.090704.151533}}.

\bibitem{Appelquist:1996qy}
G.~Appelquist, et~al., {Antinuclei Production in Pb + Pb Collisions at 158$A$
  GeV/$c$}, Phys. Lett. B376 (1996) 245--250.
\newblock \href {http://dx.doi.org/10.1016/0370-2693(96)00415-7}
  {\path{doi:10.1016/0370-2693(96)00415-7}}.

\bibitem{Ambrosini:1997bf}
G.~Ambrosini, et~al., {Baryon and Antibaryon Production in Lead-Lead Collisions
  at 158$A$ GeV/$c$}, Phys. Lett. B417 (1998) 202--210.
\newblock \href {http://dx.doi.org/10.1016/S0370-2693(97)01383-X}
  {\path{doi:10.1016/S0370-2693(97)01383-X}}.

\bibitem{Armstrong:2000gd}
T.~A. Armstrong, et~al., {Antideuteron Yield at the AGS and Coalescence
  Implications}, Phys. Rev. Lett. 85 (2000) 2685--2688.
\newblock \href {http://dx.doi.org/10.1103/PhysRevLett.85.2685}
  {\path{doi:10.1103/PhysRevLett.85.2685}}.

\bibitem{Bearden:2000we}
I.~G. Bearden, et~al., {Antideuteron Production in 158$A$ GeV/$c$ Pb + Pb
  Collisions}, Phys. Rev. Lett. 85 (2000) 2681--2684.
\newblock \href {http://dx.doi.org/10.1103/PhysRevLett.85.2681}
  {\path{doi:10.1103/PhysRevLett.85.2681}}.

\bibitem{Armstrong:2000gz}
T.~A. Armstrong, et~al., {Measurements of Light Nuclei Production in 11.5$A$
  GeV/$c$ Au + Pb Heavy-Ion Collisions}, Phys. Rev. C61 (2000) 064908.
\newblock \href {http://dx.doi.org/10.1103/PhysRevC.61.064908}
  {\path{doi:10.1103/PhysRevC.61.064908}}.

\bibitem{Adler:2001bp}
C.~Adler, et~al., {Midrapidity Antiproton to Proton Ratio from Au + Au
  Collisions at $\sqrt{s_{NN}} = 130$ GeV}, Phys. Rev. Lett. 86 (2001) 4778,
  [Erratum: Phys. Rev. Lett. 90 (2003) 119903].
\newblock \href {http://dx.doi.org/10.1103/PhysRevLett.86.4778}
  {\path{doi:10.1103/PhysRevLett.86.4778}}.

\bibitem{Adler:2001uy}
C.~Adler, et~al., {Antideuteron and Anti-$^3$He Production in $\sqrt{s_{NN}} =
  130$ GeV Au + Au Collisions}, Phys. Rev. Lett. 87 (2001) 262301, [Erratum:
  Phys. Rev. Lett. 87 (2001) 279902].
\newblock \href {http://dx.doi.org/10.1103/PhysRevLett.87.262301}
  {\path{doi:10.1103/PhysRevLett.87.262301}}.

\bibitem{Agakishiev:2011ib}
H.~Agakishiev, et~al., {Observation of the Antimatter Helium-4 Nucleus}, Nature
  473 (2011) 353, [Erratum: Nature 475 (2011) 412].
\newblock \href {http://dx.doi.org/10.1038/nature10079}
  {\path{doi:10.1038/nature10079}}.

\bibitem{Andronic:2010qu}
A.~Andronic, P.~Braun-Munzinger, J.~Stachel, H.~St{\"o}cker, {Production of
  Light Nuclei, Hypernuclei and Their Antiparticles in Relativistic Nuclear
  Collisions}, Phys. Lett. B697 (2011) 203--207.
\newblock \href {http://dx.doi.org/10.1016/j.physletb.2011.01.053}
  {\path{doi:10.1016/j.physletb.2011.01.053}}.

\bibitem{Acharya:2017fvb}
S.~Acharya, et~al., {Production of deuterons, tritons, $^{3}$He nuclei and
  their antinuclei in pp collisions at $\mathbf{\sqrt{{\textit s}}}$ = 0.9,
  2.76 and 7 TeV}, Phys. Rev. C97~(2) (2018) 024615.
\newblock \href {http://dx.doi.org/10.1103/PhysRevC.97.024615}
  {\path{doi:10.1103/PhysRevC.97.024615}}.

\bibitem{Greiner:2010zzb}
W.~Greiner, {New Dimensions of the Periodic System: Superheavy, Superneutronic,
  Superstrange, Antimatter nuclei}, AIP Conf. Proc. 1323 (2010) 109--118.
\newblock \href {http://dx.doi.org/10.1063/1.3537839}
  {\path{doi:10.1063/1.3537839}}.

\bibitem{Greiner:1985ce}
W.~Greiner, B.~Muller, J.~Rafelski, {Quantum Electrodynamics of Strong Fields},
  Springer, Berlin, 1985.

\bibitem{Muller:1984zz}
U.~M{\"u}ller, G.~Soff, J.~Reinhardt, T.~de~Reus, B.~M{\"u}ller, W.~Greiner,
  {Electron Emission and Positron Production in Deep Inelastic Heavy-Ion
  Reactions}, Phys. Rev. C30 (1984) 1199--1207.
\newblock \href {http://dx.doi.org/10.1103/PhysRevC.30.1199}
  {\path{doi:10.1103/PhysRevC.30.1199}}.

\bibitem{Greiner:2003xg}
W.~Greiner, {Superheavy Matter, Strange Matter, Antimatter and the Structure of
  the Vacuum}, Acta Phys. Hung. A17 (2003) 357--367.
\newblock \href {http://dx.doi.org/10.1556/APH.17.2003.2-4.19}
  {\path{doi:10.1556/APH.17.2003.2-4.19}}.

\bibitem{Auerbach:1986wrs}
N.~Auerbach, A.~S. Goldhaber, M.~B. Johnson, L.~D. Miller, A.~Picklesimer,
  {Probing for Dirac Vacuum Structure in Nuclei}, Phys. Lett. B182 (1986)
  221--225.
\newblock \href {http://dx.doi.org/10.1016/0370-2693(86)90078-X}
  {\path{doi:10.1016/0370-2693(86)90078-X}}.

\bibitem{Sorge:1989dy}
H.~Sorge, H.~St{\"o}cker, W.~Greiner, {Poincare Invariant Hamiltonian Dynamics:
  Modeling Multi-Hadronic Interactions in a Phase Space Approach}, Annals Phys.
  192 (1989) 266--306.
\newblock \href {http://dx.doi.org/10.1016/0003-4916(89)90136-X}
  {\path{doi:10.1016/0003-4916(89)90136-X}}.

\bibitem{Sorge:1990fw}
H.~Sorge, A.~von Keitz, R.~Mattiello, H.~St{\"o}cker, W.~Greiner, {String
  dynamics in hadronic matter}, Z. Phys. C47 (1990) 629--634.
\newblock \href {http://dx.doi.org/10.1007/BF01552329}
  {\path{doi:10.1007/BF01552329}}.

\bibitem{Sorge:1992ej}
H.~Sorge, M.~Berenguer, H.~St{\"o}cker, W.~Greiner, {Color Rope Formation and
  Strange Baryon Production in Ultrarelativistic Heavy Ion Collisions}, Phys.
  Lett. B289 (1992) 6--11.
\newblock \href {http://dx.doi.org/10.1016/0370-2693(92)91353-B}
  {\path{doi:10.1016/0370-2693(92)91353-B}}.

\bibitem{Bass:1998ca}
S.~A. Bass, et~al., {Microscopic Models for Ultrarelativistic Heavy Ion
  Collisions}, Prog. Part. Nucl. Phys. 41 (1998) 255--369, [Prog. Part. Nucl.
  Phys.41,225(1998)].
\newblock \href {http://dx.doi.org/10.1016/S0146-6410(98)00058-1}
  {\path{doi:10.1016/S0146-6410(98)00058-1}}.

\bibitem{Bleicher:1999xi}
M.~Bleicher, et~al., {Relativistic Hadron-Hadron Collisions in the
  Ultrarelativistic Quantum Molecular Dynamics Model}, J. Phys. G25 (1999)
  1859--1896.
\newblock \href {http://dx.doi.org/10.1088/0954-3899/25/9/308}
  {\path{doi:10.1088/0954-3899/25/9/308}}.

\bibitem{Lin:2004en}
Z.-W. Lin, C.~M. Ko, B.-A. Li, B.~Zhang, S.~Pal, {A Multi-Phase Transport Model
  for Relativistic Heavy Ion Collisions}, Phys. Rev. C72 (2005) 064901.
\newblock \href {http://dx.doi.org/10.1103/PhysRevC.72.064901}
  {\path{doi:10.1103/PhysRevC.72.064901}}.

\bibitem{Wheaton:2004qb}
S.~Wheaton, J.~Cleymans, {THERMUS: A Thermal Model Package for ROOT}, Comput.
  Phys. Commun. 180 (2009) 84--106.
\newblock \href {http://dx.doi.org/10.1016/j.cpc.2008.08.001}
  {\path{doi:10.1016/j.cpc.2008.08.001}}.

\bibitem{Torrieri:2004zz}
G.~Torrieri, S.~Steinke, W.~Broniowski, W.~Florkowski, J.~Letessier,
  J.~Rafelski, {SHARE: Statistical Hadronization with Resonances}, Comput.
  Phys. Commun. 167 (2005) 229--251.
\newblock \href {http://dx.doi.org/10.1016/j.cpc.2005.01.004}
  {\path{doi:10.1016/j.cpc.2005.01.004}}.

\bibitem{Torrieri:2006xi}
G.~Torrieri, S.~Jeon, J.~Letessier, J.~Rafelski, {SHAREv2: Fluctuations and a
  Comprehensive Treatment of Decay Feed-Down}, Comput. Phys. Commun. 175 (2006)
  635--649.
\newblock \href {http://dx.doi.org/10.1016/j.cpc.2006.07.010}
  {\path{doi:10.1016/j.cpc.2006.07.010}}.

\bibitem{Petran:2013dva}
M.~Petran, J.~Letessier, J.~Rafelski, G.~Torrieri, {SHARE with CHARM}, Comput.
  Phys. Commun. 185 (2014) 2056--2079.
\newblock \href {http://dx.doi.org/10.1016/j.cpc.2014.02.026}
  {\path{doi:10.1016/j.cpc.2014.02.026}}.

\bibitem{Cleymans:2011pe}
J.~Cleymans, S.~Kabana, I.~Kraus, H.~Oeschler, K.~Redlich, N.~Sharma,
  {Antimatter Production in Proton-Proton and Heavy-Ion Collisions at
  Ultrarelativistic Energies}, Phys. Rev. C84 (2011) 054916.
\newblock \href {http://dx.doi.org/10.1103/PhysRevC.84.054916}
  {\path{doi:10.1103/PhysRevC.84.054916}}.

\bibitem{BraunMunzinger:2003zd}
P.~Braun-Munzinger, K.~Redlich, J.~Stachel, {Particle production in heavy ion
  collisions} (2003) 491--599{, Published in Hwa, R.C. (ed.) et al.: Quark
  Gluon Plasma}.
\newblock \href {http://dx.doi.org/10.1142/9789812795533_0008}
  {\path{doi:10.1142/9789812795533_0008}}.

\bibitem{Andronic:2016nof}
A.~Andronic, P.~Braun-Munzinger, K.~Redlich, J.~Stachel, {Hadron Yields, the
  Chemical Freeze-out and the QCD Phase Diagram}, J. Phys. Conf. Ser. 779~(1)
  (2017) 012012.
\newblock \href {http://dx.doi.org/10.1088/1742-6596/779/1/012012}
  {\path{doi:10.1088/1742-6596/779/1/012012}}.

\bibitem{Sun:2015ulc}
K.-J. Sun, L.-W. Chen, {Antimatter $^4_{\Lambda}$H Hypernucleus Production and
  the $^3_{\Lambda}$H/$^3$He Puzzle in Relativistic Heavy-Ion Collisions},
  Phys. Rev. C93 (2016) 064909.
\newblock \href {http://dx.doi.org/10.1103/PhysRevC.93.064909}
  {\path{doi:10.1103/PhysRevC.93.064909}}.

\bibitem{Albergo:2002gi}
S.~Albergo, et~al., {Light nuclei production in heavy ion collisions at
  relativistic energies}, Phys. Rev. C65 (2002) 034907.
\newblock \href {http://dx.doi.org/10.1103/PhysRevC.65.034907}
  {\path{doi:10.1103/PhysRevC.65.034907}}.

\bibitem{Anticic:2004yj}
T.~Anticic, et~al., {Energy and centrality dependence of deuteron and proton
  production in Pb + Pb collisions at relativistic energies}, Phys. Rev. C69
  (2004) 024902.
\newblock \href {http://dx.doi.org/10.1103/PhysRevC.69.024902}
  {\path{doi:10.1103/PhysRevC.69.024902}}.

\bibitem{Adler:2004uy}
S.~S. Adler, et~al., {Deuteron and Antideuteron Production in Au + Au
  Collisions at $\sqrt{s_{NN}} = 200$ GeV}, Phys. Rev. Lett. 94 (2005) 122302.
\newblock \href {http://dx.doi.org/10.1103/PhysRevLett.94.122302}
  {\path{doi:10.1103/PhysRevLett.94.122302}}.

\bibitem{Acharya:2017dmc}
S.~Acharya, et~al., {Measurement of deuteron spectra and elliptic flow in Pb-Pb
  collisions at $\sqrt{s_{\mathrm {NN}}}$ = 2.76 TeV at the LHC}, Eur. Phys. J.
  C77~(10) (2017) 658.
\newblock \href {http://dx.doi.org/10.1140/epjc/s10052-017-5222-x}
  {\path{doi:10.1140/epjc/s10052-017-5222-x}}.

\bibitem{Steinheimer:2012tb}
J.~Steinheimer, K.~Gudima, A.~Botvina, I.~Mishustin, M.~Bleicher,
  H.~St{\"o}cker, {Hypernuclei, Dibaryon and Antinuclei Production in High
  Energy Heavy Ion Collisions: Thermal Production Versus Coalescence}, Phys.
  Lett. B714 (2012) 85--91.
\newblock \href {http://dx.doi.org/10.1016/j.physletb.2012.06.069}
  {\path{doi:10.1016/j.physletb.2012.06.069}}.

\bibitem{Xue:2012gx}
L.~Xue, Y.~G. Ma, J.~H. Chen, S.~Zhang, {Production of Light (Anti)nuclei,
  (Anti)hypertriton and Di-$\Lambda$ in Central Au+Au Collisions at Energies
  Available at the BNL Relativistic Heavy Ion Collider}, Phys. Rev. C85 (2012)
  064912, [Erratum: Phys. Rev. C92 (2015) 059901].
\newblock \href {http://dx.doi.org/10.1103/PhysRevC.92.059901}
  {\path{doi:10.1103/PhysRevC.92.059901}}.

\bibitem{Liu:2006my}
H.-D. Liu, Z.~Xu, {Universal Antibaryon Density in $e^+e^-$, $\gamma p$, $pp$,
  $pA$ and $AA$ Collisions. }\href {http://arxiv.org/abs/nucl-ex/0610035}
  {\path{arXiv:nucl-ex/0610035}}.

\bibitem{Steinheimer:2013lza}
J.~Steinheimer, Z.~Xu, P.~Rau, C.~Sturm, H.~St{\"o}cker, {From d-Bars to
  Antimatter and Hyperclusters}, in: {Exciting Interdisciplinary Physics:
  Quarks and Gluons / Atomic Nuclei / Relativity and Cosmology / Biological
  Systems}, 2013, pp. 275--289.
\newblock \href {http://dx.doi.org/10.1007/978-3-319-00047-3_24}
  {\path{doi:10.1007/978-3-319-00047-3_24}}.

\bibitem{Armstrong:1999xw}
T.~A. Armstrong, et~al., {Mass Dependence of Light Nucleus Production in
  Ultrarelativistic Heavy-Ion Collisions}, Phys. Rev. Lett. 83 (1999)
  5431--5434.
\newblock \href {http://dx.doi.org/10.1103/PhysRevLett.83.5431}
  {\path{doi:10.1103/PhysRevLett.83.5431}}.

\bibitem{Alexopoulos:2000jk}
T.~Alexopoulos, et~al., {Cross Sections for Deuterium, Tritium, and Helium
  Production in $\bar{p}p$ Collisions at $\sqrt{s} = 1.8$ TeV}, Phys. Rev. D62
  (2000) 072004.
\newblock \href {http://dx.doi.org/10.1103/PhysRevD.62.072004}
  {\path{doi:10.1103/PhysRevD.62.072004}}.

\bibitem{Albrecht:1989ag}
H.~Albrecht, et~al., {Study of Antideuteron Production in $e^+ e^-$
  Annihilation at 10 GeV Center-of-Mass Energy}, Phys. Lett. B236 (1990)
  102--108.
\newblock \href {http://dx.doi.org/10.1016/0370-2693(90)90602-3}
  {\path{doi:10.1016/0370-2693(90)90602-3}}.

\bibitem{Aktas:2004pq}
A.~Aktas, et~al., {Measurement of Anti-Deuteron Photoproduction and a Search
  for Heavy Stable Charged Particles at HERA}, Eur. Phys. J. C36 (2004)
  413--423.
\newblock \href {http://dx.doi.org/10.1140/epjc/s2004-01978-x}
  {\path{doi:10.1140/epjc/s2004-01978-x}}.

\bibitem{Schael:2006fd}
S.~Schael, et~al., {Deuteron and Anti-Deuteron Production in $e^+ e^-$
  Collisions at the $Z$ Resonance}, Phys. Lett. B639 (2006) 192--201.
\newblock \href {http://dx.doi.org/10.1016/j.physletb.2006.06.043}
  {\path{doi:10.1016/j.physletb.2006.06.043}}.

\bibitem{Rochester:1947}
G.~D. Rochester, C.~C. Butler, Evidence for the existence of new unstable
  elementary particles, Nature 160 (1947) 855--857.
\newblock \href {http://dx.doi.org/10.1038/160855a0}
  {\path{doi:10.1038/160855a0}}.

\bibitem{Seriff:1950}
A.~J. Seriff, R.~B. Leighton, C.~Hsiao, E.~W. Cowan, C.~D. Anderson,
  Cloud-chamber observations of the new unstable cosmic-ray particles, Phys.
  Rev. 78 (1950) 290--291.
\newblock \href {http://dx.doi.org/10.1103/PhysRev.78.290}
  {\path{doi:10.1103/PhysRev.78.290}}.

\bibitem{Hopper:1950}
V.~D. Hopper, S.~Biswas, Evidence concerning the existence of the new unstable
  elementary neutral particle, Phys. Rev. 80 (1950) 1099--1100.
\newblock \href {http://dx.doi.org/10.1103/PhysRev.80.1099}
  {\path{doi:10.1103/PhysRev.80.1099}}.

\bibitem{Barnes:1964}
V.~E. Barnes, et~al., Observation of a hyperon with strangeness minus three,
  Phys. Rev. Lett. 12 (1964) 204--206.
\newblock \href {http://dx.doi.org/10.1103/PhysRevLett.12.204}
  {\path{doi:10.1103/PhysRevLett.12.204}}.

\bibitem{Danysz:1953}
M.~Danysz, J.~Pniewski, {Delayed Disintegration of a Heavy Nuclear Fragment:
  I}, Phil. Mag. 44 (1953) 348.
\newblock \href {http://dx.doi.org/10.1080/14786440308520318}
  {\path{doi:10.1080/14786440308520318}}.

\bibitem{Alberico:2001jb}
W.~M. Alberico, G.~Garbarino, {Weak decay of $\Lambda$ Hypernuclei}, Phys.
  Rept. 369 (2002) 1--109.
\newblock \href {http://dx.doi.org/10.1016/S0370-1573(02)00199-0}
  {\path{doi:10.1016/S0370-1573(02)00199-0}}.

\bibitem{Chen:2009ku}
J.~H. Chen, {Observation of Hypertritons in Au + Au Collisions at
  $\sqrt{s_{NN}} = 200$ GeV}, Nucl. Phys. A830 (2009) 761C--764C.
\newblock \href {http://dx.doi.org/10.1016/j.nuclphysa.2009.10.001}
  {\path{doi:10.1016/j.nuclphysa.2009.10.001}}.

\bibitem{Chen:2010zzc}
J.~H. Chen, {Hypernucleus Physics at RHIC}, Nucl. Phys. A835 (2010) 117--120.
\newblock \href {http://dx.doi.org/10.1016/j.nuclphysa.2010.01.183}
  {\path{doi:10.1016/j.nuclphysa.2010.01.183}}.

\bibitem{Abelev:2010rv}
B.~I. Abelev, et~al., {Observation of an Antimatter Hypernucleus}, Science 328
  (2010) 58--62.
\newblock \href {http://dx.doi.org/10.1126/science.1183980}
  {\path{doi:10.1126/science.1183980}}.

\bibitem{Adam:2015yta}
J.~Adam, et~al., {$^{3}_{\Lambda}\mathrm H$ and $^{3}_{\bar{\Lambda}}
  \overline{\mathrm H}$ Production in Pb-Pb collisions at $\sqrt{s_{\rm NN}} =$
  2.76 TeV}, Phys. Lett. B754 (2016) 360--372.
\newblock \href {http://dx.doi.org/10.1016/j.physletb.2016.01.040}
  {\path{doi:10.1016/j.physletb.2016.01.040}}.

\bibitem{BraunMunzinger:2007zz}
P.~Braun-Munzinger, J.~Stachel, {The Quest for the Quark-Gluon Plasma}, Nature
  448 (2007) 302--309.
\newblock \href {http://dx.doi.org/10.1038/nature06080}
  {\path{doi:10.1038/nature06080}}.

\bibitem{Koch:1986ud}
P.~Koch, B.~Muller, J.~Rafelski, {Strangeness in Relativistic Heavy-Ion
  Collisions}, Phys. Rept. 142 (1986) 167--262.
\newblock \href {http://dx.doi.org/10.1016/0370-1573(86)90096-7}
  {\path{doi:10.1016/0370-1573(86)90096-7}}.

\bibitem{Kamada:1997rv}
H.~Kamada, J.~Golak, K.~Miyagawa, H.~Witala, W.~Gloeckle, {$\pi$-Mesonic Decay
  of the Hypertriton}, Phys. Rev. C57 (1998) 1595--1603.
\newblock \href {http://dx.doi.org/10.1103/PhysRevC.57.1595}
  {\path{doi:10.1103/PhysRevC.57.1595}}.

\bibitem{Heinz:1985pm}
U.~W. Heinz, P.~R. Subramanian, H.~St{\"o}cker, W.~Greiner, {Formation of
  Antimatter Clusters in the Hadronization Phase Transition}, J. Phys. G12
  (1986) 1237.
\newblock \href {http://dx.doi.org/10.1088/0305-4616/12/11/013}
  {\path{doi:10.1088/0305-4616/12/11/013}}.

\bibitem{Greiner:1988pc}
C.~Greiner, D.-H. Rischke, H.~St{\"o}cker, P.~Koch, {The Creation of Strange
  Quark Matter Droplets as a Unique Signature for Quark-Gluon Plasma Formation
  in Relativistic Heavy Ion Collisions}, Phys. Rev. D38 (1988) 2797--2807.
\newblock \href {http://dx.doi.org/10.1103/PhysRevD.38.2797}
  {\path{doi:10.1103/PhysRevD.38.2797}}.

\bibitem{Lattimer:2006xb}
J.~M. Lattimer, M.~Prakash, {Neutron Star Observations: Prognosis for Equation
  of State Constraints}, Phys. Rept. 442 (2007) 109--165.
\newblock \href {http://dx.doi.org/10.1016/j.physrep.2007.02.003}
  {\path{doi:10.1016/j.physrep.2007.02.003}}.

\bibitem{Burgio:2011wt}
G.~F. Burgio, H.~J. Schulze, A.~Li, {Hyperon Stars at Finite Temperature in the
  Brueckner Theory}, Phys. Rev. C83 (2011) 025804.
\newblock \href {http://dx.doi.org/10.1103/PhysRevC.83.025804}
  {\path{doi:10.1103/PhysRevC.83.025804}}.

\bibitem{Lonardoni:2014bwa}
D.~Lonardoni, A.~Lovato, S.~Gandolfi, F.~Pederiva, {Hyperon Puzzle: Hints from
  Quantum Monte Carlo Calculations}, Phys. Rev. Lett. 114~(9) (2015) 092301.
\newblock \href {http://dx.doi.org/10.1103/PhysRevLett.114.092301}
  {\path{doi:10.1103/PhysRevLett.114.092301}}.

\bibitem{Demorest:2010bx}
P.~B. Demorest, T.~Pennucci, S.~M. Ransom, M.~S.~E. Roberts, J.~W.~T. Hessels,
  {A Two-Solar-Mass Neutron Star Measured Using Shapiro Delay}, Nature 467
  (2010) 1081--1083.
\newblock \href {http://dx.doi.org/10.1038/nature09466}
  {\path{doi:10.1038/nature09466}}.

\bibitem{Antoniadis:2013pzd}
J.~Antoniadis, et~al., {A Massive Pulsar in a Compact Relativistic Binary},
  Science 340 (2013) 6131.
\newblock \href {http://dx.doi.org/10.1126/science.1233232}
  {\path{doi:10.1126/science.1233232}}.

\bibitem{Dexheimer:2008ax}
V.~Dexheimer, S.~Schramm, {Proto-Neutron and Neutron Stars in a Chiral SU(3)
  Model}, Astrophys. J. 683 (2008) 943--948.
\newblock \href {http://dx.doi.org/10.1086/589735} {\path{doi:10.1086/589735}}.

\bibitem{Weissenborn:2011ut}
S.~Weissenborn, D.~Chatterjee, J.~Schaffner-Bielich, {Hyperons and Massive
  Neutron Stars: Vector Repulsion and SU(3) Symmetry}, Phys. Rev. C85 (2012)
  065802, [Erratum: Phys. Rev. C90 (2014) 019904].
\newblock \href {http://dx.doi.org/10.1103/PhysRevC.85.065802,
  10.1103/PhysRevC.90.019904} {\path{doi:10.1103/PhysRevC.85.065802,
  10.1103/PhysRevC.90.019904}}.

\bibitem{Steiner:2012rk}
A.~W. Steiner, M.~Hempel, T.~Fischer, {Core-Collapse Supernova Equations of
  State Based on Neutron Star Observations}, Astrophys. J. 774 (2013) 17.
\newblock \href {http://dx.doi.org/10.1088/0004-637X/774/1/17}
  {\path{doi:10.1088/0004-637X/774/1/17}}.

\bibitem{Lopes:2013cpa}
L.~L. Lopes, D.~P. Menezes, {Hypernuclear Matter in a Complete SU(3) Symmetry
  Group}, Phys. Rev. C89 (2014) 025805.
\newblock \href {http://dx.doi.org/10.1103/PhysRevC.89.025805}
  {\path{doi:10.1103/PhysRevC.89.025805}}.

\bibitem{Gusakov:2014ota}
M.~E. Gusakov, P.~Haensel, E.~M. Kantor, {Physics Input for Modelling
  Superfluid Neutron Stars with Hyperon Cores}, Mon. Not. Roy. Astron. Soc. 439
  (2014) 318--333.
\newblock \href {http://dx.doi.org/10.1093/mnras/stt2438}
  {\path{doi:10.1093/mnras/stt2438}}.

\bibitem{Gomes:2014aka}
R.~O. Gomes, V.~Dexheimer, S.~Schramm, C.~A.~Z. Vasconcellos, {Many-Body Forces
  in the Equation of State of Hyperonic Matter}, Astrophys. J. 808 (2015) 8.
\newblock \href {http://dx.doi.org/10.1088/0004-637X/808/1/8}
  {\path{doi:10.1088/0004-637X/808/1/8}}.

\bibitem{Dutra:2014qga}
M.~Dutra, O.~Louren\c{c}o, S.~S. Avancini, B.~V. Carlson, A.~Delfino, D.~P.
  Menezes, C.~Provid{\^e}ncia, S.~Typel, J.~R. Stone, {Relativistic Mean-Field
  Hadronic Models under Nuclear Matter Constraints}, Phys. Rev. C90~(5) (2014)
  055203.
\newblock \href {http://dx.doi.org/10.1103/PhysRevC.90.055203}
  {\path{doi:10.1103/PhysRevC.90.055203}}.

\bibitem{Bizarro:2015wxa}
D.~Bizarro, A.~Rabhi, C.~Provid{\^e}ncia, {Effect of the Symmetry Energy and
  Hyperon Interaction on Neutron Stars}\href {http://arxiv.org/abs/1502.04952}
  {\path{arXiv:1502.04952}}.

\bibitem{Rayet:1966}
M.~Rayet, R.~H. Dalitz, {The Lifetime of $^3$H$_\Lambda$}, Nuovo Cimento
  A(1971-1996)46 (1966) 786--794.
\newblock \href {http://dx.doi.org/10.1007/BF02857527}
  {\path{doi:10.1007/BF02857527}}.

\bibitem{Juric:1973zq}
M.~Juric, et~al., {A New Determination of the Binding-Energy Values of the
  Light Hypernuclei ($A \leq 15$)}, Nucl. Phys. B52 (1973) 1--30.
\newblock \href {http://dx.doi.org/10.1016/0550-3213(73)90084-9}
  {\path{doi:10.1016/0550-3213(73)90084-9}}.

\bibitem{Prem:1964}
R.~J. Prem, P.~H. Steinberg, {Lifetimes of Hypernuclei $_\Lambda{\rm H}^3$,
  $_\Lambda{\rm H}^4$, $_\Lambda{\rm He}^5$}, Phys. Rev. 136 (1964)
  B1803--B1806.
\newblock \href {http://dx.doi.org/10.1103/PhysRev.136.B1803}
  {\path{doi:10.1103/PhysRev.136.B1803}}.

\bibitem{Phillips:1969uy}
R.~E. Phillips, J.~Schneps, {Lifetimes of Light Hyperfragments. II}, Phys. Rev.
  180 (1969) 1307--1318.
\newblock \href {http://dx.doi.org/10.1103/PhysRev.180.1307}
  {\path{doi:10.1103/PhysRev.180.1307}}.

\bibitem{Bohm:1970se}
G.~Bohm, et~al., {On the Lifetime of the $^3_\Lambda$H Hypernucleus}, Nucl.
  Phys. B16 (1970) 46--52, [Erratum: Nucl. Phys. B16 (1970) 523].
\newblock \href {http://dx.doi.org/10.1016/0550-3213(70)90265-8,
  10.1016/0550-3213(70)90335-4} {\path{doi:10.1016/0550-3213(70)90265-8,
  10.1016/0550-3213(70)90335-4}}.

\bibitem{Keyes:1970ck}
G.~Keyes, M.~Derrick, T.~Fields, L.~G. Hyman, J.~G. Fetkovich, J.~McKenzie,
  B.~Riley, I.~T. Wang, {Properties of $_\Lambda$H$^3$}, Phys. Rev. D1 (1970)
  66--77.
\newblock \href {http://dx.doi.org/10.1103/PhysRevD.1.66}
  {\path{doi:10.1103/PhysRevD.1.66}}.

\bibitem{Keyes:1974ev}
G.~Keyes, J.~Sacton, J.~H. Wickens, M.~M. Block, {A Measurement of the Lifetime
  of the $^3_\Lambda$H Hypernucleus}, Nucl. Phys. B67 (1973) 269--283.
\newblock \href {http://dx.doi.org/10.1016/0550-3213(73)90197-1}
  {\path{doi:10.1016/0550-3213(73)90197-1}}.

\bibitem{Rappold:2013fic}
C.~Rappold, et~al., {Hypernuclear Spectroscopy of Products from $^6$Li
  Projectiles on a Carbon Target at 2$A$ GeV}, Nucl. Phys. A913 (2013)
  170--184.
\newblock \href {http://dx.doi.org/10.1016/j.nuclphysa.2013.05.019}
  {\path{doi:10.1016/j.nuclphysa.2013.05.019}}.

\bibitem{Adamczyk:2017buv}
L.~Adamczyk, et~al., {Measurement of $^3_\Lambda$H Lifetime in Au + Au
  Collisions at the Relativistic Heavy-Ion Collider~}\href
  {http://arxiv.org/abs/1710.00436} {\path{arXiv:1710.00436}}.

\bibitem{Saito:2016ulw}
T.~R. Saito, et~al., {Summary of the HypHI Phase 0 Experiment and Future Plans
  with FRS at GSI (FAIR Phase 0)}, Nucl. Phys. A954 (2016) 199--212.
\newblock \href {http://dx.doi.org/10.1016/j.nuclphysa.2016.05.011}
  {\path{doi:10.1016/j.nuclphysa.2016.05.011}}.

\bibitem{Rappold:2014jqa}
C.~Rappold, et~al., {On the Measured Lifetime of Light Hypernuclei
  $^3_\Lambda$H and $^4_\Lambda$H}, Phys. Lett. B728 (2014) 543--548.
\newblock \href {http://dx.doi.org/10.1016/j.physletb.2013.12.037}
  {\path{doi:10.1016/j.physletb.2013.12.037}}.

\bibitem{Bhamathi:1969ny}
G.~Bhamathi, K.~Prema, {A Note on the Decay Rates of $^3$H$_\Lambda$ and
  $^4$H$_\Lambda$}, Nuovo Cimento A63 (1969) 555--558.
\newblock \href {http://dx.doi.org/10.1007/BF02756232}
  {\path{doi:10.1007/BF02756232}}.

\bibitem{Mansour:1979}
H.~M.~M. Mansour, K.~Higgins, {The Decay Rate of $^3$H$_\Lambda$}, Nuovo
  Cimento A51 (1979) 180--186.
\newblock \href {http://dx.doi.org/10.1007/BF02775419}
  {\path{doi:10.1007/BF02775419}}.

\bibitem{Kolesnikov:1988uy}
N.~N. Kolesnikov, V.~A. Kopylov, {Meson Decays of Hypertritium}, Sov. Phys. J.
  31 (1988) 210--213.
\newblock \href {http://dx.doi.org/10.1007/BF00898225}
  {\path{doi:10.1007/BF00898225}}.

\bibitem{Congleton:1992kk}
J.~G. Congleton, {A Simple Model of the Hypertriton}, J. Phys. G18 (1992)
  339--357.
\newblock \href {http://dx.doi.org/10.1088/0954-3899/18/2/015}
  {\path{doi:10.1088/0954-3899/18/2/015}}.

\bibitem{DAVIS20053}
D.~H. Davis, {50 years of hypernuclear physics. I. The early experiments},
  Nucl. Phys. A754 (2005) 3--13.
\newblock \href {http://dx.doi.org/10.1016/j.nuclphysa.2005.01.002}
  {\path{doi:10.1016/j.nuclphysa.2005.01.002}}.

\bibitem{Dalitz:2005mc}
R.~H. Dalitz, {50 years of hypernuclear physics. II. The later years}, Nucl.
  Phys. A754 (2005) 14--24.
\newblock \href {http://dx.doi.org/10.1016/j.nuclphysa.2005.01.016}
  {\path{doi:10.1016/j.nuclphysa.2005.01.016}}.

\bibitem{Gal:2016boi}
A.~Gal, E.~V. Hungerford, D.~J. Millener, {Strangeness in Nuclear Physics},
  Rev. Mod. Phys. 88~(3) (2016) 035004.
\newblock \href {http://dx.doi.org/10.1103/RevModPhys.88.035004}
  {\path{doi:10.1103/RevModPhys.88.035004}}.

\bibitem{Agnello:2016jlx}
M.~Agnello, E.~Botta, T.~Bressani, S.~Bufalino, A.~Feliciello, {On the Use of
  the ($\pi^-,\,K^0$) Reaction on Nuclear Targets for the Precise Determination
  of the Lifetime of the Hydrogen Hyperisotopes and Other Neutron-Rich
  $\Lambda$-Hypernuclei}, Nucl. Phys. A954 (2016) 176--198.
\newblock \href {http://dx.doi.org/10.1016/j.nuclphysa.2016.04.033}
  {\path{doi:10.1016/j.nuclphysa.2016.04.033}}.

\bibitem{Liu:2017rjm}
P.~Liu, J.-H. Chen, Y.-G. Ma, S.~Zhang, {Production of Light Nuclei and
  Hypernuclei at High Intensity Accelerator Facility Energy Region}, Nucl. Sci.
  Tech. 28~(4) (2017) 55, [Erratum: Nucl. Sci. Tech. 28 (2017) 89].
\newblock \href {http://dx.doi.org/10.1007/s41365-017-0224-9,
  10.1007/s41365-017-0207-x} {\path{doi:10.1007/s41365-017-0224-9,
  10.1007/s41365-017-0207-x}}.

\bibitem{STAR:SN0598}
\href{https://drupal.star.bnl.gov/STAR/starnotes/public/sn0598}{{STAR
  Collaboration, Studying the Phase Diagram of QCD Matter at RHIC, Beam Energy
  Scan Phase-II Whitepaper}} STAR Note SN0598 (2014).
\newline\urlprefix\url{https://drupal.star.bnl.gov/STAR/starnotes/public/sn0598}

\bibitem{Senger:2017oqn}
P.~Senger, {QCD Matter Physics at FAIR}, Nucl. Phys. A967 (2017) 892--895.
\newblock \href {http://dx.doi.org/10.1016/j.nuclphysa.2017.06.056}
  {\path{doi:10.1016/j.nuclphysa.2017.06.056}}.

\bibitem{Steinheimer:2011hoa}
J.~Steinheimer, H.~St{\"o}cker, Z.~Xu, {From d-bars to Antimatter- \&
  Hyperclusters from the Stars to FAIR in Europe}, Pontif. Acad. Sci. Scr.
  Varia 119 (2011) 203--217.

\bibitem{Xue:2011ej}
L.~Xue, {Observation of the Antimatter Helium-4 Nucleus at RHIC}, J. Phys. G38
  (2011) 124072.
\newblock \href {http://dx.doi.org/10.1088/0954-3899/38/12/124072}
  {\path{doi:10.1088/0954-3899/38/12/124072}}.

\bibitem{Sharma:2011ya}
N.~Sharma, {Production of Nuclei and Antinuclei in $pp$ and Pb-Pb Collisions
  with ALICE at the LHC}, J. Phys. G38 (2011) 124189.
\newblock \href {http://dx.doi.org/10.1088/0954-3899/38/12/124189}
  {\path{doi:10.1088/0954-3899/38/12/124189}}.

\bibitem{Blum:2017}
K.~Blum, K.~C.~Y. Ng, R.~Sato, M.~Takimoto, Cosmic rays, antihelium, and an old
  navy spotlight, Phys. Rev. D 96 (2017) 103021.
\newblock \href {http://dx.doi.org/10.1103/PhysRevD.96.103021}
  {\path{doi:10.1103/PhysRevD.96.103021}}.

\bibitem{Blum:2017iwq}
K.~Blum, R.~Sato, E.~Waxman, {Cosmic-ray Antimatter}\href
  {http://arxiv.org/abs/1709.06507} {\path{arXiv:1709.06507}}.

\bibitem{Sun:2015jta}
K.-J. Sun, L.-W. Chen, {Production of Antimatter $^{5,6}$Li Nuclei in Central
  Au + Au Collisions at $\sqrt{s_{NN}} = 200$ GeV}, Phys. Lett. B751 (2015)
  272--277.
\newblock \href {http://dx.doi.org/10.1016/j.physletb.2015.10.056}
  {\path{doi:10.1016/j.physletb.2015.10.056}}.

\bibitem{Gabrielse:1999kc}
G.~Gabrielse, A.~Khabbaz, D.~S. Hall, C.~Heimann, H.~Kalinowsky, W.~Jhe,
  {Precision Mass Spectroscopy of the Antiproton and Proton Using
  Simultaneously Trapped Particles}, Phys. Rev. Lett. 82 (1999) 3198--3201.
\newblock \href {http://dx.doi.org/10.1103/PhysRevLett.82.3198}
  {\path{doi:10.1103/PhysRevLett.82.3198}}.

\bibitem{Hori:2011zz}
M.~Hori, et~al., {Two-Photon Laser Spectroscopy of Antiprotonic Helium and the
  Antiproton-to-Electron Mass Ratio}, Nature 475 (2011) 484--488.
\newblock \href {http://dx.doi.org/10.1038/nature10260}
  {\path{doi:10.1038/nature10260}}.

\bibitem{Akindinov:2015hva}
A.~Akindinov, et~al., {The Highest Precision Proof of CPT Invariance in the
  Nuclear Binding Masses}, Int. J. Mod. Phys. A30~(24) (2015) 1550170.
\newblock \href {http://dx.doi.org/10.1142/S0217751X15501705}
  {\path{doi:10.1142/S0217751X15501705}}.

\bibitem{Adam:2015pna}
J.~Adam, et~al., {Precision Measurement of the Mass Difference Between Light
  nuclei and Antinuclei}, Nature Phys. 11~(10) (2015) 811--814.
\newblock \href {http://dx.doi.org/10.1038/nphys3432}
  {\path{doi:10.1038/nphys3432}}.

\bibitem{Denisov:1971im}
S.~P. Denisov, S.~V. Donskov, {\relax Yu}.~P. Gorin, V.~A. Kachanov, V.~M.
  Kutjin, A.~I. Petrukhin, {\relax Yu}.~D. Prokoshkin, E.~A. Razuvaev, R.~S.
  Shuvalov, D.~A. Stojanova, {Measurements of Antideuteron Absorption and
  Stripping Cross Sections at the Momentum 13.3 GeV/$c$}, Nucl. Phys. B31
  (1971) 253--260.
\newblock \href {http://dx.doi.org/10.1016/0550-3213(71)90229-X}
  {\path{doi:10.1016/0550-3213(71)90229-X}}.

\bibitem{Kessler:1999zz}
E.~G. Kessler, Jr., M.~S. Dewey, R.~D. Deslattes, A.~Henins, H.~G. Borner,
  M.~Jentschel, C.~Doll, H.~Lehmann, {The Deuteron Binding Energy and the
  Neutron Mass}, Phys. Lett. A255 (1999) 221.
\newblock \href {http://dx.doi.org/10.1016/S0375-9601(99)00078-X}
  {\path{doi:10.1016/S0375-9601(99)00078-X}}.

\bibitem{Mohr:2012tt}
P.~J. Mohr, B.~N. Taylor, D.~B. Newell, {CODATA Recommended Values of the
  Fundamental Physical Constants: 2010}, Rev. Mod. Phys. 84 (2012) 1527--1605.
\newblock \href {http://dx.doi.org/10.1103/RevModPhys.84.1527}
  {\path{doi:10.1103/RevModPhys.84.1527}}.

\bibitem{Cresti:1986eg}
M.~Cresti, L.~Peruzzo, G.~Sartori, {Measurement of the Antineutron Mass}, Phys.
  Lett. B177 (1986) 206--210.
\newblock \href {http://dx.doi.org/10.1016/0370-2693(86)91058-0}
  {\path{doi:10.1016/0370-2693(86)91058-0}}.

\bibitem{DelAguila:1987ic}
F.~Del~Aguila, J.~A. Gonzalez, M.~Quiros, {String Goniometry by Neutral
  Currents}, Phys. Lett. B197 (1987) 89, [Erratum: Phys. Lett. B200 (1988)
  587].
\newblock \href {http://dx.doi.org/10.1016/0370-2693(87)90347-9}
  {\path{doi:10.1016/0370-2693(87)90347-9}}.

\bibitem{Adamczyk:2015hza}
L.~Adamczyk, et~al., {Measurement of Interaction between Antiprotons}, Nature
  527 (2015) 345--348.
\newblock \href {http://dx.doi.org/10.1038/nature15724}
  {\path{doi:10.1038/nature15724}}.

\bibitem{HanburyBrown:1954amm}
R.~Hanbury~Brown, R.~Q. Twiss, {A New Type of Interferometer for Use in Radio
  Astronomy}, Phil. Mag. Ser. 7 45 (1954) 663--682.

\bibitem{Ma:2015b}
Y.~G. Ma, D.~Q. Fang, S.~X. Yan, et~al., {Different mechanism of two-proton
  emission from proton-rich nuclei $^{23}\mathrm{Al}$ and $^{22}\mathrm{Mg}$},
  Phys. Lett. B 743 (2015) 306--309.
\newblock \href {http://dx.doi.org/10.1016/j.physletb.2015.02.066}
  {\path{doi:10.1016/j.physletb.2015.02.066}}.

\bibitem{Ma:2017c}
D.~Q. Fang, Y.~G. Ma, S.~X. Yan, et~al., {Proton-proton correlations in
  distinguishing the two-proton emission mechanism of $^{23}\mathrm{Al}$ and
  $^{22}\mathrm{Mg}$}, Phys. Rev. C 94 (2017) 044621.
\newblock \href {http://dx.doi.org/10.1103/PhysRevC.94.044621}
  {\path{doi:10.1103/PhysRevC.94.044621}}.

\bibitem{Yang:2016NST}
J.~Yang, W.~N. Zhang, {Interferometry analyses of pion and kaon for the
  granular sources for Au + Au collisions at 200 GeV}, Nucl. Sci. Tech. 27
  (2016) 147.
\newblock \href {http://dx.doi.org/10.1007/s41365-016-0145-z}
  {\path{doi:10.1007/s41365-016-0145-z}}.

\bibitem{Zbroszczyk:2008jja}
H.~Zbroszczyk,
  \href{http://drupal.star.bnl.gov/STAR/theses/ph-d/hanna-zbroszczyk}{{Studies
  of Baryon-Baryon Correlations in Relativistic Nuclear Collisions Registered
  at the STAR Experiment}}, Ph.D. thesis, Warsaw U. of Tech. (2008).
\newline\urlprefix\url{http://drupal.star.bnl.gov/STAR/theses/ph-d/hanna-zbroszczyk}

\bibitem{Lednicky:1981su}
R.~Lednicky, V.~L. Lyuboshits, {Final State Interaction Effect on Pairing
  Correlations Between Particles with Small Relative Momenta}, Sov. J. Nucl.
  Phys. 35 (1982) 770, [Yad. Fiz.35,1316(1981)].

\bibitem{Adams:2005ws}
J.~Adams, et~al., {Proton-$\Lambda$ Correlations in Central Au + Au Collisions
  at $\sqrt{s_{NN}}= 200$ GeV}, Phys. Rev. C74 (2006) 064906.
\newblock \href {http://dx.doi.org/10.1103/PhysRevC.74.064906}
  {\path{doi:10.1103/PhysRevC.74.064906}}.

\bibitem{Adamczyk:2014vca}
L.~Adamczyk, et~al., {$\Lambda\Lambda$ Correlation Function in Au + Au
  Collisions at $\sqrt{s_{NN}}=$ 200 GeV}, Phys. Rev. Lett. 114~(2) (2015)
  022301.
\newblock \href {http://dx.doi.org/10.1103/PhysRevLett.114.022301}
  {\path{doi:10.1103/PhysRevLett.114.022301}}.

\bibitem{Chojnacki:2011hb}
M.~Chojnacki, A.~Kisiel, W.~Florkowski, W.~Broniowski, {THERMINATOR 2: THERMal
  heavy IoN generATOR 2}, Comput. Phys. Commun. 183 (2012) 746--773.
\newblock \href {http://dx.doi.org/10.1016/j.cpc.2011.11.018}
  {\path{doi:10.1016/j.cpc.2011.11.018}}.

\bibitem{Mathelitsch:1984hq}
L.~Mathelitsch, B.~J. Verwest, {Effective Range Parameters in Nucleon-Nucleon
  Scattering}, Phys. Rev. C29 (1984) 739--746.
\newblock \href {http://dx.doi.org/10.1103/PhysRevC.29.739}
  {\path{doi:10.1103/PhysRevC.29.739}}.

\bibitem{Slaus89}
I.~\v{S}laus, Y.~Akaishi, H.~Tanaka, Neutron-neutron effective range
  parameters, Phys. Rep. 173 (1989) 257--300.
\newblock \href {http://dx.doi.org/10.1016/0370-1573(89)90127-0}
  {\path{doi:10.1016/0370-1573(89)90127-0}}.

\bibitem{Mott:1965}
N.~F. Mott, H.~S.~W. Massey, {Theory of Atomic Collisions, Third Edition},
  Oxford University Press, Oxford, United Kingdom, 1965.

\bibitem{Rom:2006}
T.~Rom, T.~Best, D.~Van~Oosten, U.~Schneider, S.~F{\"o}lling, B.~Paredes,
  I.~Bloch, {Free Fermion Antibunching in a Degenerate Atomic Fermi Gas
  Released from an Optical Lattice}, Nature 444 (2006) 733--736.
\newblock \href {http://dx.doi.org/10.1038/nature05319}
  {\path{doi:10.1038/nature05319}}.

\bibitem{Ambrosi:2017wek}
G.~Ambrosi, et~al., {Direct Detection of a Break in the Teraelectronvolt
  Cosmic-Ray Spectrum of Electrons and Positrons}, Nature 552 (2017) 63--66.
\newblock \href {http://dx.doi.org/10.1038/nature24475}
  {\path{doi:10.1038/nature24475}}.

\bibitem{Adriani:2014pza}
O.~Adriani, et~al., {The PAMELA Mission: Heralding a New Era in Precision
  Cosmic Ray Physics}, Phys. Rept. 544 (2014) 323--370.
\newblock \href {http://dx.doi.org/10.1016/j.physrep.2014.06.003}
  {\path{doi:10.1016/j.physrep.2014.06.003}}.

\bibitem{Aizu:1961zz}
H.~Aizu, Y.~Fujimoto, S.~Hasegawa, M.~Koshiba, I.~Mito, J.~Nishimura, K.~Yokoi,
  M.~Schein, {Heavy Nuclei in the Primary Cosmic Radiation at Prince Albert,
  Canada. II}, Phys. Rev. 121 (1961) 1206--1218.
\newblock \href {http://dx.doi.org/10.1103/PhysRev.121.1206}
  {\path{doi:10.1103/PhysRev.121.1206}}.

\bibitem{Smoot:1975bv}
G.~F. Smoot, A.~Buffington, C.~D. Orth, {Search for Cosmic Ray Antimatter},
  Phys. Rev. Lett. 35 (1975) 258--261.
\newblock \href {http://dx.doi.org/10.1103/PhysRevLett.35.258}
  {\path{doi:10.1103/PhysRevLett.35.258}}.

\bibitem{Badhwar:1978hz}
G.~D. Badhwar, R.~L. Golden, J.~L. Lacy, J.~E. Zipse, R.~R. Daniel, S.~A.
  Stephens, {Relative Abundance of Anti-Protons and Anti-Helium in the Primary
  Cosmic Radiation}, Nature 274 (1978) 137--139.
\newblock \href {http://dx.doi.org/10.1038/274137b0}
  {\path{doi:10.1038/274137b0}}.

\bibitem{Buffington:1981zz}
A.~Buffington, S.~M. Schindler, C.~R. Pennypacker, {A Measurement of the
  Cosmic-Ray Antiproton Flux and a Search for Antihelium}, Astrophys. J. 248
  (1981) 1179--1193.
\newblock \href {http://dx.doi.org/10.1086/159247} {\path{doi:10.1086/159247}}.

\bibitem{Golden:1997iz}
R.~L. Golden, S.~J. Stochaj, S.~A. Stephens, A.~A. Moiseev, J.~F. Ormes, R.~E.
  Streitmatter, T.~Bowen, A.~Moats, J.~Lloyd-Evans, {Search for Anti-Helium in
  the Cosmic Rays}, Astrophys. J. 479 (1997) 992--996.
\newblock \href {http://dx.doi.org/10.1086/303886} {\path{doi:10.1086/303886}}.

\bibitem{Alcaraz:2000ss}
J.~Alcaraz, et~al., {Search for Anti-Helium in Cosmic Rays}, Phys. Lett. B461
  (1999) 387--396.
\newblock \href {http://dx.doi.org/10.1016/S0370-2693(99)00874-6}
  {\path{doi:10.1016/S0370-2693(99)00874-6}}.

\bibitem{Aguilar:2013qda}
M.~Aguilar, et~al., {First Result from the Alpha Magnetic Spectrometer on the
  International Space Station: Precision Measurement of the Positron Fraction
  in Primary Cosmic Rays of 0.5-350 GeV}, Phys. Rev. Lett. 110 (2013) 141102.
\newblock \href {http://dx.doi.org/10.1103/PhysRevLett.110.141102}
  {\path{doi:10.1103/PhysRevLett.110.141102}}.

\bibitem{Adriani:2008zr}
O.~Adriani, et~al., {An Anomalous Positron Abundance in Cosmic Rays with
  Energies 1.5-100 GeV}, Nature 458 (2009) 607--609.
\newblock \href {http://dx.doi.org/10.1038/nature07942}
  {\path{doi:10.1038/nature07942}}.

\bibitem{Weidenspointner2018}
G.~Weidenspointner, G.~Skinner, P.~Jean, J.~Kn{\"o}dlseder, P.~von Ballmoos,
  G.~Bignami, R.~Diehl, A.~W. Strong, B.~Cordier, S.~Schanne, C.~Winkler, {An
  Asymmetric Distribution of Positrons in the Galactic Disk Revealed by
  $\gamma$-Rays}, Nature 451 (2018) 159--162.
\newblock \href {http://dx.doi.org/10.1038/nature06490}
  {\path{doi:10.1038/nature06490}}.

\bibitem{Serpico:2011wg}
P.~D. Serpico, {Astrophysical Models for the Origin of the Positron `Excess'},
  Astropart. Phys. 39-40 (2012) 2--11.
\newblock \href {http://dx.doi.org/10.1016/j.astropartphys.2011.08.007}
  {\path{doi:10.1016/j.astropartphys.2011.08.007}}.

\bibitem{Delahaye:2008ua}
T.~Delahaye, F.~Donato, N.~Fornengo, J.~Lavalle, R.~Lineros, P.~Salati,
  R.~Taillet, {Galactic Secondary Positron Flux at the Earth}, Astron.
  Astrophys. 501 (2009) 821--833.
\newblock \href {http://dx.doi.org/10.1051/0004-6361/200811130}
  {\path{doi:10.1051/0004-6361/200811130}}.

\bibitem{Moskalenko:1997gh}
I.~V. Moskalenko, A.~W. Strong, {Production and Propagation of Cosmic Ray
  Positrons and Electrons}, Astrophys. J. 493 (1998) 694--707.
\newblock \href {http://dx.doi.org/10.1086/305152} {\path{doi:10.1086/305152}}.

\bibitem{Aguilar:2014fea}
M.~Aguilar, et~al., {Precision Measurement of the $(e^+ + e^-)$ Flux in Primary
  Cosmic Rays from 0.5 GeV to 1 TeV with the Alpha Magnetic Spectrometer on the
  International Space Station}, Phys. Rev. Lett. 113 (2014) 221102.
\newblock \href {http://dx.doi.org/10.1103/PhysRevLett.113.221102}
  {\path{doi:10.1103/PhysRevLett.113.221102}}.

\bibitem{Abdollahi:2017nat}
S.~Abdollahi, et~al., {Cosmic-Ray Electron-Positron Spectrum from 7 GeV to 2
  TeV with the Fermi Large Area Telescope}, Phys. Rev. D95~(8) (2017) 082007.
\newblock \href {http://dx.doi.org/10.1103/PhysRevD.95.082007}
  {\path{doi:10.1103/PhysRevD.95.082007}}.

\bibitem{Aharonian:2008aa}
F.~Aharonian, et~al., {The Energy Spectrum of Cosmic-Ray Electrons at TeV
  Energies}, Phys. Rev. Lett. 101 (2008) 261104.
\newblock \href {http://dx.doi.org/10.1103/PhysRevLett.101.261104}
  {\path{doi:10.1103/PhysRevLett.101.261104}}.

\bibitem{Coombes:1976hi}
R.~Coombes, et~al., {Detection of $\pi-\mu$ Coulomb Bound States}, Phys. Rev.
  Lett. 37 (1976) 249--252.
\newblock \href {http://dx.doi.org/10.1103/PhysRevLett.37.249}
  {\path{doi:10.1103/PhysRevLett.37.249}}.

\bibitem{Aronson:1982bz}
S.~H. Aronson, R.~H. Bernstein, G.~J. Bock, R.~D. Cousins, J.~F. Greenhalgh,
  D.~Hedin, M.~Schwartz, T.~K. Shea, G.~B. Thomson, B.~Winstein, {Measurement
  of the Rate of Formation of $\pi-\mu$ Atoms in $K^0_L$ Decay}, Phys. Rev.
  Lett. 48 (1982) 1078--1081.
\newblock \href {http://dx.doi.org/10.1103/PhysRevLett.48.1078}
  {\path{doi:10.1103/PhysRevLett.48.1078}}.

\bibitem{Lamb1947}
W.~E. Lamb, R.~C. Retherford, {Fine Structure of the Hydrogen Atom by a
  Microwave Method}, Phys. Rev. 72 (1947) 241--243.
\newblock \href {http://dx.doi.org/10.1103/PhysRev.72.241}
  {\path{doi:10.1103/PhysRev.72.241}}.

\bibitem{Pohl:2010zza}
R.~Pohl, et~al., {The Size of the Proton}, Nature 466 (2010) 213--216.
\newblock \href {http://dx.doi.org/10.1038/nature09250}
  {\path{doi:10.1038/nature09250}}.

\bibitem{Antognini:1900ns}
A.~Antognini, et~al., {Proton Structure from the Measurement of $^2S$-$^2P$
  Transition Frequencies of Muonic Hydrogen}, Science 339 (2013) 417--420.
\newblock \href {http://dx.doi.org/10.1126/science.1230016}
  {\path{doi:10.1126/science.1230016}}.

\bibitem{Baym:1993ae}
G.~Baym, G.~Friedman, R.~J. Hughes, B.~V. Jacak, {Production of Muon-Meson
  Atoms in Ultrarelativistic Heavy Ion Collisions}, Phys. Rev. D48 (1993)
  R3957--R3959.
\newblock \href {http://dx.doi.org/10.1103/PhysRevD.48.R3957}
  {\path{doi:10.1103/PhysRevD.48.R3957}}.

\bibitem{Kapusta:1998fh}
J.~I. Kapusta, A.~Mocsy, {Hydrogen-Like Atoms from Ultrarelativistic Nuclear
  Collisions}, Phys. Rev. C59 (1999) 2937--2940.
\newblock \href {http://dx.doi.org/10.1103/PhysRevC.59.2937}
  {\path{doi:10.1103/PhysRevC.59.2937}}.

\bibitem{Xin:2015noa}
K.~Xin, {Search for Muonic Atoms at RHIC}, J. Phys. Conf. Ser. 612~(1) (2015)
  012032.
\newblock \href {http://dx.doi.org/10.1088/1742-6596/612/1/012032}
  {\path{doi:10.1088/1742-6596/612/1/012032}}.

\bibitem{Lednicky:1995vk}
R.~Lednicky, V.~L. Lyuboshits, B.~Erazmus, D.~Nouais, {How to Measure Which
  Sort of Particle Was Emitted Earlier and Which Later}, Phys. Lett. B373
  (1996) 30--34.
\newblock \href {http://dx.doi.org/10.1016/0370-2693(96)00124-4}
  {\path{doi:10.1016/0370-2693(96)00124-4}}.

\bibitem{Voloshin:1997jh}
S.~Voloshin, R.~Lednicky, S.~Panitkin, N.~Xu, {Relative Space-Time Asymmetries
  in Pion and Nucleon Production in Noncentral Nucleus-Nucleus Collisions at
  High Energies}, Phys. Rev. Lett. 79 (1997) 4766--4769.
\newblock \href {http://dx.doi.org/10.1103/PhysRevLett.79.4766}
  {\path{doi:10.1103/PhysRevLett.79.4766}}.

\bibitem{DIRAC:2016rpv}
B.~Adeva, et~al., {Observation of $\pi^- K^+$ and $\pi^+ K^-$ Atoms}, Phys.
  Rev. Lett. 117~(11) (2016) 112001.
\newblock \href {http://dx.doi.org/10.1103/PhysRevLett.117.112001}
  {\path{doi:10.1103/PhysRevLett.117.112001}}.

\bibitem{Huang:2013xta}
H.~Z. Huang, {STAR Upgrade Plan for the Coming Decade}, Nucl. Phys. A904-905
  (2013) 921--924.
\newblock \href {http://dx.doi.org/10.1016/j.nuclphysa.2013.02.165}
  {\path{doi:10.1016/j.nuclphysa.2013.02.165}}.

\bibitem{STAR:SN0644}
\href{https://drupal.star.bnl.gov/STAR/starnotes/public/sn0644}{{STAR
  Collaboration, Technical Design Report for the iTPC Upgrade}} STAR Note
  SN0644 (2015).
\newline\urlprefix\url{https://drupal.star.bnl.gov/STAR/starnotes/public/sn0644}

\bibitem{Martinengo:2017fuc}
P.~Martinengo, {The new Inner Tracking System of the ALICE experiment}, Nucl.
  Phys. A967 (2017) 900--903.
\newblock \href {http://dx.doi.org/10.1016/j.nuclphysa.2017.05.069}
  {\path{doi:10.1016/j.nuclphysa.2017.05.069}}.

\bibitem{Abelevetal:2014cna}
B.~Abelev, et~al., {Upgrade of the ALICE Experiment: Letter Of Intent}, J.
  Phys. G41 (2014) 087001.
\newblock \href {http://dx.doi.org/10.1088/0954-3899/41/8/087001}
  {\path{doi:10.1088/0954-3899/41/8/087001}}.

\bibitem{Abelevetal:2014dna}
B.~Abelev, et~al., {Technical Design Report for the Upgrade of the ALICE Inner
  Tracking System}, J. Phys. G41 (2014) 087002.
\newblock \href {http://dx.doi.org/10.1088/0954-3899/41/8/087002}
  {\path{doi:10.1088/0954-3899/41/8/087002}}.

\bibitem{Ma:2013xn}
Y.-G. Ma, J.-H. Chen, L.~Xue, {A Brief Review of Antimatter Production}, Front.
  Phys. (Beijing) 7 (2012) 637--646.
\newblock \href {http://dx.doi.org/10.1007/s11467-012-0273-9}
  {\path{doi:10.1007/s11467-012-0273-9}}.

\bibitem{Wu:2018}
B.~Wu, J.~C. Yang, J.~W. Xia, et~al., {The design of the Spectrometer Ring at
  the HIAF}, Nucl. Instrum. Meth. A 881 (2018) 27--35.
\newblock \href {http://dx.doi.org/https://doi.org/10.1016/j.nima.2017.08.017}
  {\path{doi:https://doi.org/10.1016/j.nima.2017.08.017}}.

\bibitem{Li:2017}
C.~Li, P.~Wen, J.~Li, et~al., Production of heavy neutron-rich nuclei with
  radiative beams in mutinucleon transfer reactions, Nucl. Sci. Tech. 28 (2017)
  110.
\newblock \href {http://dx.doi.org/10.1007/s41365-017-0266-z}
  {\path{doi:10.1007/s41365-017-0266-z}}.

\bibitem{Kekelidze:2017tgp}
V.~Kekelidze, A.~Kovalenko, R.~Lednicky, V.~Matveev, I.~Meshkov, A.~Sorin,
  G.~Trubnikov, {Feasibility Study of Heavy-Ion Collision Physics at NICA
  JINR}, Nucl. Phys. A967 (2017) 884--887.
\newblock \href {http://dx.doi.org/10.1016/j.nuclphysa.2017.06.031}
  {\path{doi:10.1016/j.nuclphysa.2017.06.031}}.

\bibitem{Sakaguchi:2017ggo}
T.~Sakaguchi, {Study of high baryon density QCD matter at J-PARC-HI}, Nucl.
  Phys. A967 (2017) 896--899.
\newblock \href {http://dx.doi.org/10.1016/j.nuclphysa.2017.05.081}
  {\path{doi:10.1016/j.nuclphysa.2017.05.081}}.

\end{thebibliography}
\end{document}